\documentclass[twocolumn]{aa}
\usepackage{graphicx}
\usepackage{rotating}
\usepackage{longtable}
\usepackage{subfigure}
\usepackage{lscape}
\usepackage{txfonts}
\begin{document}
   \title{{\it XMM-Newton} observations of the Lockman Hole IV: spectra
of the brightest AGN \thanks{Based on observations obtained with XMM-Newton, 
an ESA science mission with instruments and contributions directly funded by 
ESA Member States and NASA}
   } \titlerunning{X-ray spectra LH}

   \subtitle{}
   \author{
     S. Mateos  \inst{1,2}
    \and
     X. Barcons \inst{1}
    \and
     F. J. Carrera  \inst{1}
    \and
     M. T. Ceballos \inst{1}
    \and
    G. Hasinger \inst{3}
    \and
         I. Lehmann \inst{3}
    \and
         A. C. Fabian  \inst{4}
    \and
         A. Streblyanska \inst{3}
    }

%     \fnmsep\thanks{Just to show the usage
%          of the elements in the author field}
           \authorrunning{S. Mateos et al.}

   \offprints{S. Mateos, \email{sm279@star.le.ac.uk}}
   \institute{Instituto de F\'\i sica de Cantabria (CSIC-UC), 39005 Santander, Spain
     \and X-ray Astronomy Group, Department of Physics and Astronomy, Leicester University, Leicester LE1 7RH, UK
     \and Max-Planck-Institut f\"{u}r Extraterrestrische Physik, Giessenbachstrasse, Garching D-85748, Germany.
      \and Institute of Astronomy, University of Cambridge, Madingley Road, Cambridge CB3 OHA, UK. }
 \date{29 June 2005}

   \abstract{This paper presents the results of a detailed X-ray spectral analysis
     of a sample of 123 X-ray sources detected with {\it XMM-Newton}
in the {\it Lockman Hole} field.
This is the deepest observation carried out with {\it XMM-Newton} with
more that 600 ksec of good EPIC-pn data.
We have spectra with good signal to noise ($>$500 source counts) for all objects down to 0.2-12 keV fluxes of
${\sim 5\times10^{-15}\,{\rm erg\,cm^{-2}\,s^{-1}}}$ (flux limit of 
${\sim 6\times10^{-16}\,{\rm erg\,cm^{-2}\,s^{-1}}}$ in the 0.5-2 and 2-10 keV bands). At the time of the analysis, we
had optical spectroscopic identifications for 60\% of the
sources, 46 being optical type-1 AGN and 28 optical type-2 AGN.
Using a single power law model our sources' average spectral slope 
hardens at faint 0.5-2 keV fluxes but not at faint 2-10 keV fluxes. We have been able to explain
this effect in terms of an increase in X-ray absorption at faint fluxes.
We did not find in our data
any evidence for the existence of a population of faint intrinsically harder sources. The average
spectral slope of our sources is $\sim$1.9,
with an intrinsic dispersion of $\sim$0.28. 
We detected X-ray absorption (F-test significance $\ge$95\%) in 37\% of the sources,
$\sim$10\% in type-1 AGN (rest-frame ${\rm N_H \sim\,1.6\,\times 10^{21}-1.2\,\times 10^{22}\,cm^{-2}}$) 
and $\sim77\%$ (rest-frame ${\rm N_H \sim\,1.5\,\times10^{21}-4\,\times 10^{23}\,cm^{-2}}$) in type-2 AGN.
Using X-ray fluxes corrected for absorption,
the fraction of absorbed objects and the absorbing column density distribution did not
vary with X-ray flux.
Our type-1 and type-2 AGN do not appear to have
different continuum shapes, but the distribution of intrinsic
 (rest-frame) absorbing column densities is different among both classes.
A significant fraction of our type-2 AGN (5 out of 28) were found to display 
no substantial absorption (${\rm N_H<10^{21}\,cm^{-2}}$). 
We discuss possible interpretations to this in terms of 
Compton-thick AGN and intrinsic Broad Line Region properties.
An emission line compatible with Fe K$\alpha$ was detected in 8 sources (1 type-1 AGN, 5 type-2 AGN 
and 2 unidentified) with rest frame equivalent widths 120-1000 eV. However weak 
broad components can be easily missed in other sources 
by the relatively noisy data. 
The AGN continuum or intrinsic 
absorption did not depend on X-ray luminosity and/or redshift.
Soft excess emission was detected in 18 objects, but only in 9 
(including 4 type-1 AGN and 4 type-2 AGN)
could we fit this spectral component with a black body model. 
The measured 0.5-2 keV luminosities of the fitted
black body were not significantly different in type-1 and type-2 AGN, 
although the temperatures of the black body
were slightly higher in type-2 AGN ($\langle {\rm kT}\rangle$=$0.26\pm0.08$) than 
in type-1 AGN ($\langle {\rm kT}\rangle$=$0.09\pm0.01$). 
For 9 sources (including 1 type-1 AGN and 3 type-2 AGN) a
scattering model provided a better fit of the soft excess emission. 
We found that the integrated contribution from our sources to the X-ray 
background in the 2-7 keV band is softer ($\Gamma=1.5-1.6$) than the background itself, implying that fainter sources need to be more absorbed.
    } \maketitle
%
%________________________________________________________________

\section{Introduction} %***************************************************************************
\label{Introduction} The extragalactic
X-ray background (XRB) at energies above $\sim$ 0.2 keV is made up of
the integrated emission from point sources, mostly Active Galactic Nuclei (AGN). Synthesis models of 
the XRB (e.g. Setti \& Woltjer, ~\cite{Setti1989}; 
Gilli et al., ~\cite{Gilli2001}; Ueda et al., ~\cite{Ueda2003}), 
based on unification schemes of AGN, can reproduce
the spectral shape of the XRB with the superposition of a
mixture of absorbed and unabsorbed AGN. There is much
observational evidence supporting the unified model of AGN
(Antonucci, ~\cite{Antonucci1993}). For
example the discovery of large columns of X-ray absorbing gas in
type-2 AGN (Awaki et al., ~\cite{Awaki1991}; Risaliti et al., ~\cite{Risaliti1999}),
and the lack of
this absorbing material in type-1 AGN.

With the launch of the {\it Chandra} and {\it XMM-Newton} observatories, our
knowledge of the nature and cosmic evolution of AGN has increased
significantly. However the amount of observational results
that cannot be explained in terms of the unified model of AGN is
also significant. There is a substantial
number of type-1 AGN for which X-ray absorption has been detected
(Mittaz et al., \cite{Mittaz1999}; Fiore et al., \cite{Fiore2001};
Page et al., \cite{Page2001}; Schartel et al., \cite{schartel}; 
Tozzi et al., \cite{Tozzi2001b};
Mainieri et al., \cite{Mainieri2002}; Brusa et al., \cite{Brusa2003};
Page et al., \cite{Page2003}; Carrera et al., \cite{Carrera2004}; 
Perola et al., \cite{Perola2004};
Mateos et al., \cite{Mateos2005}),
as well as Seyfert 2 galaxies unabsorbed in X-rays
(Pappa et al., \cite{Pappa2001}; Panessa et al., \cite{Panessa2002};
Barcons et al., \cite{Barcons2003}; Mateos et al., \cite{Mateos2005}).
The origin of the X-ray absorption in type-1 AGN is still not clear.
Possible explanations include cold gas in the host galaxy.

For all these objects their optical and X-ray properties cannot be
explained in terms of an orientation effect only.

In order to gain insight into these problems and to understand better the
X-ray emission and cosmic evolution of the AGN that make up most
of the XRB, spectral analysis of large samples of AGN detected in
medium and deep X-ray surveys have been or are being conducted
(Mainieri et al., ~\cite{Mainieri2002}; Piconcelli et al., ~\cite{Piconcelli2002}, 
~\cite{Piconcelli2003}; Georgantopoulos et al., ~\cite{Georgantopoulos2004};
 Caccianiga et al., ~\cite{Caccianiga2004}; Perola et al., ~\cite{Perola2004}; 
Della Ceca et al., ~\cite{Ceca2004}; Mateos et al., ~\cite{Mateos2005}).
However,
at the moment, only a small number of these studies have carried out 
a proper spectral analysis of the spectra of each individual
source, and in many cases some assumptions had to be made prior to
the spectral analysis (frequently on the spectral slope). These
studies provide observational constrains with large uncertainties.

The {\it Lockman Hole} field is one of the sky regions best studied at X-ray
wavelengths, because the Galactic absorbing column density in this
direction is minimal ($5.7\times 10^{19} {\rm cm}^{-2}$, see Lockman et al.,
\cite{Lockman1986}).

{\it XMM-Newton} has carried out its deepest observation in the
direction of the {\it Lockman Hole} field. These observations have
allowed us to extract good quality ($>$500 0.2-12 keV counts) X-ray spectra for
objects down to 0.2-12 keV fluxes
 of ${\rm \sim 5\times10^{-15}\,erg\,cm^{-2}\,sec^{-1}}$ (the flux limit is 
${\rm \sim 6\times10^{-16}\,erg\,cm^{-2}\,sec^{-1}}$ in the 0.5-2 and 2-10 keV energy bands). We have used the {\it
XMM-Newton} observations in the {\it Lockman Hole} to carry out a
detailed analysis of the X-ray emission of the 123 brightest
objects detected in the field. The results from the analysis
of a sample of fainter objects will be described in a forthcoming paper.

Using a subset of these observations, Hasinger et al. (\cite{Hasinger2001}) presented the
source detection and properties of X-ray sources.
Mainieri et al. (\cite{Mainieri2002}) conducted a X-ray spectral analysis of the objects
detected in the field. Worsley et al. (\cite{Worsley2004}, \cite{Worsley2005}) 
used the total observation of the field
to calculate the fraction of unresolved XRB in different energy
bands. Finally Streblyanska et al. (\cite{Streblyanska2005}) have conducted
a detailed study of the Fe K$\alpha$ emission in the stacked spectra of
type-1 and type-2 AGN. They found in their analysis indications for 
broad relativistic lines in both type-1 and type-2 AGN.

This paper is organised as follows: Sec.~\ref{observations}
describes the X-ray data that we used for the analysis;
 Sec.~\ref{source list} describes how we built our sample
of objects; Sec.~\ref{spectral products} explains how the 
time averaged spectra were extracted for each individual object;
in Sec.~\ref{optical identifications} we show the current
status of the optical identification process;
Sec.~\ref{X-ray spectral analysis} describes the models
that we used to fit the X-ray emission of our sources; we show
the results of the analysis in Sec.~\ref{SPL_fitting},
Sec.~\ref{APL_fitting} and Sec.~\ref{best_fit};
the dependence of spectral parameters with luminosity and
redshift is show in Sec.~\ref{dependence};
In Sec.~\ref{unab_agn2} we discuss possible explanations 
for the lack of X-ray absorption signatures found in the 
spectra of five of the sources in our sample of type-2 AGN; 
In Sec.~\ref{EX_XRB} we compare the integrated emission of our sources
with the cosmic X-ray background in the 2-7 keV band; 
the results of our analysis are summarised in Sec.~\ref{conclusions}.

 Throughout this paper we
have adopted the {\it WMAP} derived cosmology with
${\rm H_0=70\,km\,s}^{-1}\,{\rm Mpc}^{-1}$,
$\Omega_M=0.3$ and $\Omega_{\Lambda}=0.7$.  All errors are
computed with a delta chi-square of 2.706, equivalent to 90\%
confidence region for a single parameter.

%*********************************************************************************************************************************
\section{{\it XMM-Newton} observations}
\label{observations} 
\begin{figure}
    \hbox{
    \includegraphics[angle=90,width=0.50\textwidth]{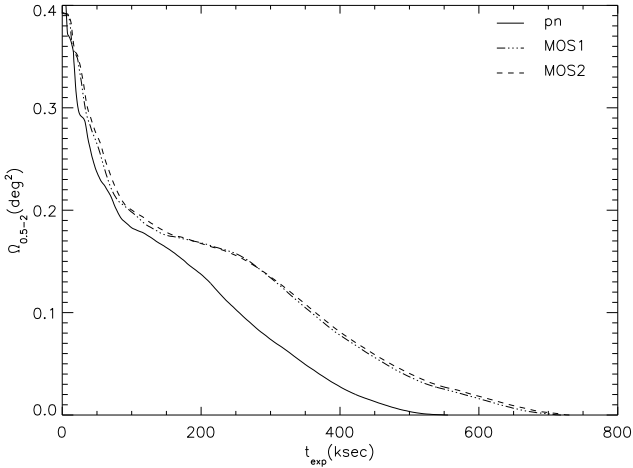}}
    \caption{Solid angle($\Omega$\,(t)) covered by MOS1, MOS2 and pn detectors as a function
      of the effective exposure time (after removal of background flares) in
      the 0.5-2 keV band. $\Omega$\,(t) is the solid angle 
(in deg$^2$) in the 0.5-2 keV exposure map of each EPIC camera with
an exposure $\ge$t.
    }
    \label{solid_angles}
\end{figure}

The deepest {\it XMM-Newton} observation has been carried
out in the direction of the {\it Lockman Hole} field, that is centred on the
sky position RA: 10:52:43 and DEC: +57:28:48 (J2000). The data were obtained by
adding 17 {\it XMM-Newton} observations obtained from 2000 to 2002. 
A summary of the observations used in this analysis is given in
Table~\ref{tab_observations}. The {\it Lockman Hole} was also observed during
revolution 071, however at the time of this analysis there was no Observation
Data File (ODF) available, and hence, we could not reprocess the data.
The first column in Table~\ref{tab_observations} shows the revolution
number and observation identifier. 
The second column shows the phase of the 
observation (i.e. PV for observations during the EPIC-Payload
Verification Phase, and AO1 and AO2 for observations during the first and
second Announcement of Opportunity). 
The third and fourth columns list the coordinates of the field used for each observation. 
The next column lists the observation dates and 
the last three columns show the filters
that were used during each observation for each X-ray detector together with
the exposure times after removal of periods of high background. 
The 17 {\it XMM-Newton} observations gave a total exposure (after removal of periods of
high background) of $\sim$850 ksec for MOS1 and MOS2 detectors and $\sim$650
ksec for pn. Some AO2 observations have an offset of more 
than $\sim$25 arcmin with respect to the other observations.
Because of this offset between observations, the total exposure time in the centre of
the {\it Lockman Hole} field was reduced significantly, however, the total 
solid angle covered by the observations increased substantially.
The solid angle of the observation as a function of the effective 
exposure time \footnote{The solid angle for a value of {\it t} was obtained
by summing the number of pixels in M1, M2 and pn exposure maps of the total observation
with a value $\ge$ t. Because the exposure maps include the energy dependent mirror vignetting function, the solid angle is given as a function of the effective exposure time on 
each camera} is plotted in Fig.~\ref{solid_angles}.

\begin{table*}[!ht]
\caption{Summary of {\it XMM-Newton} observations in the {\it Lockman Hole}}
\begin{center}
{\large  %añade esto
\begin{tabular}{ccccccccccc}
\hline
{\rm rev/obs. id} & obs phase  & RA &DEC & obs. date & & ${\rm Filter^a/GTI^b}$  \\
& & (J2000)& (J2000) & &EPIC-pn & EPIC-MOS1 & EPIC-MOS2 \\
\hline
\hline
 070\,/\,0123700101  & PV & 10\,\,52\,\,43.0 & +57\,\,28\,\,48 & 2000-04-27 & Th/34  & Th/34  & Tck/33 \\
 073\,/\,0123700401  & PV & 10\,\,52\,\,43.0 & +57\,\,28\,\,48 & 2000-05-02 & Th/14  & Th/14  & Tck/14 \\
 074\,/\,0123700901  & PV & 10\,\,52\,\,41.8 & +57\,\,28\,\,59 & 2000-05-05 & Th/5  & Th/8  & Tck/5 \\
 081\,/\,0123701001  & PV & 10\,\,52\,\,41.8 & +57\,\,28\,\,59 & 2000-05-19 & Th/27  & Th/36  & Tck/28 \\ \\[0.5ex]

 345\,/\,0022740201  & AO1 & 10\,\,52\,43.0 & +57\,\,28\,\,48 & 2001-10-27  & M/40  & M/37  & M/24 \\
 349\,/\,0022740301  & AO1 & 10\,\,52\,43.0 & +57\,\,28\,\,48 & 2001-11-04  & M/35  & M/34  & M/31 \\ \\[0.5ex]

 522\,/\,0147510101  & AO2 & 10\,\,51\,\,03.4 & +57\,\,27\,\,50 & 2002-10-15 &M/79  & M/81  & M/55 \\
 523\,/\,0147510801  & AO2 & 10\,\,51\,\,27.7 & +57\,\,28\,\,07 & 2002-10-17 &M/55  & M/56  & M/46 \\
 524\,/\,0147510901  & AO2 & 10\,\,52\,\,43.0 & +57\,\,28\,\,48 & 2002-10-19 &M/55  & M/57  & M/50 \\
 525\,/\,0147511001  & AO2 & 10\,\,52\,\,08.1 & +57\,\,28\,\,29 & 2002-10-21 &M/78  & M/79  & M/61 \\
 526\,/\,0147511101  & AO2 & 10\,\,53\,\,17.9 & +57\,\,29\,\,07 & 2002-10-23 &M/45  & M/52  & M/27 \\
 527\,/\,0147511201  & AO2 & 10\,\,53\,\,58.3 & +57\,\,29\,\,29 & 2002-10-25 &M/30  & M/34  & M/23 \\
 528\,/\,0147511301  & AO2 & 10\,\,54\,\,29.5 & +57\,\,29\,\,46 & 2002-10-27 &M/28  & M/33  & M/14 \\ \\[0.5ex]
 544\,/\,0147511601  & AO2 & 10\,\,52\,\,43.0 & +57\,\,28\,\,48 & 2002-11-27 &M/104  & M/103  & M/68 \\
 547\,/\,0147511701  & AO2 & 10\,\,52\,\,40.6 & +57\,\,28\,\,29 & 2002-12-04 &M/98  & M/98  & M/89 \\
 548\,/\,0147511801  & AO2 & 10\,\,52\,\,45.3 & +57\,\,29\,\,07 & 2002-12-06 &M/86  & M/86  & M/72 \\
\hline
%signal rms & 112.3 & \textcolor{green}{0.69} & \textcolor{green}{5.37} &\textcolor{green}{55.8}& \textcolor{green}{0.66}&\textcolor{green}{0.32} \\
\end{tabular}
\label{tab_observations}
\begin{list}{}{}
\item[$^{\mathrm{a}}$] Blocking filter: Th: Thin at 40nm A1; M: Medium at 80nm A1;
               Tck: Thick at 200nm A1
\item[$^{\mathrm{b}}$] Exposure time (in ksec) per observation and
detector obtained after removal of background flares
\end{list}
}  % y añade esto
\end{center}
\end{table*}

%*********************************************************************************************************************************
\section{X-ray source list} %***************************************************************************
\label{source list} We have used the {\it XMM-Newton} Science
Analysis Software (SAS, Gabriel et al. \cite{Gabriel2004}) version v5.4,
the latest public version of the SAS at the time of study, 
to analyse the X-ray
observations. Spurious noise events not created by X-rays were 
filtered from the event files using the {\it XMM-Newton} 
pipeline standard filtering. We have cleaned the event files 
of the individual observations with periods of high background due 
to soft high background flares. 
For each observation and detector 
we built light curves (histograms of counts as
a function of time). The light curves were visually inspected to search for time intervals
affected by high flaring background periods. Thresholds in count rate were determined separately for each observation 
and camera. 

Events covering patterns 0-12 for MOS and 0-4 for pn data were selected.

Cleaned event files were used to create images, background
maps and exposure maps for each detector and for each of the 
{\it XMM-Newton} standard energy bands (0.2-0.5,
0.5-2, 2-4.5, 4.5-7.5  and 7.5-12 keV). We have run the {\it XMM-Newton}
source detection algorithm, {\tt eboxdetect-emldetect},
simultaneously in the five energy bands.
Our motivation was to reach the maximum sensitivity in each
individual band in order to best detect objects with spectra
peaking at different energies (as it is the case for AGN with
different absorbing column densities). However, it is important to
note that the sensitivity of the {\it XMM-Newton} X-ray
detectors is a strong function of energy, with the maximum
sensitivity reached between 0.5 and 4.5 keV. This implies that we will
best detect objects with X-ray spectra peaking within this
interval of energy.

Due to the large offset between different observations of the
field (see Sec. \ref{observations}), we did not merge the event
files, because in the merging process important
information from the individual observations is lost (e.g., bad
columns in the detectors). Moreover, it is not possible to create
exposure maps or background maps from the merged event files.
Therefore we extracted images, exposure maps and background maps, for each individual
observation, detector and energy band, and then, we combined
them to obtain the total observation of the field for each
 X-ray detector and energy band.

We decided to run the source detection algorithm independently for
each detector. Because the pn data give the deepest
observation of the field (the MOS1 and MOS2 detectors only receive
about half of the radiation from the X-ray telescopes, the other half
goes to the Reflection Grating Spectrometers), we used the pn
source list to build our catalogue of sources. However, we cross-correlated the
sources detected with the pn with the ones detected with each MOS detector.
We found that only one faint object that was detected with the MOS1 and MOS2
detectors was not detected with the pn (probably because it
was very close to a brighter object). We added this object
to our source list. We have carried out a visual screening
of the objects to remove spurious detections (e.g. we have
detections in hot pixels that were still present in the data).
The final number of objects detected after visual screening
in the integrated {\it XMM-Newton} observation of the {\it Lockman Hole}
is 268.

\begin{figure}
    \hbox{
    \includegraphics[angle=90,width=0.50\textwidth]{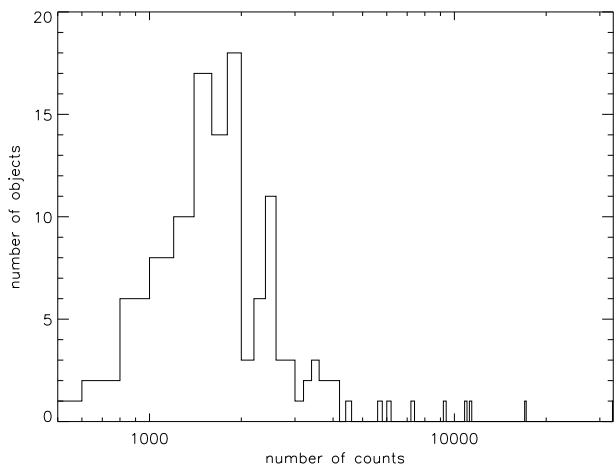}}
    \caption{Distribution of the number of background-subtracted
      counts in the 0.2-12 keV
      spectra for the sources in our sample. We used the counts of the
      time averaged spectra with better signal to noise (MOS or pn)
      }
    \label{n_counts_dist}
\end{figure}

In this paper we show the results obtained from the spectral analysis of the
123 sources with the best spectral quality (more than 500 MOS+pn
background subtracted counts in the 0.2-12 keV band). Because we were interested in
studying the X-ray spectral properties of AGN, we excluded from
the sample the objects identified as clusters of galaxies
or stars. The results from the analysis of the fainter objects will be
presented in a forthcoming paper. Fig.~\ref{n_counts_dist} shows
the distribution of (background subtracted) counts
for the sources that we have studied.
In the following we will refer to the sample of 123 brightest sources 
as our list of objects.

In order to allow comparison with previous surveys conducted in the soft,
0.5-2 keV, and hard, 2-10 keV, bands, we have checked whether our objects were
detected in any of these bands. We have the likelihoods of detection for the
soft band, because it was one of the energy bands used for the source
detection. We have calculated the likelihoods in the hard (2-10 keV) band,
combining \footnote{Detection likelihoods in the 2-10 keV band were 
obtained following the description of the {\tt emldetect} task, see 
${\rm http://xmm.vilspa.esa.es/external/xmm\_sw\_cal/sas\_frame.shtml
}$} the ones obtained in the bands 2-4.5 keV, 4.5-7.5 keV and 7.5-12
keV. Indeed, the combination of these values will give us the detection likelihood in
band 2-12 keV. However, the effective area of the X-ray telescopes
decreases rapidly at energies above 5 keV, hence we do not expect the value of
the detection likelihood in the 2-12 keV band to differ substantially from the
value in the 2-10 keV band. By selecting sources with 0.5-2 and 2-10 keV detection 
likelihoods above 10 we found that the number 
of objects detected in both the soft and hard bands was 117 (out of 123). Three objects 
were only detected in the hard band and three only in the soft band. 
These numbers indicate that the source
population that we are studying is detected in both bands, and therefore
our list of sources does not differ significantly from what would have been
detected in a hard or soft band survey at similar fluxes.

\section{X-ray spectral products}
\label{spectral products}
An automated procedure has been used to 
obtain for each individual object the spectrum of the total observation, 
hereafter, the time averaged spectrum of the sources. In addition a background 
spectrum and a response calibration matrix have been generated by combining 
individual products from each exposure.

First we have extracted the
spectra of each object for each detector (MOS1, MOS2 and pn) and observation.
We used the coordinates of the objects, RA and Dec, that we obtained from the
source detection process, and the SAS task {\tt region} to define the source
and background extraction regions. The first was defined as a circle with a
radius ($\it r_s$) that varied depending on the position of the source within
the detector. We extracted the background for each object in an annulus
centred on the source position, with inner radius $r_s$ and outer radius 
3$\times r_s$. The task {\tt region} checks the source and background regions for
overlap with neighbouring sources. If overlapping exists, then, the size of
the source region is reduced until it is removed. For the background regions,
if neighbouring objects fall inside the background region they are masked out.
The task {\tt region} also checks that the extraction regions do not extend
outside the edges of the field of view. The radius of extraction of spectra
varies from source to source, but typically was $\sim$14-20 arcsec. Once the
regions were defined, we used the SAS task {\tt evselect} to extract from 
event files the spectra of each object. Calibration matrices ({\tt arf} and
{\tt rmf}) for each spectrum were obtained with the SAS tasks 
{\tt arfgen} and {\tt rmfgen}.

We did not use the spectra from observations where the
objects were near the borders of the FOV, or near CCD gaps or bad
columns, because in these cases we found that the spectral
products, in particular the response matrices were often incorrect.
To find and remove these cases we visually checked the images of each
observation and detector.

We have obtained a MOS and pn time averaged spectra for
each object. As we see in Table \ref{tab_observations},
different filters were used for the observations.
The filters affect in a different way the X-ray spectra at low
energies. To take into account this effect when combining the 
spectral products of each source, we have weighted the data 
of each individual observation with the exposure time of the observation.
\footnote{Source and background spectra are obtained adding the
counts for each channel. The areas used to extract the spectra and the
response matrices were weighted with the exposure times of each individual
observation.}

The spectra were extracted in the energy
range from 0.2 to 12 keV, where the X-ray detectors are best calibrated.
 In order to use the $\chi^2$ minimisation during the
spectral fitting, we have grouped the spectra with a minimum number of
30 counts per bin.

\section{Optical identifications} %***************************************************************************
\label{optical identifications}

A large fraction of the X-ray sources that were detected with the
{\it ROSAT} satellite in the {\it Lockman Hole} have already been 
identified through optical spectroscopy (Schmidt et al., \cite{Schmidt1998}; 
Lehmann et al., \cite{Lehmann2001}). 
These sources were detected in the 0.5-2 keV energy range, 
and therefore the optical identification is expected 
to be biased against absorbed sources, whose X-ray spectra 
does not peak in the {\it ROSAT} energy interval. However, the optical 
identifications go as deep as R$\sim$24, hence 
we do not expect the identifications in our sample 
to be significantly affected by the bias against absorbed sources.

Our {\it XMM-Newton} observations cover a larger solid angle than 
the {\it ROSAT} observations and find additional sources due to the superb 
high energy response of the {\it XMM-Newton} detectors.
Some of the objects that we have analysed fall outside the solid angle covered by 
{\it ROSAT} and we do not have optical identifications. Other sources were not
detected with {\it ROSAT}. For 8 of these newly detected {\it XMM-Newton} sources
optical spectra have been obtained with the LRIS and Deimos instruments at the Keck 
telescopes in 2001, 2003, and 2004 (PI: M. Schmidt and P. Henry). The spectroscopic 
identification of these objects and of the entire
Deep {\it XMM-Newton} Survey in the {\it Lockman Hole} will be presented in a forthcoming paper 
(Lehmann et al. 2005 in preparation). 

To be consistent with the {\it ROSAT} identifications we have used the same criteria to differentiate 
between type-1 and type-2 AGN as described in detail by Schmidt et al.~(\cite{Schmidt1998}). 
Sources were classified depending on the properties of their UV/optical emission lines. 
Objects with UV/optical emission lines with $FWHM>1500\,{\rm km\,s^{-1}}$ in their optical spectra 
were classified as type-1 AGN. Sources that do not exhibit broad emission lines but that 
show Ne emission lines ([${\rm Ne\,V}$] and/or strong [${\rm Ne\,III}$]) were classified as type-2 AGN.
Classification of Narrow Line Seyfert 1 galaxies (NLSy1) was only possible 
for bright nearby objects. Hence, we cannot be sure that the optical classification 
criteria used to separate the type-1 and type-2 AGN in our sample has excluded all 
NLSy1 from the sample of type-2 AGN.
However, based on the observed X-ray properties of our sources, we do not expect this 
to affect the results of our study.

\begin{figure}
    \hbox{
    \includegraphics[angle=90,width=0.50\textwidth]{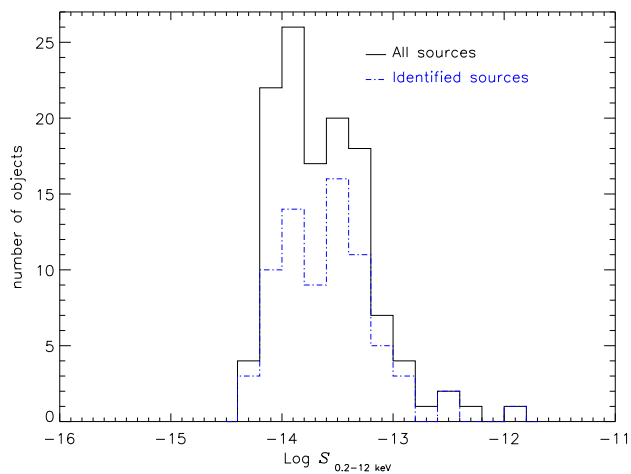}}
    \caption{Histograms of 0.2-12 keV fluxes for the whole sample of sources analysed and
      for the objects with optical identifications. The fluxes were obtained from the
      sources best fit model (See Sec.~\ref{best_fit}).
      }
    \label{flux_hist}
\end{figure}

At the time of this analysis, 74 ($\sim$ 60\%) of the sources
had optical spectroscopic identifications. Of these, 46 were
classified as type-1 AGN and 28 as type-2 AGN.

In Fig.~\ref{flux_hist} we show the distribution in 0.2-12 keV
flux \footnote {Fluxes were obtained from the objects best fit model (See
Sec.~\ref{best_fit})}
of all the objects (solid histogram) and of the identified sources
(dot-dash histogram). We see that the two distributions agree quite well, i.e.,
the identified sources do not tend to have higher X-ray fluxes.

\begin{figure*}
    \hbox{
    \includegraphics[angle=90,width=0.50\textwidth]{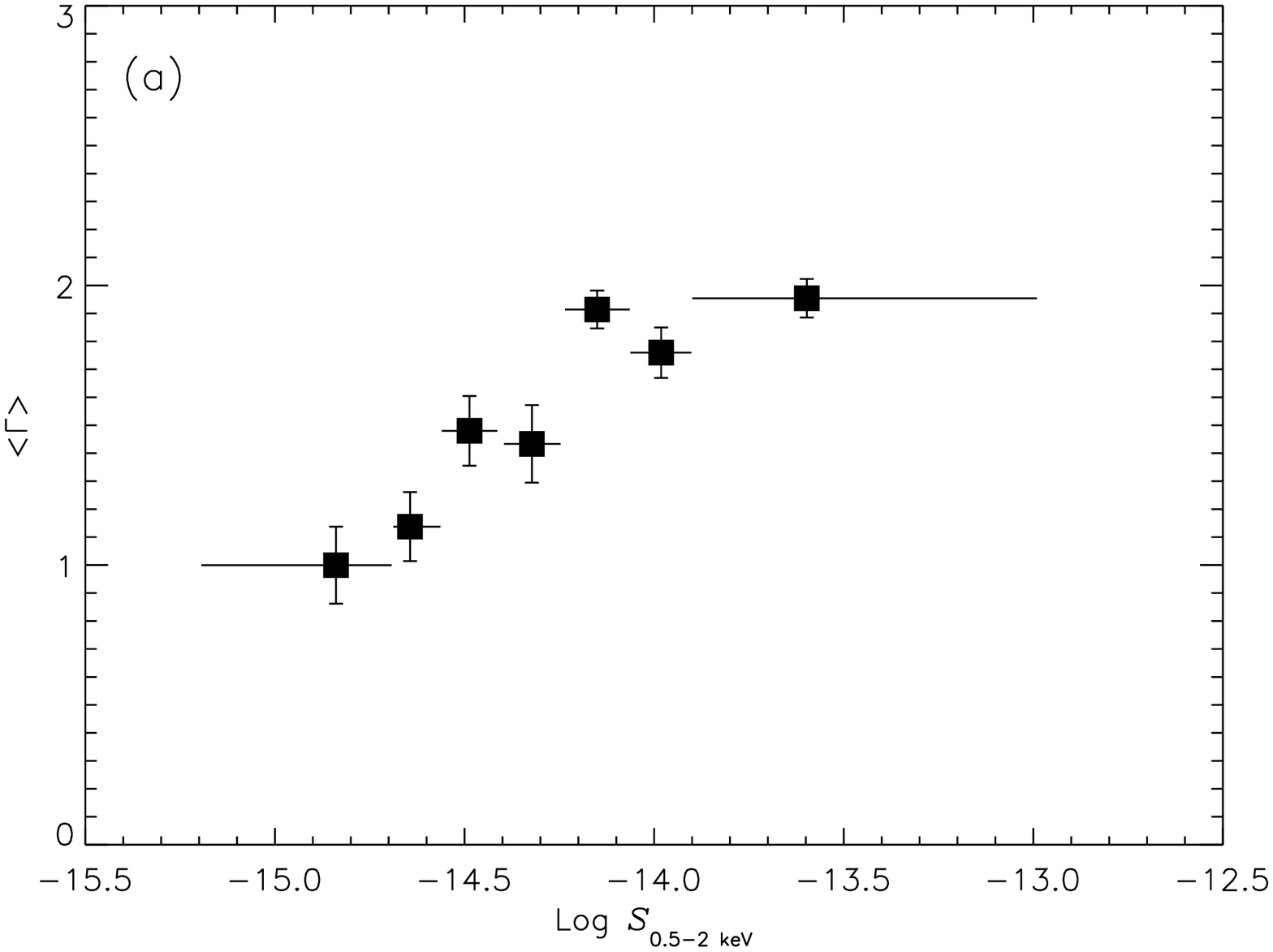}
    \includegraphics[angle=90,width=0.50\textwidth]{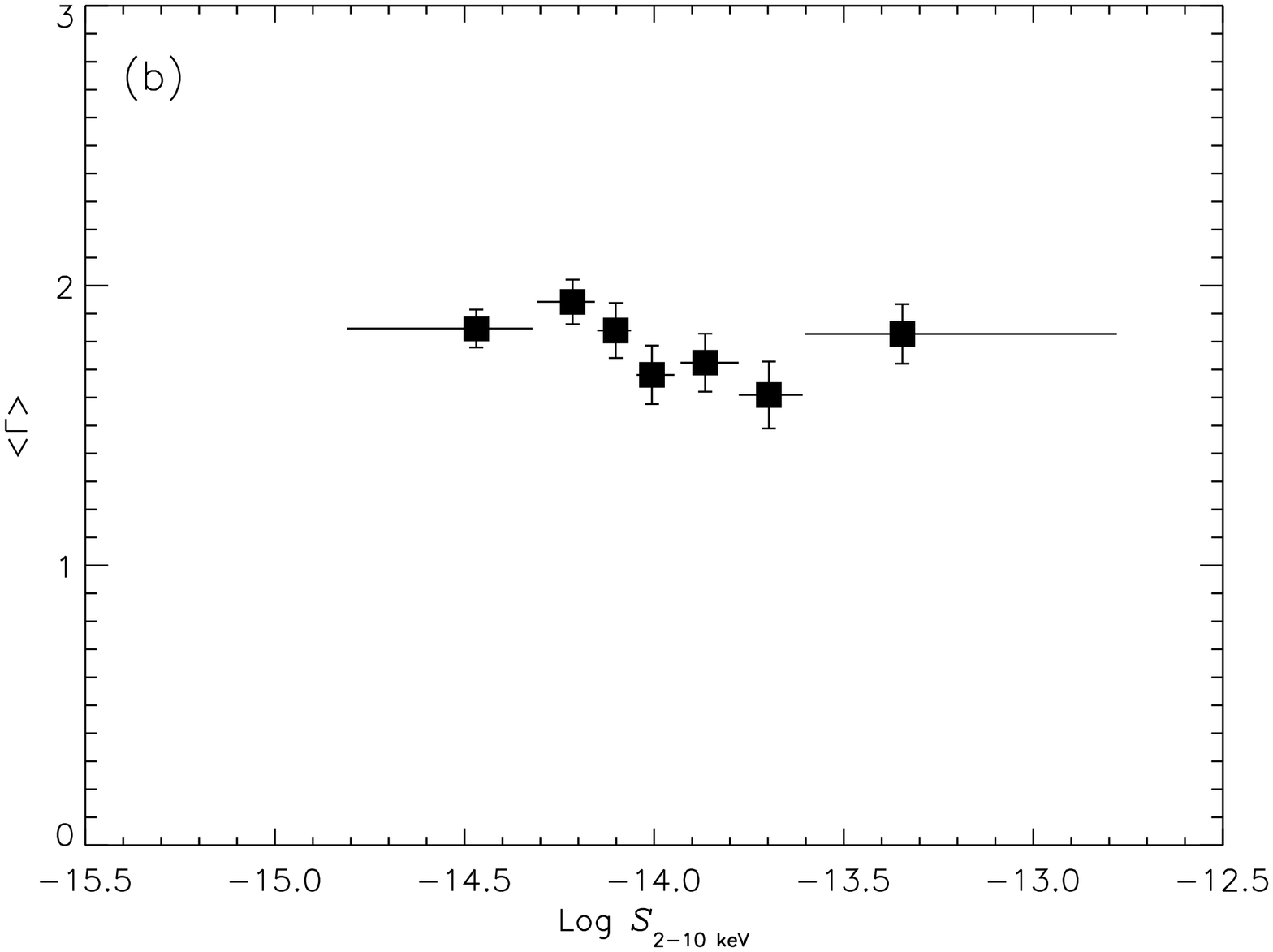}}
    \hbox{
    \includegraphics[angle=90,width=0.50\textwidth]{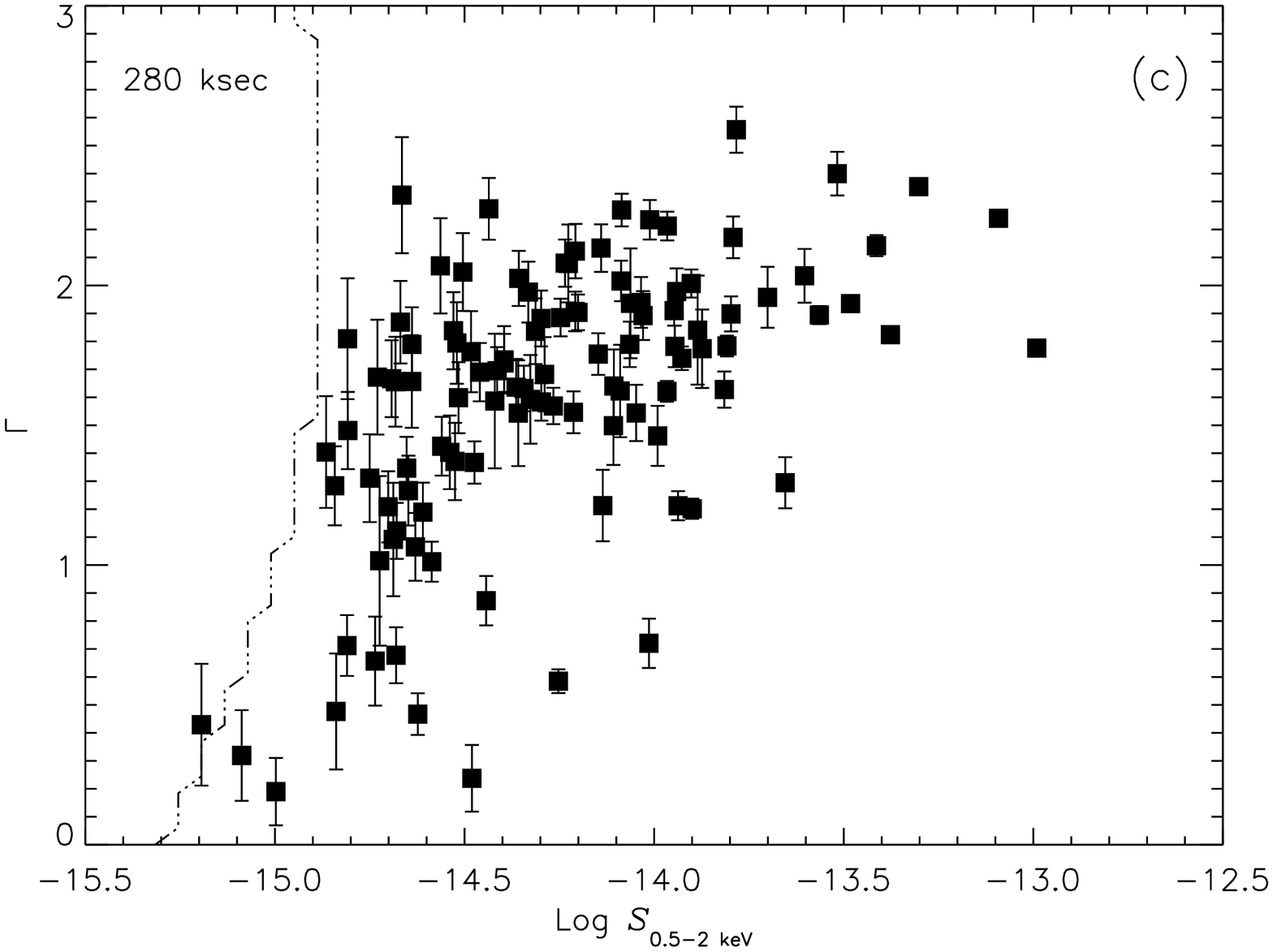}
    \includegraphics[angle=90,width=0.50\textwidth]{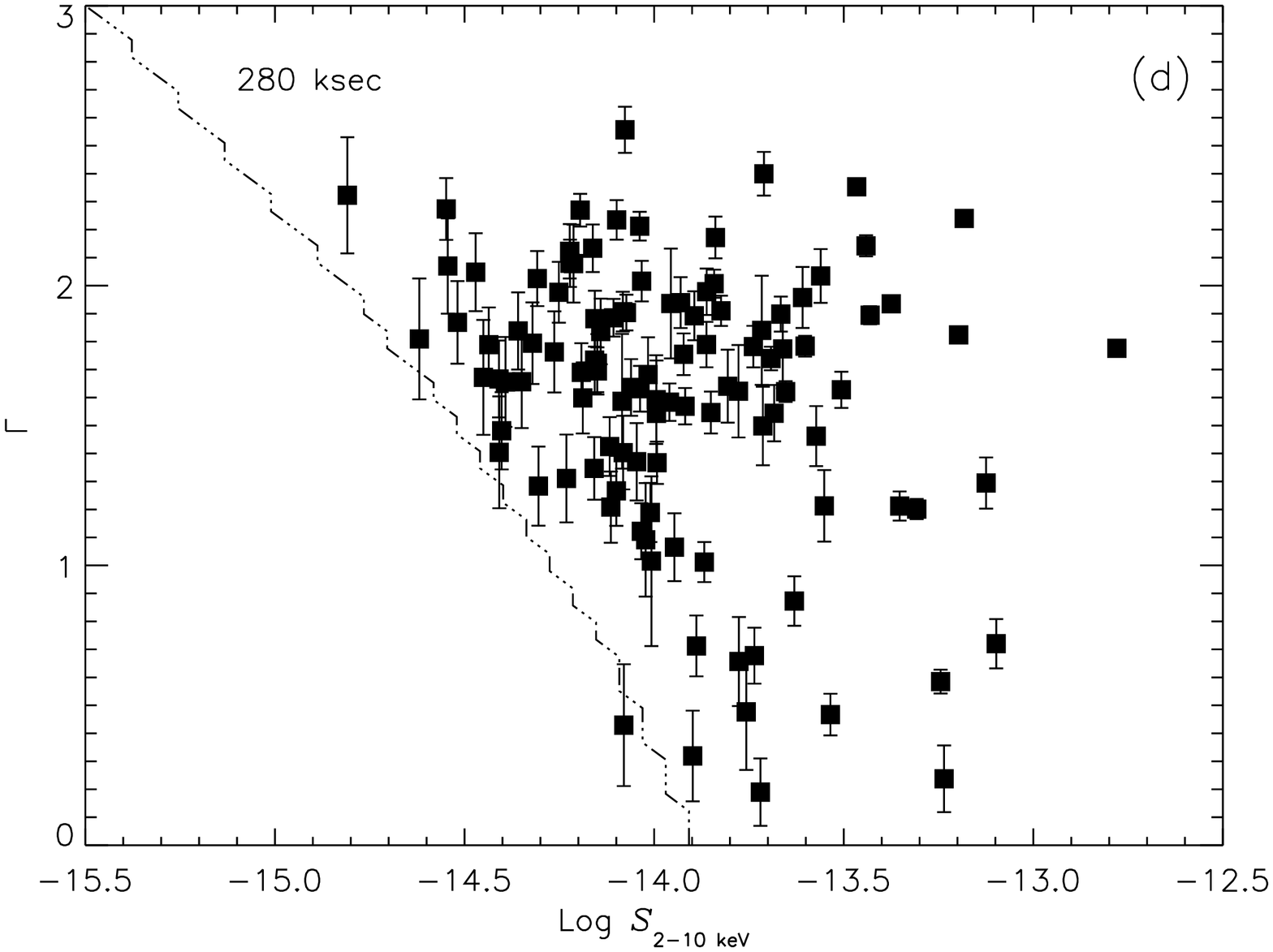}}
    \caption{Dependence of $\Gamma$ with soft (0.5-2 keV) and
      hard (2-10 keV) flux 
      when all spectra are fitted with a single power law model.
      In plots (a) and (b) we show the dependence of the weighted 
      (with the error of each individual value) $\Gamma$ 
      with the flux. The bins in flux were defined in order to have
      the same number of sources per bin. In plots (c) and (d) we
      show the values of $\Gamma$ obtained for each individual source.
      The dash-dot lines in these plots show, for an exposure time of 
      280 ksec, the limit in
      flux for detection for an object as a function of $\Gamma$ 
      (see Sec. \ref{dep gmm} for details). 
      Errors bars in (c) and (d) correspond to 
      90\% confidence.
    }
    \label{gamma_vs_flux_spl}
\end{figure*}

%*********************************************************************************************************************************
\section{X-ray spectral fitting} %***************************************************************************
\label{X-ray spectral analysis} We have used the {\tt xspec {\rm 11.3.0}}
package to fit the X-ray spectra of our objects. MOS and pn
spectra were fitted simultaneously and with the same spectral
model, including normalisation. At the time of our analysis
it was found that there was an offset
of  $\sim$1 arcmin between the optical axis of the three
 EPIC instruments and the values in the calibration files (CCF) 
\footnote{http://xmm.vilspa.esa.es/docs/documents/CAL-SRN-0156-1-3.ps.gz}. 
This could be introducing discrepancies
in the fluxes measured by the MOS and pn due to an incorrect
vignetting correction. These flux discrepancies could be as
high as $\pm14\%$. However,
we did not find a significant improvement in the quality of our
fits when different normalisations were used to fit MOS and pn
spectra. For the objects where we found an offset between
MOS and pn normalisations, this offset was much
higher than the expected flux discrepancies explained before.
We interpreted this effect as a change in the flux of
the source during the observations (note that the time averaged
spectra of MOS and pn in the majority of the cases were not
necessary built with the same set of observations, because the FOV
and the positions of the gaps are different for MOS and pn, and in
different observations).

In order to compare the results of our analysis with
other studies of data with lower signal to noise, we have fitted the 
spectra of our sources with a single power law model (hereafter SPL). 
This has allowed us to study in more detail the origin of the hardening of
the average spectra of AGN with the soft X-ray flux (see e.g. Mateos et al~\cite{Mateos2005}), and investigate
if the same effect is also present when 2-10 keV fluxes are
used. To study the effect of absorption in the results obtained with the SPL we 
fitted all the spectra with an absorbed power law model (hereafter APL model). 

Using these two models it is possible to obtain 
very useful results on the average spectral properties of our sources, the broad band 
continuum shape and the X-ray absorption. However, our 
major goal is to study in detail the 0.2-12 keV X-ray emission of each individual source. Hence,  
for each object we have obtained its best fit model. The quality of our data has allowed us to
search for soft excess at low energies, the Fe K$\alpha$ emission line complex, and  
reflection components.

We have used the F-test to measure the significance of detection for each spectral component. 
We have selected a confidence level threshold of 95\% to accept an additional
spectral component as being real. 
The criteria that we used to select X-ray absorbed sources is to have an F-test significance $\ge95\%$. 
This is different from some definitions found in the literature, 
because we did not impose a lower threshold in the detected values of ${\rm N_H}$. 
For example, Ueda et al. (\cite{Ueda2003}) defined as X-ray absorbed the sources 
with absorption column density at the source 
redshift ${\rm \ge10^{22}\,cm^{-2}}$. However, it is important to note that 
with our criteria all sources selected as X-ray absorbed had values 
of ${\rm N_H}$ ${\rm \ge21\,cm^{-2}}$.

%*********************************************************************************************************************************
\section{Single power law fitting (SPL)} %***************************************************************************
\label{SPL_fitting}
We have used a single power law model to
fit the 0.2-12 keV emission of all the objects. In this model, the free
parameters are the normalisation (the same for MOS and pn spectra)
and the slope of the broad band continuum, $\Gamma$ \footnote{We use
the power law photon number index, $\Gamma$. Its relation with the energy index is $\alpha$ = $\Gamma$-1}. The power law is absorbed with a fixed
column density of $5.7\times10^{19}\,{\rm cm^{-2}}$ to include the
effect of the absorption by our Galaxy in the direction of the
{\it Lockman Hole} field.

\subsection{Dependence of $\Gamma$ with the X-ray flux}
\label{dep gmm}
The results of the fits have been used to study the dependence of
$\Gamma$ with the X-ray fluxes obtained from the SPL model.
The results are plotted in Fig.~\ref{gamma_vs_flux_spl}. In plots
(a) and (b) we show the dependence of $\langle \Gamma
\rangle$ with the 0.5-2 keV and 2-10 keV fluxes. The bin sizes
were defined in order to have the same number of objects per bin, 
and the average values were obtained weighting with the errors of each 
individual value.
When a single power law model is used, we see that the average continuum shape
becomes harder with decreasing 0.5-2 keV flux. However, it is
interesting to note that we do not see any
dependence of $\langle \Gamma \rangle$ with the 2-10 keV flux down
to $\sim\rm{3\times10^{-15}\,erg\,cm^{-2}\,s^{-1}}$.

The values of $\langle \Gamma \rangle$ were calculated with the
standard formula for the weighted mean,
\[
\langle \Gamma\rangle=\sum\,P_i\times\,\Gamma_i
\]
where the weight, $P_i$, of each individual best fit value, $\Gamma_i$, is
a function of the error in the parameter obtained from the fit, $\sigma_i$, i.e.
\[
P_i={1\,/\sigma_i^2 \over \sum\,( 1\,/ \sigma_i^2\,)}
\]
To calculate the uncertainty in $\langle \Gamma \rangle$ we have used the
error on the weighted mean (Bevington et al. \cite{Bevington1992}),
\[
\sigma^2(\langle\Gamma\rangle)={1 \over (\rm{N}-1)}\,\sum\,P_i\times( \Gamma_i-\langle \Gamma\rangle)^2
\]
that includes the measurement errors, $\sigma_i$, and the
dispersion of each $\Gamma_i$ from the estimated value $\langle
\Gamma \rangle$. Using these expressions we found that our objects
have $\langle \Gamma \rangle$=$1.79\pm0.03$ when their spectra are fitted
with a SPL (the value is 1.60$\pm$0.05 if we use the unweighted mean).

In order to understand better the origin of the hardening of
$\langle \Gamma \rangle$ with the 0.5-2 keV flux, and why we do
not see the same effect using 2-10 keV fluxes, we have plotted
in Fig.~\ref{gamma_vs_flux_spl} (figures (c) and (d)) the values of
$\Gamma$ that we obtained for each individual object. Thanks to the good quality of our data
we can see that $\langle \Gamma \rangle$ becomes harder
because at faint 0.5-2 keV fluxes a population of
faint sources is revealed with very hard ($\le$ 1) spectral slopes. We also see
that the number of faint hard objects becomes more important as we go to 
fainter fluxes. 
In the 2-10 keV band we do not see the hardening of $\langle \Gamma \rangle$
 because these hard objects are detected at all 2-10 keV fluxes.
Moreover, their number seems not to vary with the 2-10 keV flux.

We have studied whether our criteria for selection of objects (i.e. MOS+pn
background subtracted counts above 500) could be introducing any bias in our
results. In particular we have studied whether, for a given flux, we are
favouring objects with a given spectral slope. To study this, we have carried
out simulations. We first defined a grid of points in $\Gamma$ and {\it
S} (first using 0.5-2 keV flux and later with 2-10 keV flux), covering the
same range of $\Gamma$-{\it S} values as our sources. Using a pair of on-axis
response matrices, {\tt arf} and {\tt rmf}, and typical background spectra
selected from one of our objects, we have simulated a spectrum on each grid
point. With the simulated spectra we have calculated the minimum exposure time
that is needed to reach the threshold in number of counts that we have used to
select our sources (i.e. 500 MOS+pn background subtracted counts). With these
simulations we were not interested in quantifying the limits of detection as a
function of {\it S} and $\Gamma$, but to study the biases in our sample, and
whether they affect our results. Therefore we only need to do one simulation
on each grid point and then we just have to search for the points in the
$\Gamma$-$S$ grid with the same value of the exposure time.

A constant exposure time line to get 500 counts is represented with the
dot-dash lines in plots (c) and (d) on Fig.~\ref{gamma_vs_flux_spl} for an
exposure time of 280 ksec. We see that in the soft band, for a given flux, we
have the same efficiency of detection for different values of $\Gamma$ down to
1.5.
 At fluxes above ${\rm \sim 10^{-15}\,erg\,cm^{-2}\,s^{-1}}$ this bias
is not affecting the observed hardening of $\Gamma$. Only the bin
at the faintest 0.5-2 keV fluxes (plot (a)) could be affected by this bias.
Our simulations show that the hardening in $\langle \Gamma \rangle$ is an
intrinsic property
of our objects. As we said in Sec.~\ref{source list}, our objects were all detected in the soft band, therefore
this effect is a property of the 0.5-2 keV population of objects.  The objects
responsible for the hardening of $\langle \Gamma \rangle$
can be more absorbed sources or sources having intrinsically harder spectra.

In the 2-10 keV band we obtained different results from the simulations.
At the faintest fluxes we most easily detect objects with soft spectra. This is
an expected result because the effective area of the X-ray detectors in
{\it XMM-Newton} decreases rapidly at energies $\ge$5 keV and therefore
it is more difficult to detect faint objects with flat spectral
slopes. However, down to the flux level where we start to lose faint hard
objects, ${\rm \sim 6\times10^{-15}\,erg\,cm^{-2}\,s^{-1}}$, we see that there is no
dependence of $\Gamma$ with hard flux because hard objects are
detected at all flux levels.

\begin{figure*}
    \hbox{
    \includegraphics[angle=90,width=0.50\textwidth]{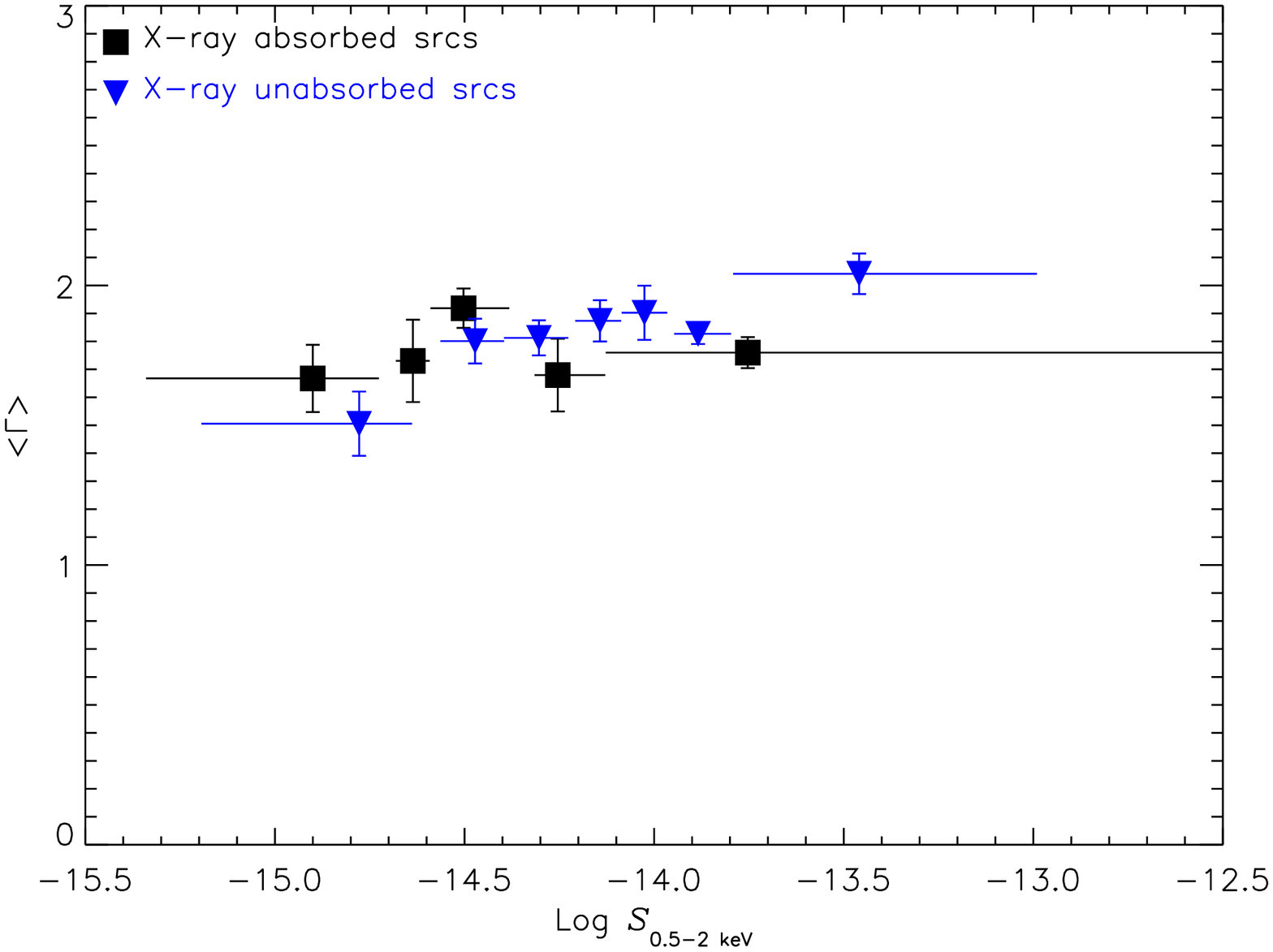}
    \includegraphics[angle=90,width=0.50\textwidth]{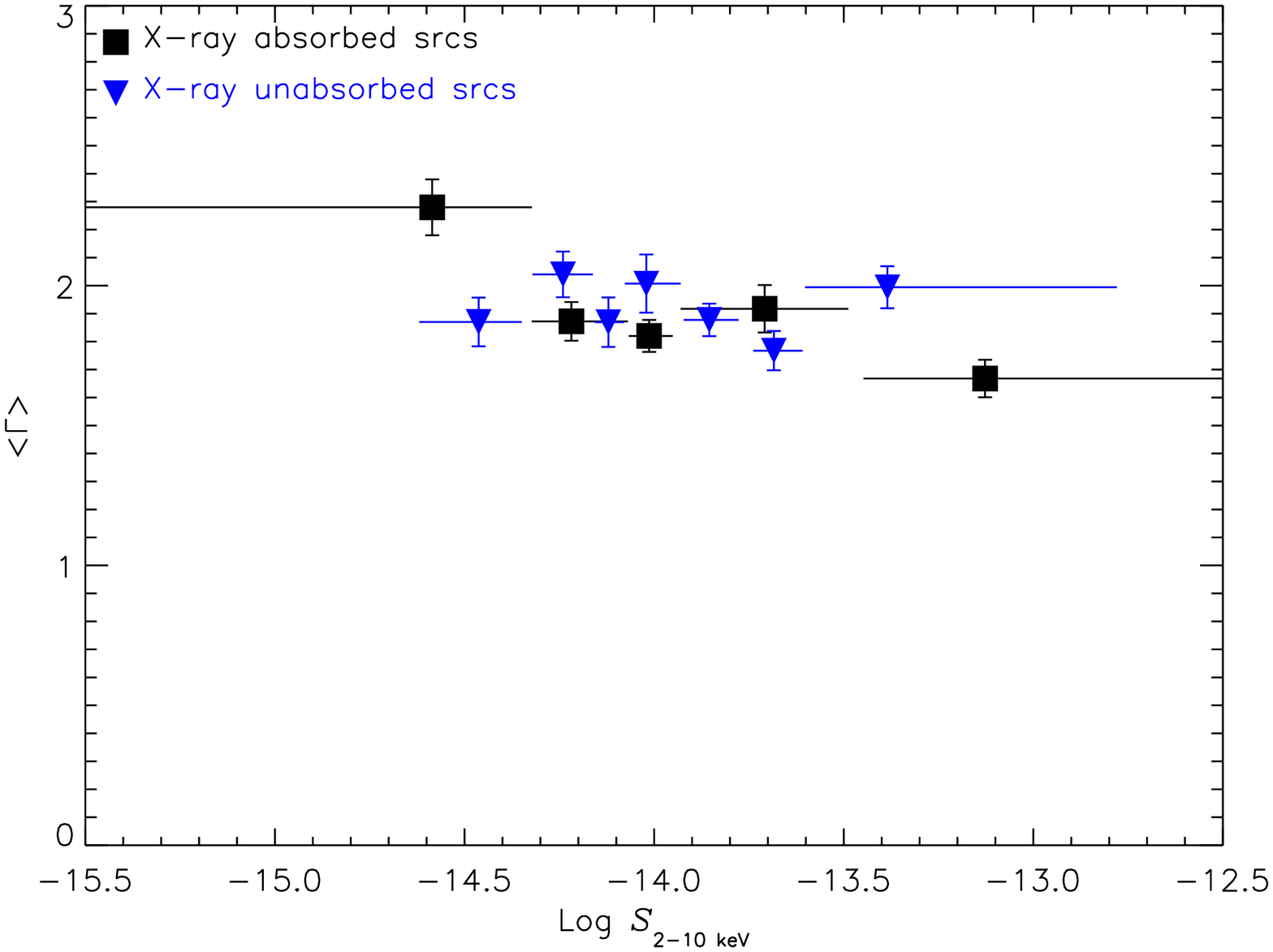}}
%    \hbox{
%    \includegraphics[angle=90,width=0.50\textwidth]{gamma_vs_sflux_lh_best_fit.ps}
%    \includegraphics[angle=90,width=0.50\textwidth]{gamma_vs_hflux_lh_best_fit.ps}}
    \caption{Dependence of $\langle \Gamma \rangle$ with 0.5-2 and 2-10 keV 
      flux for absorbed (F-test $\ge 95\%$) and unabsorbed (F-test $< 95\%$)
      sources. For each source we used $\Gamma$ and {\it S} from its best 
      fit model (single power law or absorbed power law). 
      See Sec.~\ref{APL_fitting} for details. 
    }
    \label{gamma_vs_flux_best_fit}
\end{figure*}

\section{Absorbed power law fitting (APL)}
\label{APL_fitting}
In order to study the nature of the population of faint hard sources responsible
for the hardening of $\langle \Gamma \rangle$ with the 0.5-2 keV flux we have fitted
the spectra of all the objects with an absorbed power
law model. The free parameters of this model are the normalisation, the
spectral slope of the power law component, and the intrinsic (rest-frame) absorption
of the objects that we have measured in the observer's frame 
(${\rm N_H^{obs}}$). Again, we
also included the effect of the Galactic absorption in the
direction of the {\it Lockman Hole}.
Using the APL model we obtained a value for the weighted mean 
of $\langle \Gamma \rangle = 1.87\pm 0.04$
 (the value being $\langle \Gamma \rangle = 1.95\pm0.08$ using the arithmetic mean)
for the objects where absorption was detected (F-test $\ge$ 95\%), and $\langle \Gamma \rangle = 1.95\pm 0.03$
 (the value being $\langle \Gamma \rangle = 1.82\pm0.04$ using the arithmetic mean)
for the objects where we did not detect absorption (F-test $<$ 95\%).

The dependence of $\langle \Gamma \rangle$ on 0.5-2 keV and 2-10 keV fluxes that we see fitting
the spectra of our objects with the APL model is shown in
Fig.~\ref{gamma_vs_flux_best_fit}. For the objects where
absorption was detected we used the spectral parameters
($\Gamma$ and observed {\it S}) obtained with the
APL model. For the unabsorbed sources we used the parameters from the
SPL model. We see that absorption can account for most of the hardening
of $\langle \Gamma \rangle$ with the soft
X-ray flux. We also see in Fig.~\ref{gamma_vs_flux_best_fit} 
that the same dependence of $\langle \Gamma \rangle$ 
with the X-ray flux is obtained for absorbed and unabsorbed objects. 

Mateos et al. (\cite{Mateos2005}) found that for their serendipitous X-ray 
sources (with much lower spectral quality), the average spectrum 
of unabsorbed (F-test significance $<$95\%) sources significantly hardens at faint 0.5-2 keV fluxes. 
They concluded that undetected absorption was responsible for the 
observed effect in unabsorbed sources. 
The magnitude of undetected absorption cannot be very significant in 
our {\it Lockman Hole} sources as we do not see a clear hardening of 
$\langle\Gamma\rangle$ for unabsorbed sources.  
Using $\Gamma$ and {\it S} from the APL model for unabsorbed sources we 
obtain the same result.
Hence, if we still have sources with undetected absorption, 
their absorbing column densities cannot be very high. Moreover, we do not find evidence 
for the existence of a population
of faint sources with intrinsically harder spectral slopes.

The results for the
hard fluxes do not vary significantly from what we obtained with the SPL fits.
This is exactly what we would expect if absorption produces the hardening of $\Gamma$,
because the 2-10 keV fluxes are less affected by absorption.

\begin{figure*}
    \hbox{
    \includegraphics[angle=90,width=0.50\textwidth]{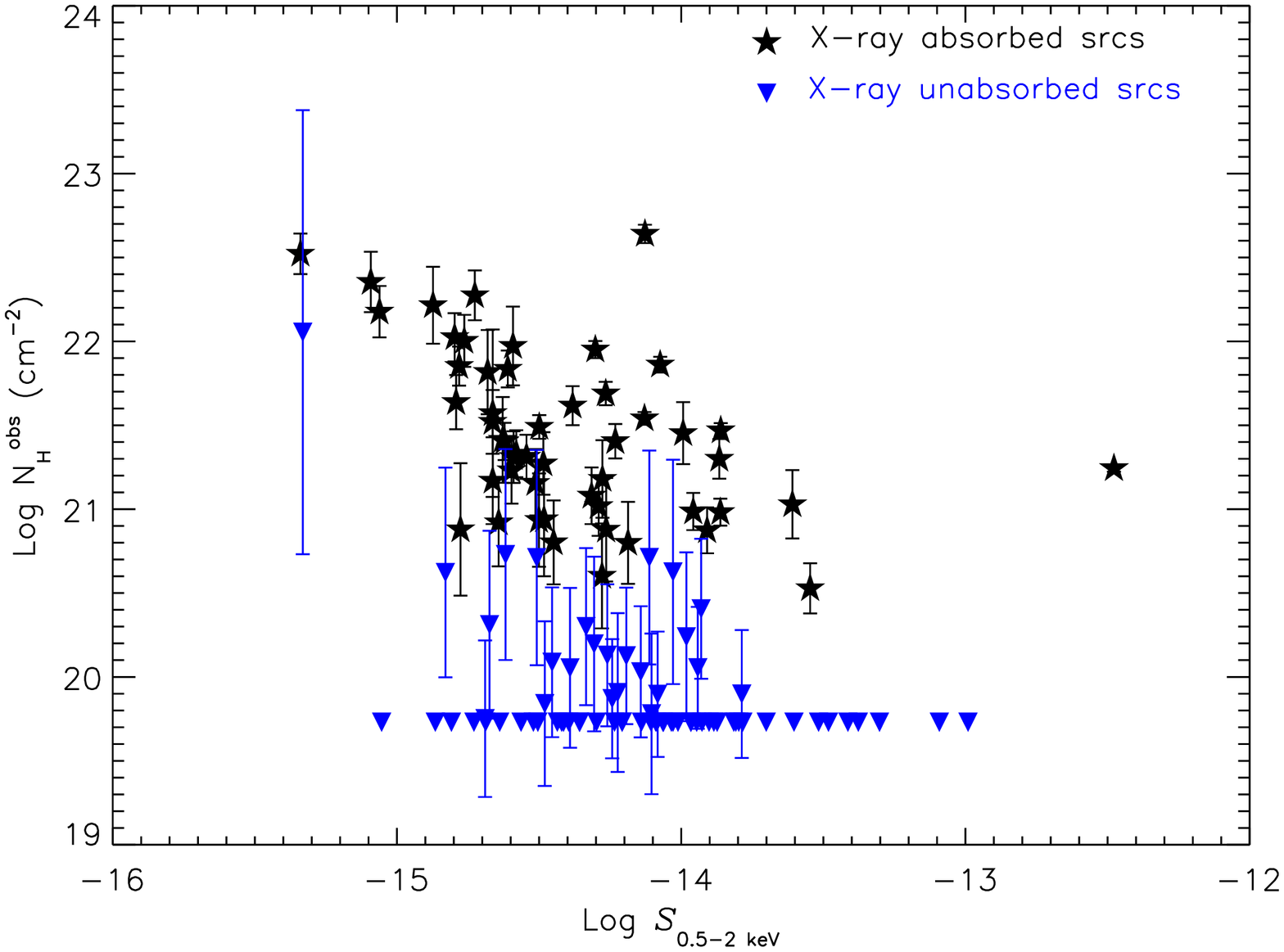}
    \includegraphics[angle=90,width=0.50\textwidth]{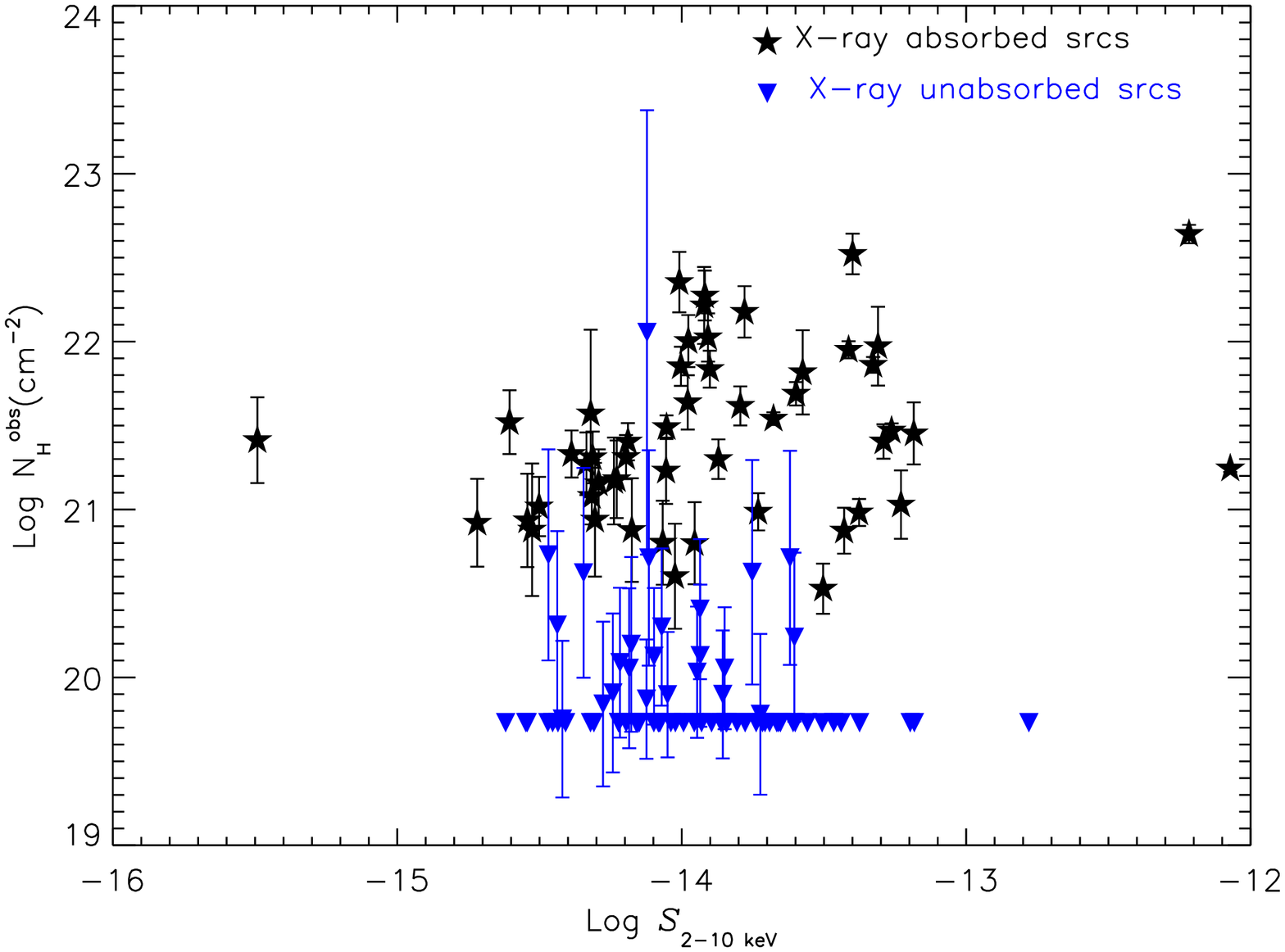}}
    \caption{Distribution of ${\rm N_H^{obs}}$ for absorbed 
      (F-test $\ge 95\%$; squares)
      and unabsorbed (F-test $<$ 95\%; triangles) objects. 
      Error bars correspond to 90\% confidence.
    }
    \label{nhobs_vs_flux}
\end{figure*}

In Fig.~\ref{nhobs_vs_flux} we plot the values of absorption (observer's frame)
that we obtained with the APL model, as a function of the 0.5-2 and 2-10 keV fluxes.
There is an obvious correlation between the absorption and the observed soft flux.
We see that the distribution of absorbing column densities does not seem to vary with the hard band
fluxes. We have studied the dependence of ${\rm N_H^{obs}}$ with the 0.5-2 keV flux using de-absorbed fluxes,
i.e. the 0.5-2 keV fluxes corrected for the effect of absorption.
We found that when de-absorbed fluxes are used, ${\rm N_H^{obs}}$ does not
vary with X-ray flux, i.e., fainter and/or more distant objects do not appear to be more
absorbed.

%\begin{figure}
%    \hbox{
%    \includegraphics[angle=90,width=0.50\textwidth]{gamma_vs_sflux_lh_best_fit_deabsorbed.ps}
%    \includegraphics[angle=90,width=0.50\textwidth]{nhobs_vs_sflux_abs2noabs_lh_deabsorbed.ps}}
%    \caption{Distribution of ${\rm N_H^{obs}}$ vs soft flux for absorbed (F-test $\ge 95\%$)
%      and unabsorbed (F-test $< 95\%$) objects. We used the flux values obtained from the APL
%      model but we corrected them for the effects of absorption.
%    }
%    \label{nhobs_deabsorbed_vs_flux}
%\end{figure}

\begin{figure*}
    \hbox{
    \includegraphics[angle=90,width=0.50\textwidth]{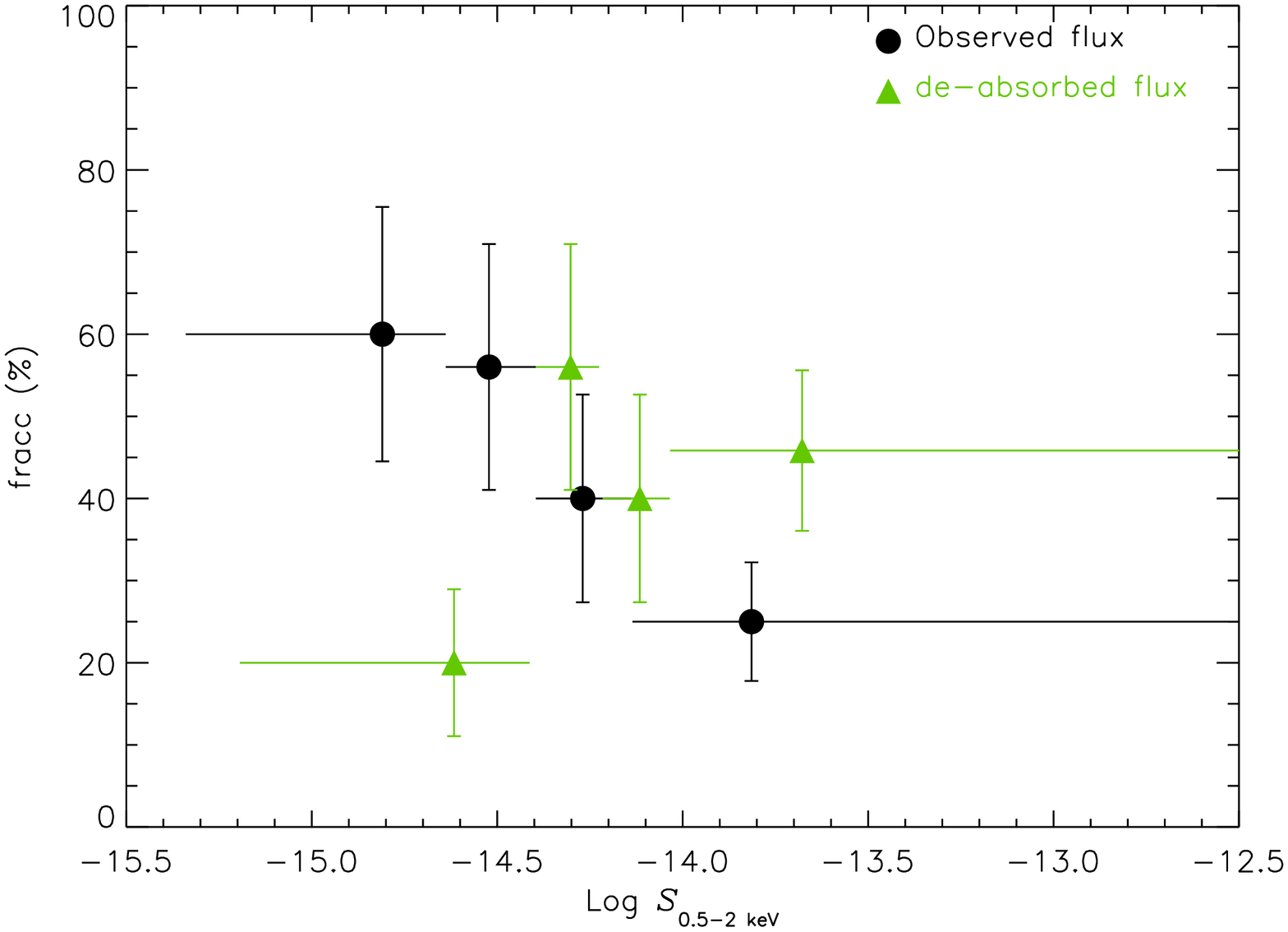}
    \includegraphics[angle=90,width=0.50\textwidth]{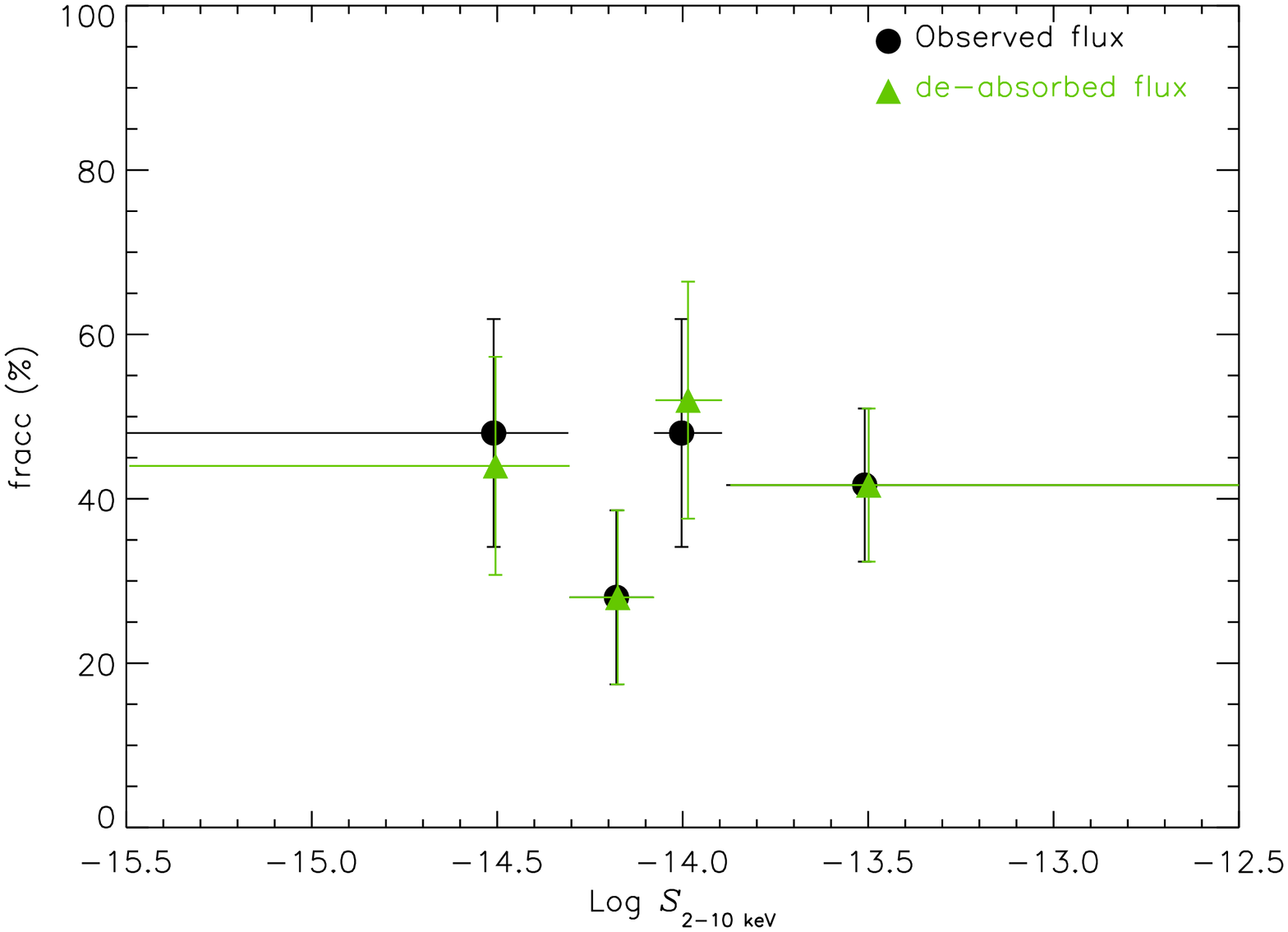}}
    \caption{Fraction of absorbed (F-test$\ge$95\%) objects vs 0.5-2 (left)
      and 2-10 (right) keV flux. For each source we used the fluxes 
      from its best fit model (single power law or absorbed power law). 
      Circles show the results that we obtained using observed fluxes, while 
      the triangles show the results using fluxes corrected for the effect 
      of absorption. See Sec.~\ref{APL_fitting} for details.
    }
    \label{abs_fracc_vs_flux}
\end{figure*}

We have studied whether the fraction of X-ray absorbed objects depends on the 
flux after correcting for the effect of absorption (the ${\rm N_H}$ 
columns measured in 
our sample of sources are not high enough as to affect 
significantly the measured 2-10 keV fluxes and hence the 
correction of 2-10 keV fluxes for the effect of absorption is not significant).
The results are plotted in Fig.\ref{abs_fracc_vs_flux}.
For comparison we have plotted the results that we obtain when absorbed fluxes are used (circles).
We do not see significant differences using absorbed or de-absorbed 2-10 keV fluxes, because as explained
before, these are not significantly affected by the absorption measured in our sources.
However important differences are seen when 0.5-2 keV fluxes are used. If we do not
correct for the effect of absorption in the 0.5-2 keV flux, we see an increase in the fraction of
absorbed objects at fainter fluxes. However, if de-absorbed fluxes are used instead, the 
fraction of absorbed objects does not vary with the X-ray flux and we obtain the same result as in the 2-10 keV band. 

Note that the fraction of absorbed sources at the faintest de-absorbed 0.5-2 keV 
fluxes ($\sim20\%$) is significantly lower than the values found at brighter fluxes.
The absorbed sources that should contribute to the bin at the faintest 
fluxes have an observed flux below the threshold applied to our objects and therefore are not included  
in our sample (remember that to select our sources we used 0.2-12 keV counts, 
i.e. $\sim$fluxes without correction for absorption). 
Another effect that could also contribute to this result is that at the faintest fluxes we may have some 
sources with undetected absorption. We do not expect this effect to be important, because 
as we see in Fig.~\ref{gamma_vs_flux_best_fit} $\langle\Gamma\rangle$ for unabsorbed sources 
does not seem to become significantly harder at the faintest 0.5-2 keV fluxes.   

\begin{table}[!ht]
\caption{Results from the X-ray spectral analysis.}
\begin{tabular}{l c c c c c c}
\hline
 Model$^{\mathrm{a}}$ & {\rm Total} & {\rm type-1 AGN} & type-2 AGN & Not id.$^{\mathrm{b}}$\\
\hline
{\rm SPL}       & 65 & 35 &  5 & 25 \\
{\rm SPL + SE}  &  4 &  4 &  0 &  0 \\
{\rm APL}       & 39 &  6 & 16 & 17 \\
{\rm APL + SE}  &  5 &  0 &  4 &  1 \\
{\rm CAPL}    &  9 &  1 &  3 &  5 \\
{\rm 2SPL}        &  1 &  0 &  0 &  1 \\
\hline
{\rm Total}     & 123& 46 & 28 &  49 \\
\hline
\end{tabular}
\begin{list}{}{}
\item[$^{\mathrm{a}}$] Best fit model: SPL: single power law; APL: Absorbed power law; SE: Soft excess;
  CAPL: partial covering; 2SPL: two power laws (see Sec.~\ref{best_fit} for details).
\item[$^{\mathrm{b}}$] Objects without optical identifications.

\end{list}
\label{ress_fits}
\end{table}

\section{Best fit model}
\label{best_fit} Up to now we have shown the results from spectral fits where
only the X-ray continuum and the intrinsic absorption were modelled. However,
there are other spectral components that can also contribute significantly to
the emission in the 0.2-12 keV energy band. There is evidence for them in
the results that we have shown previously. For example we see in
Fig.~\ref{gamma_vs_flux_best_fit} that $\langle \Gamma \rangle$ seems not to
vary with the X-ray flux when absorption is included in the fitting model.
However we still have a clear scatter in the points which cannot be explained
if $\Gamma$ does not vary significantly with flux for the objects in our
sample as our results appear to show.  However, we would expect this scatter
of $\langle \Gamma \rangle$ if other spectral components are present in the
data (e.g. soft excess emission) and they are not properly modelled. Besides
the soft excess, other spectral components that can contribute to the 0.2-12
keV emission are ionised absorption, the Fe K$\alpha$ emission complex and the
Compton reflection hump that should appear at high X-ray energies.

We have studied in detail the MOS and pn time averaged spectra of each individual object.
To model the soft excess emission we have used a black-body model. This component adds two free parameters to the fit, the temperature (in keV) and normalisation of the black-body.
For some objects we could not get a good fit of the detected soft excess with a black-body.
In all these cases we obtained a good fit using a partial covering model
(i.e., only part of the X-ray
emission from the inner most region of the AGN is absorbed) to fit the signatures
of absorption and soft excess emission.
This model introduces one new parameter to the fit with respect to the APL model,
the covering fraction of the absorber
(between 0 and 1).
The only emission line that we expect to detect
with the quality of our spectra is the Fe K$\alpha$ complex at 6.4 keV (rest frame energy for neutral iron).
To search for this component we have used a Gaussian line profile, that allows us to calculate
the centroid (in most cases we fixed the centroid to 6.4 keV), width and normalisation of the line.
Absorption signatures found in some spectra were modelled with an absorption edge 
({\tt zedge} in {\tt xspec}). This 
model introduced two free spectral parameters,
the threshold energy and the absorption depth at the threshold energy.
The most prominent signature from reflection in AGN is a change in the slope of the X-ray
continuum at energies above 10 keV (rest frame).  This component is known as
the Compton reflection hump.
We do not expect our objects to be bright enough as to detect
with high significance reflection signatures given the limited bandpass of {\it XMM-Newton}.
However, we have searched for this component in all the spectra adding a
second power law to the fits at high energies.

The best fit model for each source is the one that gave a significant 
improvement in fit over the previous one in the sequence SPL-APL (with z in the 
case of identified sources)-APL+soft excess. The improvement in the fit 
was measured by the usual F-test, taking into account the improvement 
in the $\chi^2$ value and the number of new parameters introduced. 
For some objects with soft excess emission, the detection of absorption 
was only significant after modelling of this spectral component. In particular, 
when the classification (type-1/type-2 AGN) of the identified sources is used,
 the parameters from the APL model with intrinsic absorption are always used.

Our results are listed in Table~\ref{ress_fits}. The best fit spectral
parameters obtained for each object 
are shown in Table~\ref{ress_fits_all}. 
In the present paper we identify our objects
with the numbers that will be used in a forthcoming catalogue paper 
of the {\it Lockman Hole} (H. Brunner et al. 2005, in preparation).
The results were obtained fitting MOS and pn 
spectra simultaneously. If it was not required by the data we used the same MOS and 
pn model normalisations. For the sources for which we used different MOS and pn 
normalisations we show fluxes and luminosities obtained with the spectrum with best quality.

For 65 objects ($\sim$53\%) the best fit model was a single power law. Among
these objects there were 35 type-1 AGN and 5 type-2 AGN. We detected absorption in 53
objects out of 123 (43\%), including 7 type-1 AGN and 23 type-2 AGN. In
Table~\ref{tab_absorptions} we list the fraction of absorbed objects in our
sample and in the samples of type-1 and type-2 AGN. The values were corrected for
spurious detections \footnote{We accepted the spectral signatures as being
real if the significance of detection from the F-test was $\ge$95\%. Hence 
5\% of detections are expected to be spurious.} following the method described in
Mateos et al.~(\cite{Mateos2005}). We used the method described in 
Stevens et al.~(\cite{Stevens2005}) and Mateos et al.~(\cite{Mateos2005}) 
to compare the fraction
of absorbed objects in type-1 and type-2 AGN. We find that these fractions are
different (the number of absorbed objects in type-2 AGN being larger than in type-1 AGN)
with a significance of more than 99.99\%.

\begin{table}[!ht]
\caption{Results of detection of X-ray absorption.}
\begin{tabular}{l c c c c c c}
\hline
 & ${\rm N_{tot}}$ & ${\rm N_{abs}}$ & ${\it f}^{\mathrm{a}}$ &$f_{lim}^{\mathrm{b}}$\\
\hline
{\rm All sources} & 123 & 53 & 0.38 & $\ge$0.27 \\
{\rm type-1 AGN} & 46 & 7 & 0.10 & $\le$0.29\\
{\rm type-2 AGN} & 28 & 23 & 0.77 & $\ge$0.51 \\
\hline
\end{tabular}
\begin{list}{}{}
\item[$^{\mathrm{a}}$] Fraction of absorbed objects taking into account the expected 
fraction of spurious detections (see Sec.\ref{best_fit})
\item[$^{\mathrm{b}}$] 3$\sigma$ limits in the fraction of absorbed objects
\end{list}
\label{tab_absorptions}
\end{table}

\subsection{Broad band continuum}%-------------------------------------------------
\label{broad band continuum}
\begin{figure}
%    \hbox{
%    \includegraphics[angle=90,width=0.50\textwidth]{gamma_vs_sflux_lh_hand.ps}
%    \includegraphics[angle=90,width=0.50\textwidth]{gamma_vs_hflux_lh_hand.ps}}
    \hbox{
    \includegraphics[angle=90,width=0.50\textwidth]{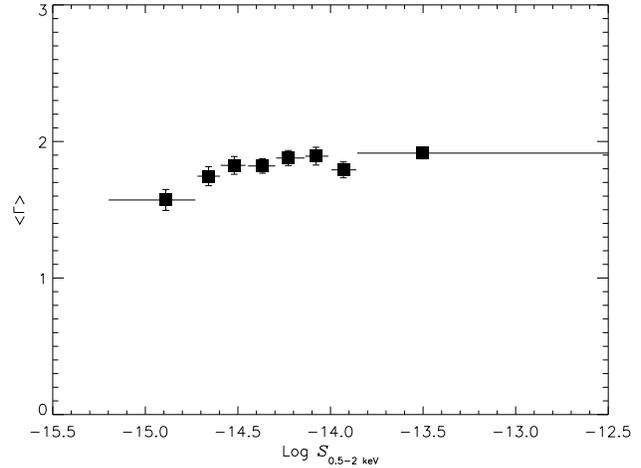}}
    \caption{Dependence $\langle \Gamma \rangle$ with 0.5-2 keV 
      flux. For each source we used the spectral parameters obtained 
      from its best fit 
      model (see Sec.~\ref{best_fit} for details).
    }
    \label{gamma_vs_flux_hand}
\end{figure}

Using the best fit spectral slopes we obtained a weighted $\langle \Gamma \rangle$
for the objects in our sample of $1.87\pm0.02$ ($1.86\pm0.02$ if the arithmetic mean
is used). In Fig.~\ref{gamma_vs_flux_hand} we show the dependence of $\langle \Gamma \rangle$ with the 0.5-2 keV flux using the best fit parameters ($\Gamma$ and {\it S}) for each object (the same 
result is obtained using 2-10 keV fluxes).
We see that when we take into account during fitting all the spectral components, the scatter
in $\langle \Gamma \rangle$ (see Fig.~\ref{gamma_vs_flux_best_fit}) is reduced significantly,
and most of the points are consistent with the obtained average value of $\Gamma$.

\begin{figure*}
    \hbox{
      \includegraphics[angle=90,width=0.50\textwidth]{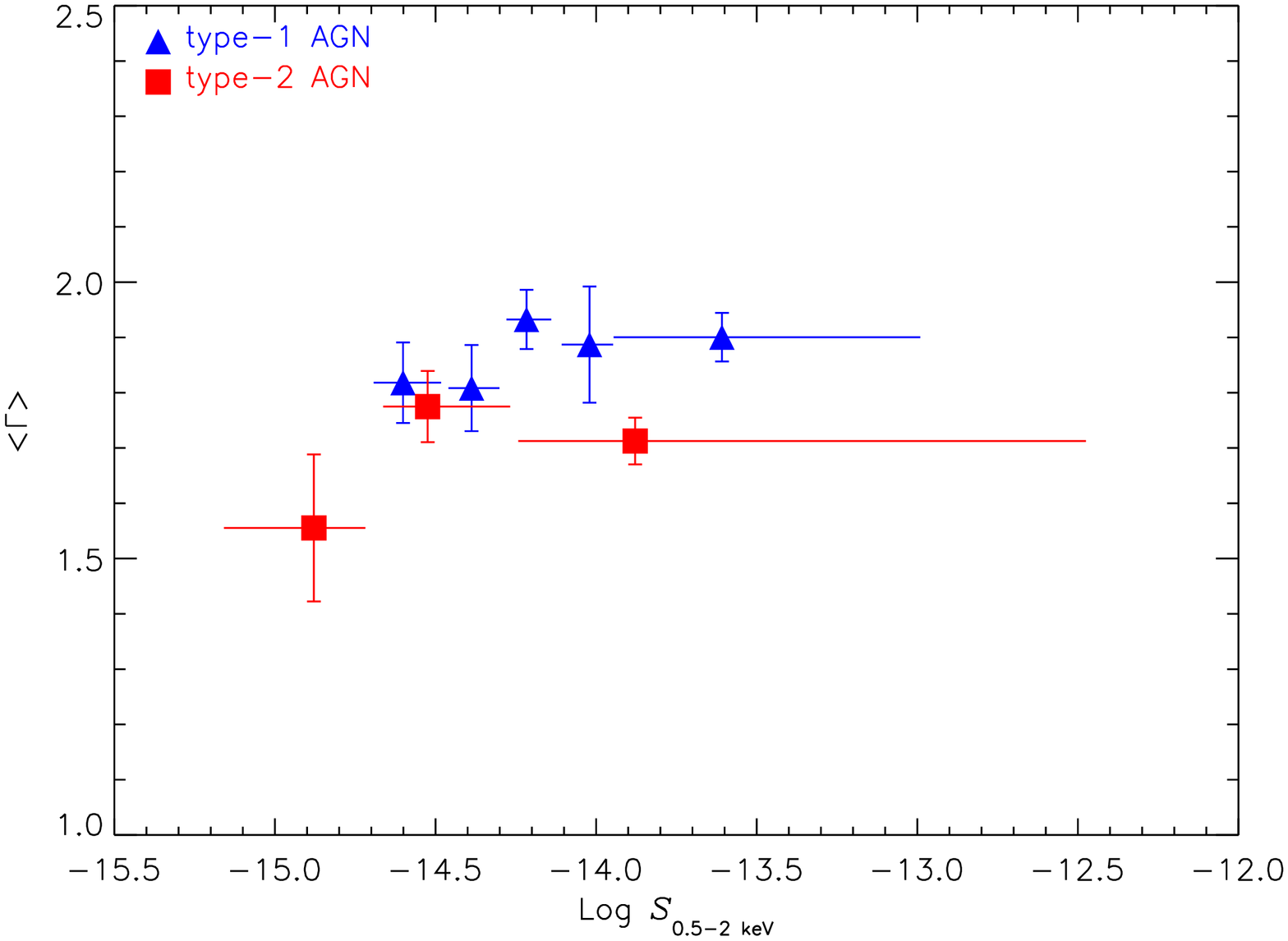}
      \includegraphics[angle=90,width=0.50\textwidth]{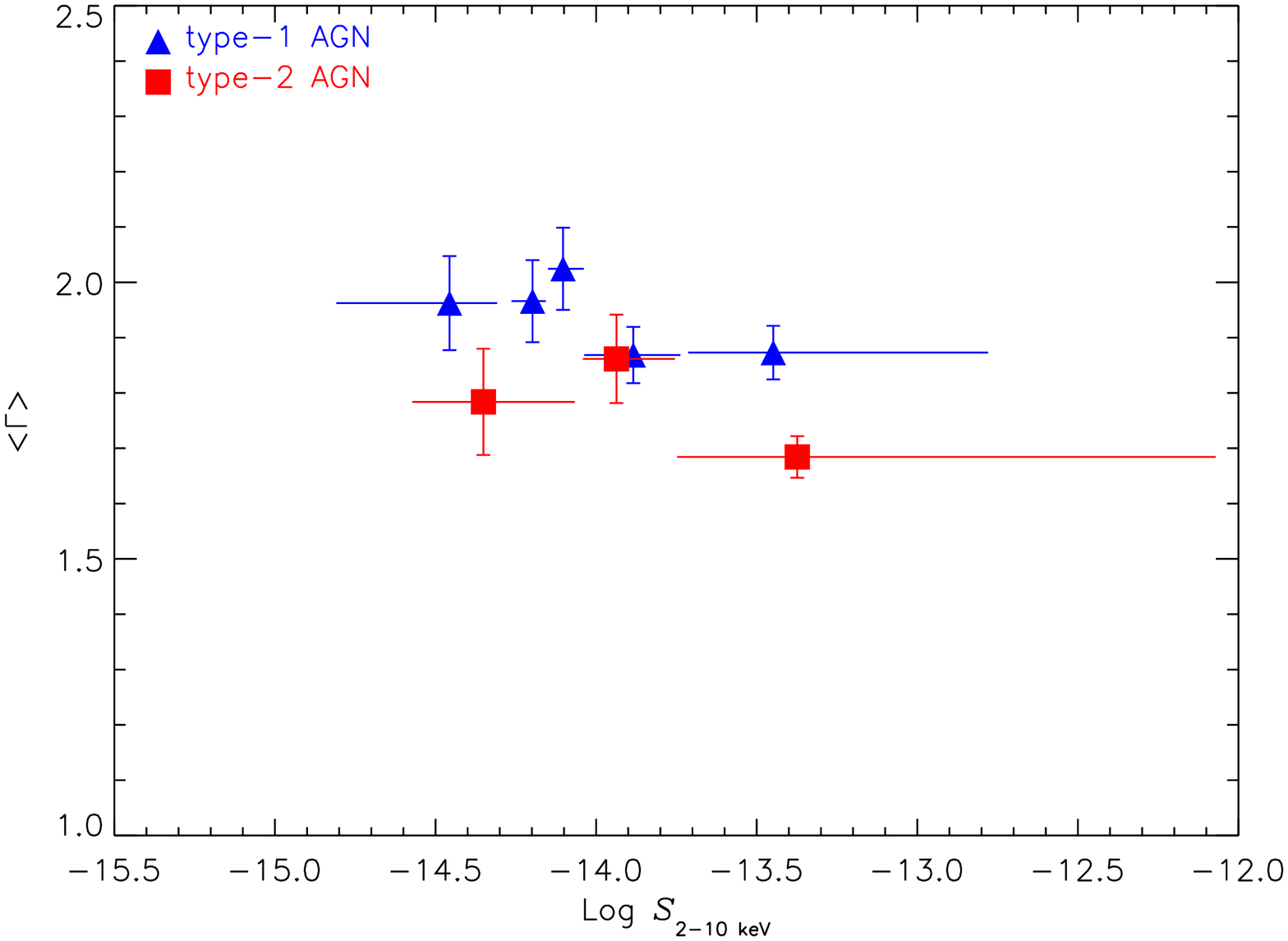}}
%   \hbox{
%    \includegraphics[angle=90,width=0.50\textwidth]{gamma_vs_sflux_agns_lh.ps}
%    \includegraphics[angle=90,width=0.50\textwidth]{gamma_vs_hflux_agns_lh.ps}}
    \caption{Dependence of $\langle \Gamma \rangle$ with 0.5-2 and 
      2-10 keV flux for type-1 and type-2 AGN. For each source we used 
      $\Gamma$ and {\it S} from its best fit model 
      (single power law or absorbed power law). See Sec.~\ref{broad band continuum} for details.
    }
    \label{gamma_vs_flux_agns}
\end{figure*}

The mean spectral slope was found to be $1.89\pm0.03$ ($1.88\pm0.03$ with the arithmetic mean) for type-1 AGN
and $1.71\pm0.03$ ($1.82\pm0.06$ with the arithmetic mean) for type-2 AGN.
In Fig.~\ref{gamma_vs_flux_agns} we compare $\langle \Gamma \rangle$ for type-1 and type-2 AGN as
a function of the X-ray fluxes. For type-1 AGN, where only 7 out of 46 are X-ray absorbed,
we obtain the same results using 0.5-2 and 2-10 keV fluxes, i.e., no dependence of
$\langle \Gamma \rangle$ with the X-ray flux. However, X-ray absorption is important in
type-2 AGN ($\ge$50\% of type-2 AGN being X-ray absorbed). The ratio of type-2 AGN/type-1 AGN increases
as we go to fainter 0.5-2 keV absorbed (i.e. not absorption corrected) fluxes,
while it remains constant with 2-10 keV flux.
We explained in detail in Sec.~\ref{nhobs_vs_flux} that this is due to the existing correlation
between ${\rm N_H}$ and the soft absorbed flux, i.e. most absorbed objects have the
faintest fluxes in the 0.5-2 keV band. We see in Fig.~\ref{gamma_vs_flux_agns} that
there is a clear dispersion in $\langle\Gamma\rangle$ for type-1 and type-2 AGN, however 
it seems that type-2 AGN tend to have lower $\langle \Gamma \rangle$ than type-1 AGN 
at the fluxes covered by our sample. 
Excluding from the sample of type-2 AGN the three sources that we 
found with no detected X-ray absorption and spectral slope significantly lower 
than the average value for type-1 AGN (see Sec.~\ref{unab_agn2} and Table~\ref{tabIIa}), we still see the same dependence of 
$\langle\Gamma\rangle$ with X-ray flux for type-2 AGN.

We have followed the procedure described in
Nandra \& Pounds~(\cite{Nandra1994}) and Maccacaro et al.~(\cite{Maccacaro1988})
to estimate the intrinsic dispersion of the photon index in type-1 and type-2 AGN, 
and to investigate whether after allowing for intrinsic dispersion in $\Gamma$, we still
find type-2 AGN to be on average flatter than type-1 AGN.
In this method it is assumed that the dispersion in $\Gamma$ values can be described well with a Gaussian function
of mean $\langle \Gamma \rangle$ and dispersion $\sigma_{\langle \Gamma \rangle}$.
The results of this analysis
are listed in Table~\ref{tab3}, where we have the values obtained using the weighted mean for
comparison. We have found that there is an intrinsic dispersion in $\Gamma$ of $\sim 0.2$ in type-1 AGN and type-2 AGN,
and that the value of the dispersion is similar in both samples of objects.
It is interesting to note that the results obtained with the weighted 
mean and the Maximum Likelihood method are consistent within each other.

\begin{figure}
    \hbox{
    \includegraphics[angle=-90,width=0.50\textwidth]{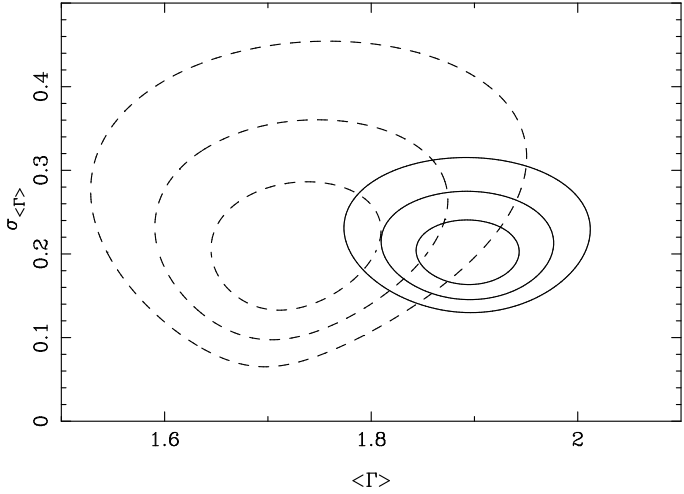}}
    \caption{Contour diagrams for the value of the average spectral slope and intrinsic dispersion of
      our samples of type-1 AGN (solid lines) and type-2 AGN (dashed lines) obtained from the Maximum Likelihood
      analysis (see Sec.~\ref{best_fit}). The contours are defined as $\Delta\chi^2$=2.3, 6.17 and 11.8
      corresponding to standard 1, 2 and 3$\sigma$ confidence regions for two parameters.
     }
    \label{gamma_agns_likelihood}
\end{figure}

In Fig.~\ref{gamma_agns_likelihood} we show the contours in $\langle \Gamma
\rangle$-$\sigma_{\langle \Gamma \rangle}$ space for a $\Delta \chi^2$ of 2.3,
6.17 and 11.8 that correspond to 1, 2 and 3$\sigma$ for two parameters. The
significance of type-2 AGN being on average flatter than type-1 AGN is only at
$1.62\sigma$ (using the values of $\Gamma$ obtained with the ML method \footnote{ Note that 
the significance 
of $\Gamma$ being different for type-1 AGN and type-2 AGN is $\sim4\sigma$ if the values 
obtained with the weighted mean are used and no intrinsic dispersion in $\Gamma$ is 
considered}). 

However, it is important to note that if the signatures of absorption 
in the X-ray spectra are not very significant, the detected values 
of ${\rm N_H}$ will tend to be lower than the real ones 
(see Mateos et al.~\cite{Mateos2005}) and then, the fitted $\Gamma$ will be flatter. 
We expect this effect to be more important for type-2 AGN where we have more sources with 
absorption. The small difference in $\langle\Gamma\rangle$ for type-1 and type-2 AGN 
might be due to this effect. Therefore with the current data we cannot reach any strong conclusion.

\subsection{X-ray absorption} %-------------------------------------------------\label{xabs}
 
   \begin{table}[bl] \caption[]{Mean spectral photon index
   of type-1 and type-2 AGN obtained with the weighted and arithmetic means and
   with the Maximum Likelihood analysis.
   The spectral slopes from the sources' best fit model were used.} $$ \begin{array}{l c c c c c}
   \hline \noalign{\smallskip}
   & \multicolumn{2}{c} {\rm{Maximum}} & {\rm{Weighted}} & {\rm {Arithmetic}}\\
   & \multicolumn{2}{c}{\rm{Likelihood}} &
   {\rm{Mean}} & {\rm Mean} &\\ {\rm Sample} & \langle\Gamma\rangle & \sigma &
   \\ \noalign{\smallskip} \hline \hline
   \noalign{\smallskip} {\rm Whole\,\,sample} & 1.92\,_{0.18}^{0.03} & 0.28\,_{0.13}^{0.04} & 1.87\pm0.02 & 1.86\pm0.02 \\
   {\rm type-1\,AGN} & 1.89\,_{0.05}^{0.06} & 0.20\,_{0.04}^{0.04} & 1.89\pm0.03 & 1.88 \pm 0.03 \\
   {\rm type-2\,AGN} & 1.72\,_{0.08}^{0.10} & 0.20\,_{0.07}^{0.10} & 1.71\pm0.03 & 1.82\pm 0.06\\
%   {\rm Optically\,\,faint\,\,sources} & 1.81\,_{0.06}^{0.05} & 0.16\,_{0.07}^{0.07} & 1.82\pm0.03\\
   \noalign{\smallskip} \hline \end{array} $$
\begin{list}{}{}
\item[]
\end{list}
\label{tab3}
\end{table}

\begin{figure}
    \hbox{
    \includegraphics[angle=90,width=0.50\textwidth]{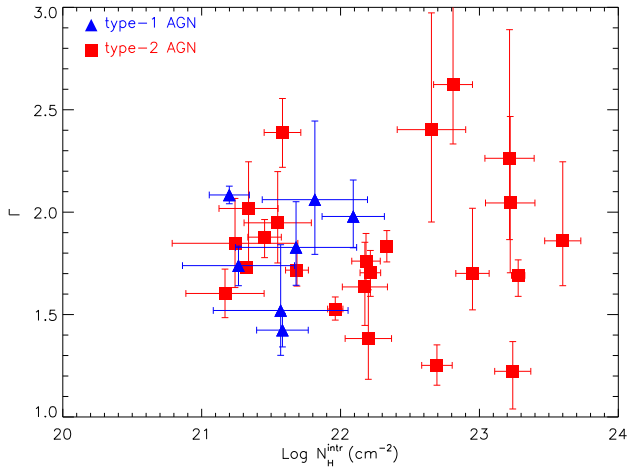}}
    \caption{$\Gamma$ vs. ${\rm N_H^{intr}}$ (rest-frame) for the type-1 and 
      type-2 AGN with detected absorption. Note
      that all type-1 AGN have column densities between 
      ${\rm 10^{21}-10^{22}\,cm^{-2}}$
      while type-2 AGN have a much wider distribution of ${\rm N_H^{intr}}$.
    }
    \label{gamma_nh}
\end{figure}

We have detected X-ray absorption in $\sim$ 37\% of the sources in our sample.
Absorption was found in $\sim 10\%$ of type-1 AGN and $\sim 77\%$ of 
type-2 AGN. We first checked that the measured $\Gamma$ 
and ${\rm N_H}$ were not correlated, and therefore
that we have obtained reliable parameters, especially the 
column density for each individual
object. The results are plotted in Fig.~\ref{gamma_nh} for 
sources with known redshifts,
where we do not
see any evident correlation between the two spectral parameters.
In the objects with large $\Gamma$, the values including the error bars
are in all cases consistent with a value of $\Gamma\sim$2.

\begin{figure}
    \hbox{
    \includegraphics[angle=90,width=0.50\textwidth]{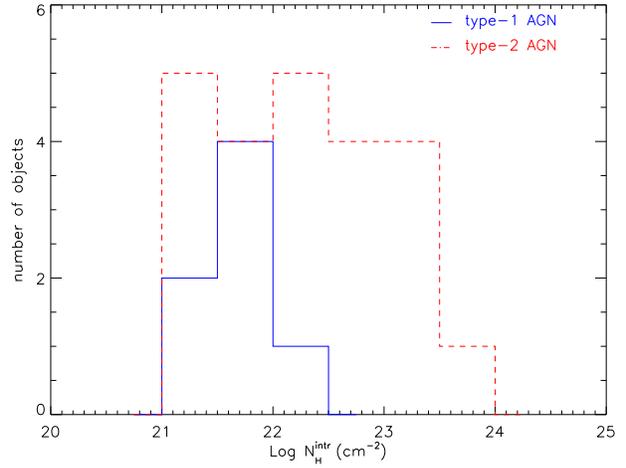}}
    \caption{Distributions of intrinsic (rest frame) absorption in 
      type-1 and type-2 AGN
      obtained from the best fit model for sources. 
    }
    \label{nhintr_distrs}
\end{figure}

Note in Fig.~\ref{gamma_nh}, that the ${\rm N_H^{intr}}$ distributions in
type-1 and type-2 AGN seem to be different. The measured column densities in 
absorbed type-1 AGN are between ${\rm 10^{21}-10^{22}\,cm^{-2}}$, while 
type-2 AGN have a much wider distribution of values, many objects 
having ${\rm N_H^{intr}\ge10^{23}\,cm^{-2}}$.
We show the distributions of {$\rm N_H^{intr}$} in type-1 and type-2 AGN in Fig.~\ref{nhintr_distrs}.
The distributions appear to be different, with type-2 AGN being in general more absorbed than
type-1 AGN. Using the KS test to compare the two distributions we obtained a probability of
them being different of $>92\%$. We will have to wait for the analysis of the faint
sample of objects before reaching a stronger conclusion.

\begin{figure}
    \hbox{
    \includegraphics[angle=90,width=0.50\textwidth]{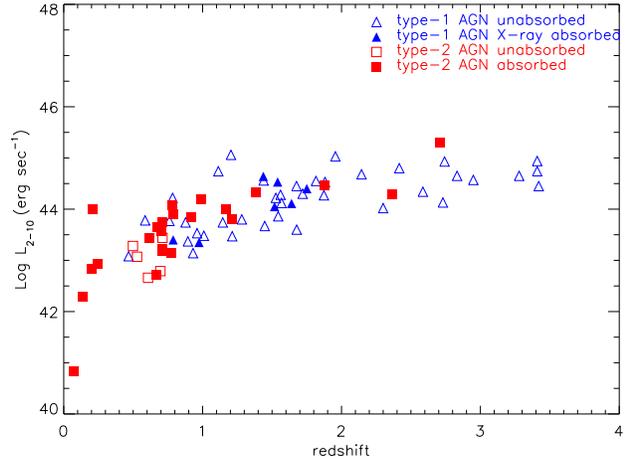}}
    \caption{2-10 keV luminosity (corrected for absorption) 
      vs. redshift for type-1 and type-2 AGN. 
      Absorbed sources are the objects where 
      we found an F-test significance of improvement of the fits $\ge95\%$.
    }
    \label{z_lumin}
\end{figure}

In terms of the unified model of AGN, X-ray absorption and optical obscuration
should be correlated. There is observational evidence that this does not
hold for all AGN, although only a few cases are based
on a proper spectral analysis. We can confirm that these discrepancies exist
in a significant fraction of our objects.
Mateos et al. (\cite{Mateos2005}) found some indications that the AGN/host galaxy contrast
effect \footnote{In the redshift interval where the distributions of type-1 and type-2 AGN overlapped,
they found that unabsorbed type-2 AGN were less luminous than type-1 AGN} might explain why broad
optical lines are not observed in unabsorbed type-2 AGN. In Fig.~\ref{z_lumin} we show the redshift distribution
of the AGN in our sample. Because our sample of AGN is much smaller we can not perform a similar
test.

\subsection{Soft excess}%-------------------------------------------------
   \begin{table*}[!ht] \caption[]{Properties of the soft excess emission in
   type-1 and type-2 AGN that was modelled with a black body.}
   $$ \begin{array}{c c c c c c c}
   \hline \noalign{\smallskip}
   {\rm{ID}} & {\rm{Class}} & {\rm redshift} &{\rm ({{\rm L_{{\rm BB}}/L_{{\rm PO}})}}} & {\rm kT} & {\rm L_{BB}} & {\rm L}\\
   & & & {\rm (0.5-2\,keV)}& {\rm (eV)} & {\rm (0.5-2\,keV)} & {\rm (2-10\,keV)}\\
   (1) & (2) & (3) & (4) & (5) & (6) & (7) \\
   \noalign{\smallskip} \hline \hline
   \noalign{\smallskip}
   {\rm 90}  & {\rm type-1\,AGN} & 0.467 & 0.363   & 105_{-20}^{+17} & 42.64 & 43.08\\
   {\rm 148} & {\rm type-1\,AGN} & 1.113 & 0.479   &  78_{- 8}^{+ 6} & 44.44 & 44.74\\
   {\rm 270} & {\rm type-1\,AGN} & 1.568 & 0.426   & 109_{-15}^{+10} & 43.80 & 44.12\\
   {\rm 342} & {\rm type-1\,AGN} & 0.586 & 0.081   &  67_{- 7}^{+ 3} & 42.79 & 43.78\\

   {\rm 259} & {\rm type-2\,AGN} & 0.792 & 0.063  & 291_{-90}^{+149} & 42.11 & 43.90\\
   {\rm 290} & {\rm type-2\,AGN} & 0.204 & 0.035  & 188_{-15}^{+ 17} & 41.32 & 42.83\\
   {\rm 424} & {\rm type-2\,AGN} & 0.707 & 0.060  & 474_{-153}^{+26} & 42.25 & 43.74\\
   {\rm 511} & {\rm type-2\,AGN} & 0.704 & 0.014  &  83_{- 38}^{+87} & 42.12 & 43.58\\

   \noalign{\smallskip} \hline \end{array} $$
%\begin{list}{}{}
%\item[$^{\mathrm{a}}$] Strength of the soft excess defined as the ratio of black body and power law luminosities in the 0.5-2 keV band.
%\item[$^{\mathrm{b}}$] Temperature of the black body in eV.
%\item[$^{\mathrm{b}}$] 2-10 keV luminosity of the power law component (corrected for soft excess and absorption).
 Columns are as follows: (1) Source X-ray identification number; 
(2) object class based on optical spectroscopy;
(3) redshift;
(4) ratio of soft excess to power law 0.5-2 keV luminosities (this ratio is frequently used to measure 
the strength of the soft excess emission);
(5) temperature of the soft excess (using a black-body model);
(6) logarithm of the 0.5-2 keV luminosity of the soft excess component; 
(7) logarithm of the 2-10 keV luminosity of the power law component (for 
absorbed sources the luminosity was absorption corrected).
%
%\end{list}
\label{tab3a}
\end{table*}

We detected soft excess emission ($\ge$95\% confidence limit from an F-test) 
in 18 (15\%) objects. The number of MOS+pn counts in the soft 
excess component vary from 100 to 1000 except for one source where 
the soft excess component has $\sim$3000 counts.
Within the sources classified as AGN we found soft excess in 5 (11\%) 
type-1 AGN and 7 (25\%) type-2 AGN. The 
significance of the fractions of
type-1 and type-2 AGN with detected soft excess emission 
being different is 97\%. 

Although our results suggest that soft excess emission might be more 
common in type-2 AGN, it is important to note that our samples of type-1 and type-2 AGN 
have different redshift distributions, and for the highest 
redshift sources (all type-1 AGN) we expect most of the signatures of soft excess 
emission to be redshifted outside the observed energy interval, making the 
detection of soft excess more difficult. Therefore, we have repeated 
the comparison using only sources in the redshift interval where we detected soft excess 
(z$<$1.568). In this case, the significance of the fractions of
type-1 and type-2 AGN with detected soft excess emission 
being different is reduced to 87\%. Hence, with our data 
we cannot confirm that soft excess emission is more common in type-2 AGN than in type-1 AGN.

In 9 sources (4 type-1 AGN, 4 type-2 AGN and 1 unidentified object) we fitted 
the soft
excess emission with a black body model (a Raymond Smith model gave an equally
good fit). The properties of the soft excess, i.e. temperature, 0.5-2 keV 
luminosity and strength, for the identified sources are
listed in Table~\ref{tab3a}. We see that the measured black-body properties do 
not 
depend on the 2-10 keV X-ray luminosity of the objects.
For the unidentified object with detected soft excess fitted with a 
black body we found an observed
black body temperature of $0.164_{+0.033}^{-0.046}$ eV.

The average temperature of the black body was found to be 0.09$\pm$0.01 keV for type-1 AGN
and 0.26$\pm$0.08 keV for type-2 AGN. The average 0.5-2 keV luminosities of the black body 
(in log units) were
${\rm 43.42\pm0.43\,erg\,s^{-1}}$ for type-1 AGN and
${\rm 44.11\pm0.44\,erg\,s^{-1}}$ for type-2 AGN. The 0.5-2 keV
luminosities of the soft excess component in type-1 and type-2 AGN were not found to be 
significantly different
(a KS test of the luminosity distributions gave a significance of them being different of only 90\%).
However the measured temperatures of the soft excess were found to be higher in type-2 AGN than in type-1 AGN. This could be 
because the soft excess in type-2 AGN might contain a fraction of scattered radiation. We also see that in most sources the temperatures of the 
black body are 
well above 60 eV \footnote{The hottest thermal emission expected from an accretion 
disc surrounding a $10^6{\rm M_\odot}$ black hole.}, and hence it is difficult to explain the origin of the 
soft excess emission in these sources as thermal emission from the accretion 
disc. Componization of cool photons in a cloud of hot electrons surrounding 
the accretion disc might be an alternative explanation. 

In 9 other objects (1 type-1 AGN, 3 type-2 AGN and 5 unidentified objects) the black body could not fit the signatures of the soft excess. An alternative method for modelling the curvature at soft energies is a
scattering or partial covering model ({\tt pcfabs} in {\tt xspec}). The model
consists of the sum of two power law components having the same spectral index, but affected by
different absorption (quantified with the covering fraction parameter). This model improved
significantly the quality of the fits, and provided a good fit of the soft excess 
emission in all sources. The average covering fraction that we obtained
was 0.82$\pm$0.06 (the maximum and minimum
values being 0.98 and 0.50). This value implies that 
the scattering fraction in these sources is rather large ($18\pm6\%$).

%\begin{figure*}
%    \hbox{
%    \includegraphics[angle=90,width=0.50\textwidth]{hist_kT_lh.ps}
%    \includegraphics[angle=90,width=0.50\textwidth]{hist_cvf_lh.ps}}
%    \caption{kT distributions
%    }
%    \label{kT_distr}
%\end{figure*}

\subsection{Reprocessed components} %-------------------------------------------------

\begin{figure}
    \hbox{
    \includegraphics[angle=-90,width=0.50\textwidth]{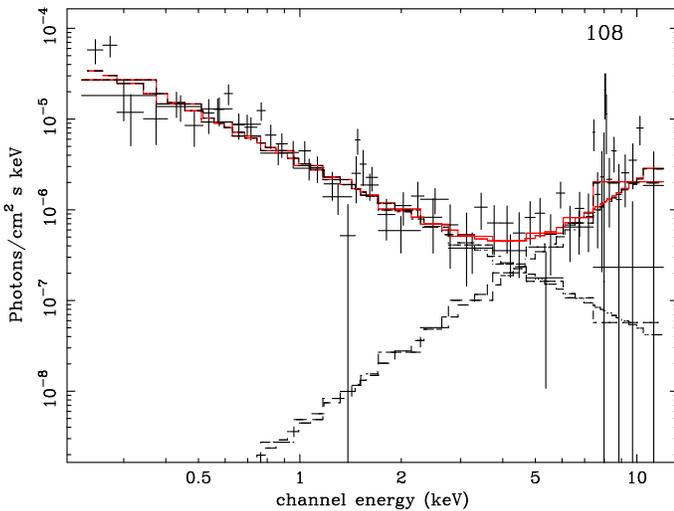}}
    \caption{Unfolded MOS and pn spectrum of the only source in the 
      sample where we detected a hardening in the X-ray continuum emission at 
      high energies (Compton reflection hump). Because this object 
      is still unidentified we fitted the hard spectral 
      component with a second power law. The F-test significance of this 
      component was found to be 99.98\%.
    }
    \label{sp_examples}
\end{figure}

   \begin{table*}[!ht] \caption[]{Parameters of the Gaussian line in the six identified sources where 
       signatures of line emission where detected with an F-test significance $\ge95\%$.
     }
   $$ \begin{array}{c c c c c c c c}
   \hline \noalign{\smallskip}
   {\rm{ID}} & {\rm{Class}} & {\rm redshift} & {\rm L_{2-10}} & {\rm Line} & \sigma & {\rm Equivalent}  
& {\rm F-test} \\
   &              &             & {\rm erg\,s^{-1}} & {\rm energy\,(keV)} & {\rm (keV)} & {\rm width\,(eV)} 
   & {\rm prob.\,(\%)} \\
   {\rm (1)} & {\rm (2)} & {\rm (3)} & {\rm (4)} & {\rm (5)} & {\rm (6)} & {\rm (7)} &{\rm (8)} \\
   \noalign{\smallskip} \hline \hline
   \noalign{\smallskip}
   {\rm 270} & {\rm type-1\,AGN}  & 1.568 & 44.12   & 6.24_{-0.08}^{+0.48} & 0.00_{-0.00}^{+0.82} & 452 & 99\\
   {\rm 21}  & {\rm type-2\,AGN} & 0.498 & 43.28   & 5.90_{-1.64}^{+0.92} & 0.56_{-0.45}^{+2.19} & 1462 & 95\\
   {\rm 172} & {\rm type-2\,AGN} & 1.170 & 44.00   & 6.40_{-0.41}^{+1.87} & 0.09_{-0.00}^{+2.93} & 360 & 99\\
   {\rm 290} & {\rm type-2\,AGN} & 0.204 & 42.83   & 6.32_{-0.61}^{+0.07} & 0.00_{-0.00}^{+0.14} & 283 & 96\\
   {\rm 326} & {\rm type-2\,AGN} & 0.780 & 44.08   & 6.59_{-0.09}^{+0.06} & 0.02_{-0.00}^{+2.12} & 224 & 99\\
   {\rm 407} & {\rm type-2\,AGN} & 0.990 & 44.20   & 6.40_{-0.43}^{+0.10} & 0.40_{-0.18}^{+0.10} & 653 & >99.99\\
   \noalign{\smallskip} \hline \end{array} $$
%\begin{list}{}{}
%\item[$^{\mathrm{a}}$] Equivalent width in the object rest frame.
%\item[$^{\mathrm{b}}$] Temperature of the black body in eV.
%\item[$^{\mathrm{b}}$] 2-10 keV luminosity of the power law component (corrected for soft excess and absorption).
 Columns are as follows: 
(1) Source X-ray identification number; 
(2) optical class based on optical spectroscopy;
(3) redshift;
(4) 2-10 keV luminosity; 
(5) and (6) rest-frame centroid energy and width of the emission line; 
(7) rest-frame equivalent width of the line;
(8) F-test significance for the detection of the line.
%\end{list}
\label{tab6a}
\end{table*}

We have searched for a flattening of the continuum at high energies 
(i.e. Compton reflection) adding a second power
law to the model.
We have found signatures of spectral hardening at high 
energies in only one object, source 108, which is still unidentified. 
The unfolded 
MOS and pn spectra of this source are shown in Fig.~\ref{sp_examples}. 
We first fitted the X-ray spectrum of this object with a single power 
law giving $\Gamma\sim$1.5 but the fit was poor, with a $\chi^2$ of 127 
for 67 degrees of freedom. We found that there was a clear excess 
emission at high energies.
We then fitted the spectrum with two power laws, and the $\chi^2$ significantly decreased 
to 98 for 65 degrees of freedom.
The F-test significance of 
improvement of the fit with the new component was 99.98\%. 
In this case we obtained a 
value of $\Gamma$ of $1.83\pm0.17$ for the continuum emission flattening 
out to $\Gamma=-2.56_{-0.19}^{+0.61}$ at high energies. 
We did not find evidence for X-ray 
absorption or emission lines in the spectrum of this source. We will have to wait until we 
have the optical identification 
of this source before saying more about the X-ray emission of this object. 

\begin{figure*}
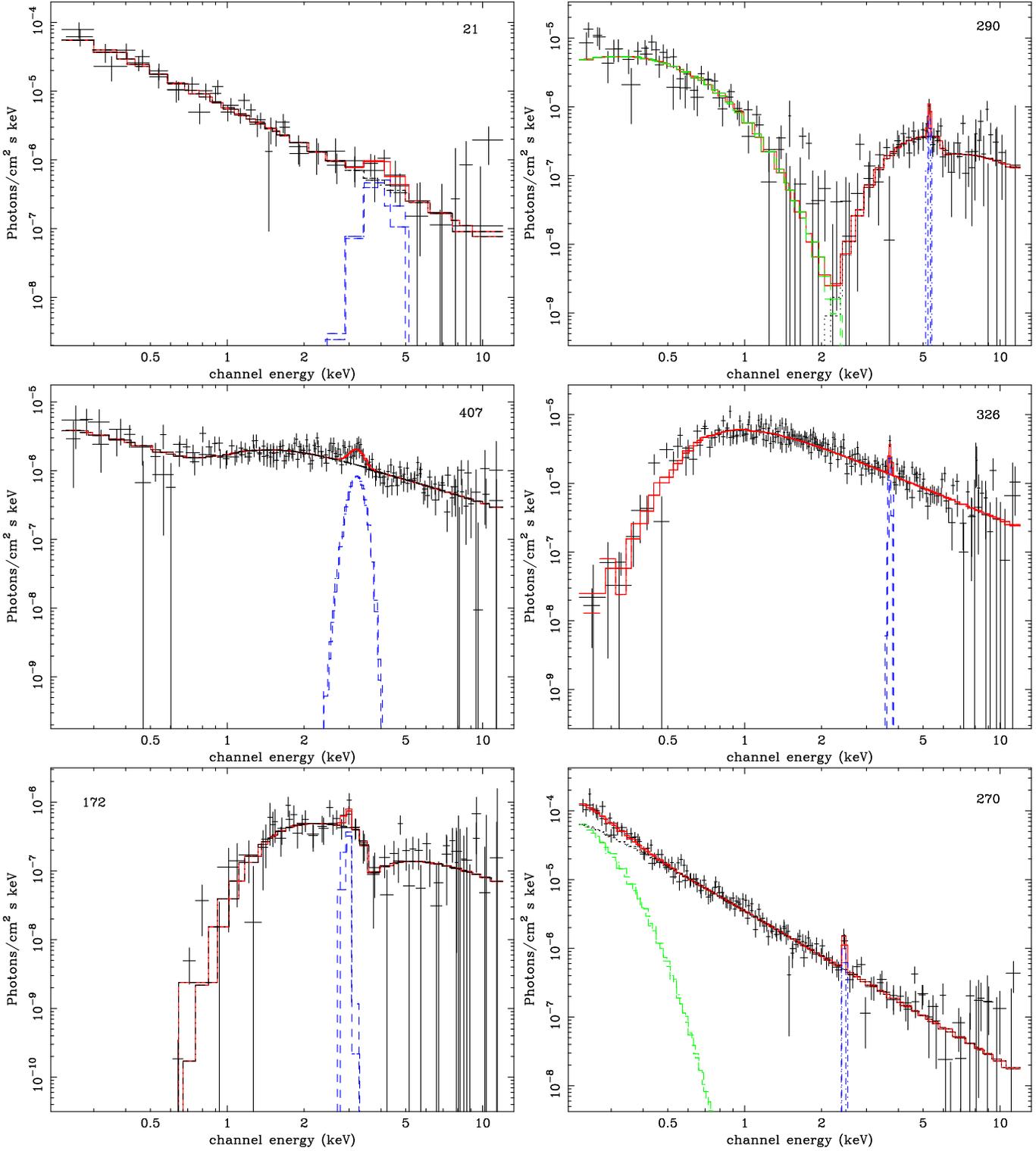

    \hbox{
    \includegraphics[angle=-90,width=0.5\textwidth]{fig22.ps}
    \includegraphics[angle=-90,width=0.5\textwidth]{fig23.ps}}
    \hbox{
    \includegraphics[angle=-90,width=0.5\textwidth]{fig24.ps}
    \includegraphics[angle=-90,width=0.5\textwidth]{fig25.ps}}
    \hbox{
    \includegraphics[angle=-90,width=0.5\textwidth]{fig26.ps}
    \includegraphics[angle=-90,width=0.5\textwidth]{fig27.ps}}
    \caption{Unfolded MOS and pn time averaged spectra of the 6 AGN 
      where we detected 
      signatures of emission line at high energies (F-test significance
      $\ge$95\%). In the X-ray spectrum of source 172 (type-2 AGN) 
      we also found an absorption edge at an energy of 
      $\sim7.56_{0.76}^{0.54}$ keV with absorption depth $\tau=1.4_{0.6}^{0.8}$ (F-test 
      significance of detection was 99\%)).
    }
    \label{sp_lines_agn}

\end{figure*}

Another signature of reprocessing that has been found in many spectra of AGN is 
an emission line around 6.4 keV. 
This is interpreted as Fe K$\alpha$ fluorescence 
from cold matter (Pounds et al., \cite{Pounds1989},\cite{Pounds1990}; 
Nandra et al., \cite{Nandra1991}; Nandra \& Pounds \cite{Nandra1994}) 
and might originate from the reprocessing of hard X-ray photons in the accretion disc (Pounds et al.
\cite{Pounds1990}). We have searched for this component in our sources using a
Gaussian model. Most time averaged spectra of our 
sources do not have enough signal to noise as to detect Fe K$\alpha$ line emission. However, 
we have been able to detect signatures of
line emission with an F-test significance $\ge95\%$ 
in the MOS and pn spectra of 8 objects (1 type-1 AGN, 5 type-2 AGN and 2 unidentified sources). In
Fig~\ref{sp_lines_agn} we show the MOS+pn unfolded time averaged spectra of these sources.
The parameters of the Gaussian line for each identified source are listed in 
Table~\ref{tab6a}.

{\bf Sources 21 and 407:} In these sources
we found a significant width in the line profile, 
which might be indicating that the line was formed in the 
inner parts of the accretion disc, and hence 
it should have a relativistic profile (with a red wing component 
due to gravitational redshift). When fitted with a Gaussian model, 
we would expect the line centroid to be found at an 
energy slightly below 6.4 keV. While in source 407 the line centroid is consistent 
with being neutral iron, in source 21 it was $\sim$5.9 keV (although 
consistent with being neutral iron within the error bars).

{\bf Sources 270 and 290:} In these sources we detected a narrow Gaussian 
line and line centroids lower (but consistent within the error bars) than 
the value for neutral iron. 

{\bf Sources 172 and 326:} In these sources we also found a significant line width, 
although in both cases it was consistent with zero at 90\% confidence. 
In source 172 the line centroid was consistent with being 
neutral iron, but in source 326 it was significantly higher (even within the 
error bars). In this source the line might be originating in an ionised accretion 
disc.

\begin{figure*}
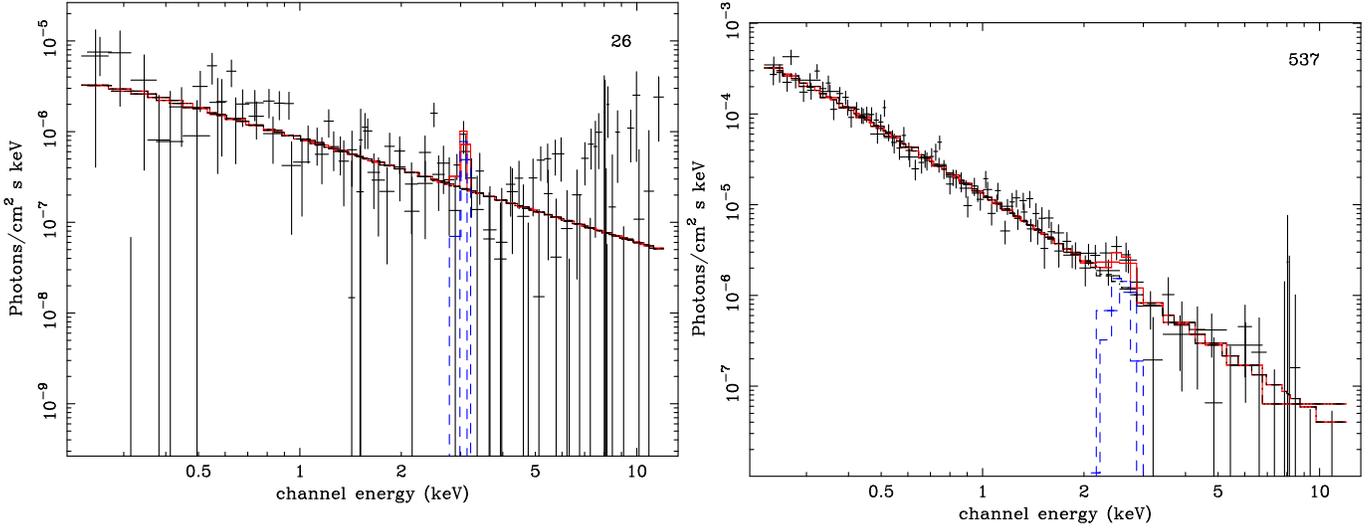

    \hbox{
    \includegraphics[angle=-90,width=0.5\textwidth]{fig28.ps}
    \includegraphics[angle=-90,width=0.5\textwidth]{fig29.ps}}
    \caption{Unfolded MOS and pn time averaged spectra of the 
      two still unidentified sources (source numbers 26 and 537) where we detected 
      signatures of emission line at high energies (F-test significance
      $\ge$95\%). 
    }
    \label{sp_lines_unid}
\end{figure*}

In all the spectra where we detected the line 
we did not have enough signal to noise in the data in order to use a 
more physical model to fit the profile of the line ({\tt 
xspec} models {\tt laor} for a Kerr black hole or {\tt diskline} for 
a Schwarzchild black hole). 
The rest frame equivalent width (EW) of the line in the type-1 AGN where we detected this 
component was found to be $\sim$452 eV. In most type-2 AGN the measured values were 
between 200 and 600 eV. However there is one source, 21 for which we found 
a rest frame EW of $\sim$1400, substantially higher than in the other type-2 AGN. 
It is important to note that in this source the F-test significance of detection 
of line was the lowest among all sources (95\%) and hence the measured value has 
the highest uncertainty.  

Using the same sample of objects, Streblyanska et al.~(\cite{Streblyanska2005})
found a clear relativistic line profile in the average rest-frame 
spectrum of type-1 and type-2 AGN. In objects with broad Fe lines, the contribution of the 
broad component is difficult to detect if there is not enough signal to noise 
since its contribution is less than 10\% above the continuum over most of the spectrum.

\section{Dependence of sources spectra with luminosity and redshift} %**************************
\label{dependence}

\begin{figure*}
    \hbox{
    \includegraphics[angle=90,width=0.50\textwidth]{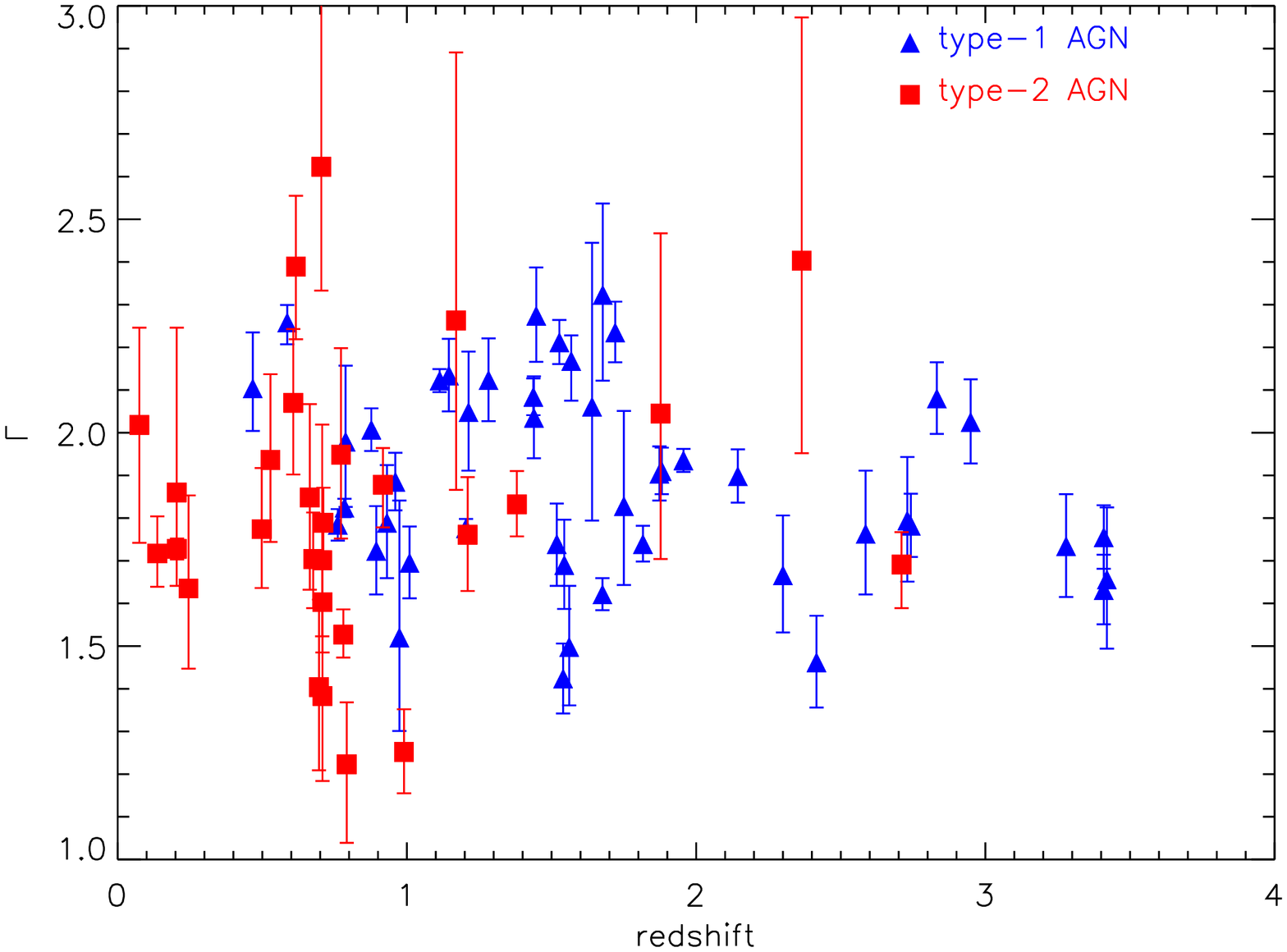}
    \includegraphics[angle=90,width=0.50\textwidth]{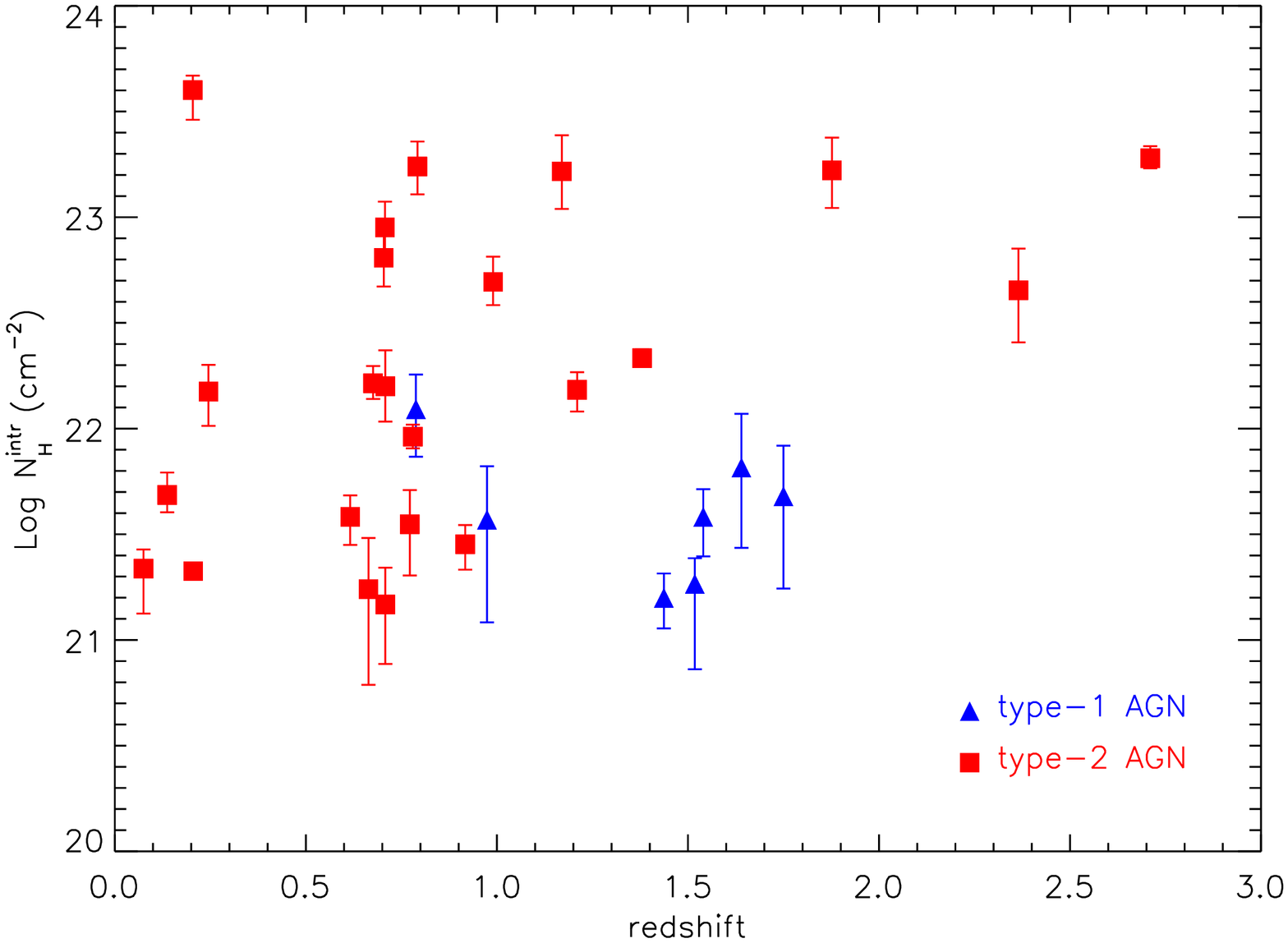}}
    \caption{Evolution with redshift of $\Gamma$ and ${\rm N_H^{intr}}$ (rest-frame) for 
      type-1 and type-2 AGN.
    Error bars correspond to 90\% confidence}
    \label{params_evol}
\end{figure*}

We show $\Gamma$ and ${\rm N_H^{intr}}$ vs. redshift for type-1 AGN and type-2 AGN in
Fig.~\ref{params_evol}.
In these plots we can see the different redshift distributions between
the type-1 and type-2 AGN in our sample. Most detected type-2 AGN have redshifts below 1, while we find 
type-1 AGN up to a redshift of $\sim$3.5.
We have applied a Spearman correlation test to search for 
evolution of $\Gamma$ and ${\rm N_H^{intr}}$ with redshift. 
We found that the correlation between $\Gamma$ and 
redshift is -0.22 for type-1 AGN and -0.04 for type-2 AGN. 
The significance of $\Gamma$ being flatter at higher redshifts is 86\%
for type-1 AGN and 15\% for type-2 AGN. 
The continuum shape of our sample of AGN does not seem to evolve with
redshift, however the number of AGN at high redshift (specially the number of 
type-2 AGN) is too small to give a strong conclusion.

The same result is obtained when searching for correlation of 
${\rm N_H^{intr}}$ with redshift, i.e. AGN at high redshift do not seem 
to be more absorbed than local ones. The apparent scarcity of high 
redshift (z$\ge$1) low ${\rm N_H}$ sources is probably a selection 
effect, since it is easier to detect highly absorbed sources at 
high redshifts.

\begin{figure*}
    \hbox{
    \includegraphics[angle=90,width=0.50\textwidth]{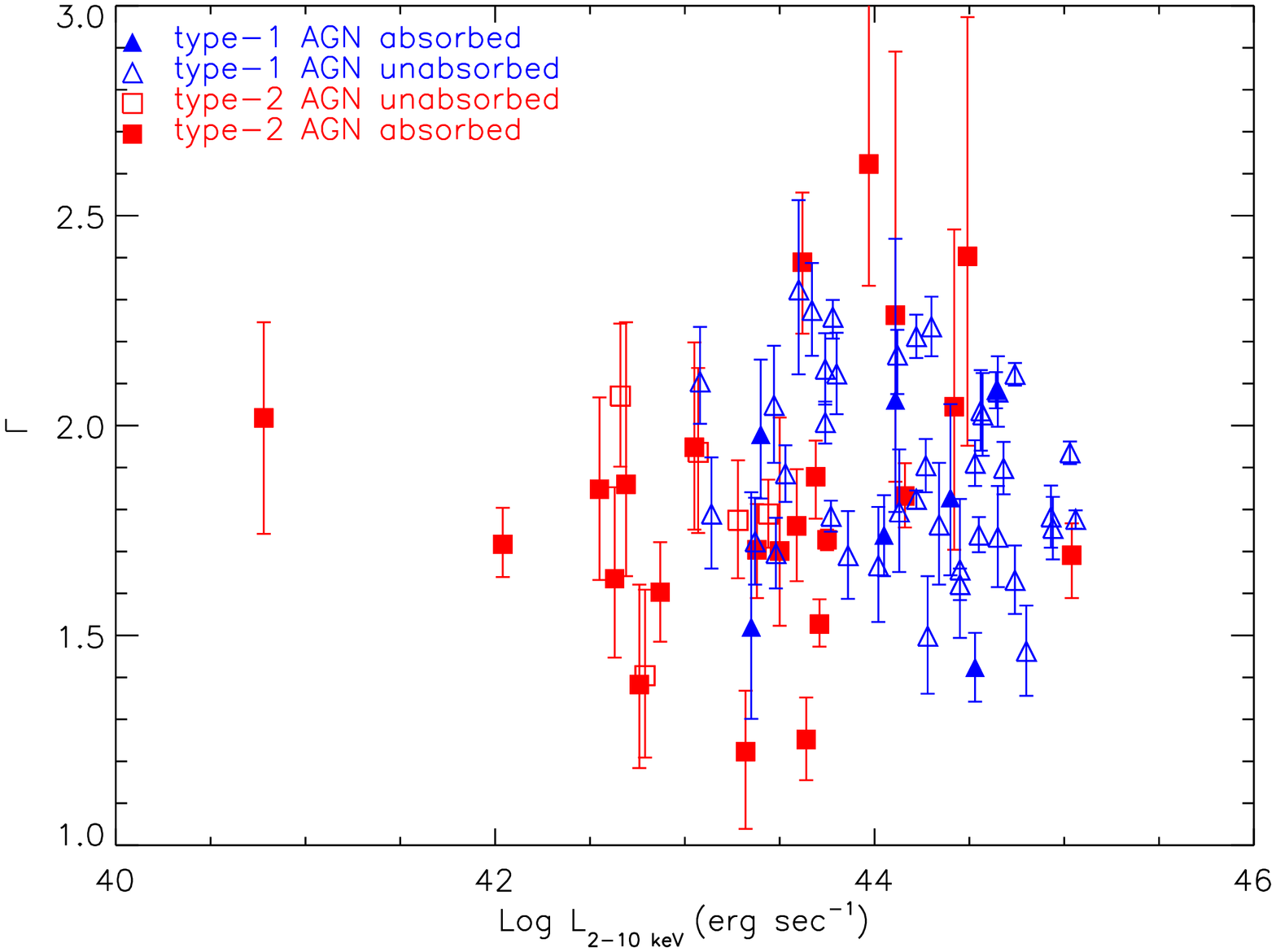}
    \includegraphics[angle=90,width=0.50\textwidth]{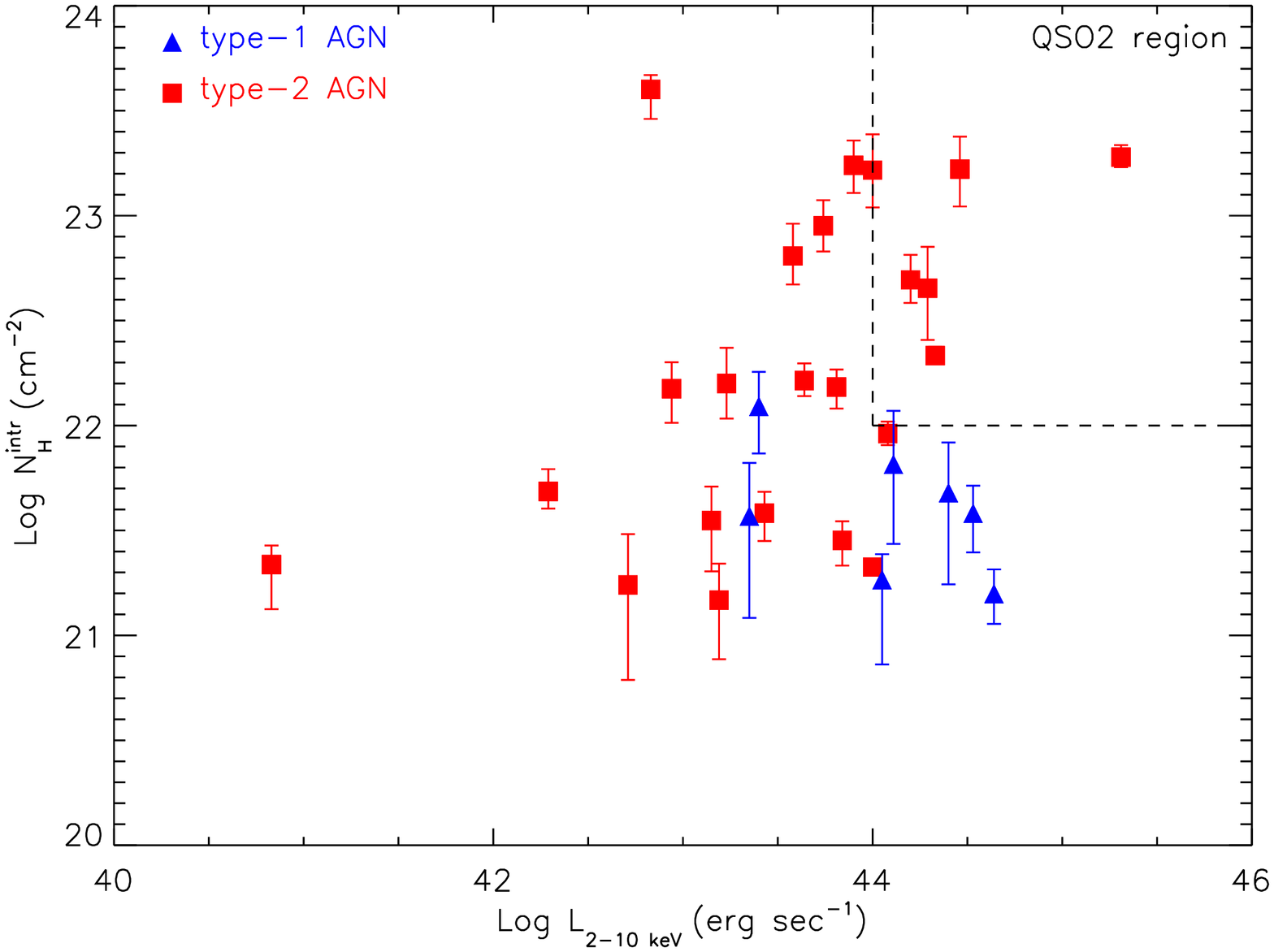}}
    \caption{Left: $\Gamma$ vs. 2-10 keV luminosity. 
      Right: ${\rm N_H^{intr}}$ vs. 2-10 keV luminosity for 
      type-1 and type-2 AGN. In the ${\rm N_H^{intr}}$ vs. 2-10 keV 
      plot we show the QSO2 region (defined as 
      ${\rm L_{2-10}\ge10^{44}erg\,s^{-1}}$ 
      and ${\rm N_H\ge10^{22}\,cm^{-2}}$) where we have 6 QSO2 
      candidates. Error bars correspond to 90\% confidence}
    \label{params_lumin}
\end{figure*}

We have studied the dependence of $\Gamma$ and ${\rm N_H^{intr}}$ with the 2-10 keV X-ray 
luminosity (we use the 2-10 keV luminosity because these values are less affected by X-ray absorption).
The results are plotted in Fig~\ref{params_lumin}. We do not see any correlation between the spectral
slope and the column density of our AGN with 2-10 keV X-ray luminosity.
Note in Fig.~\ref{params_lumin} that our sample contains
6 objects (all optically identified as type-2 AGN) that fall within the ``standard'' QSO2 region, i.e.,
${\it L_X}{\rm \ge 10^{44}\,erg\,s^{-1}}$ and ${\rm N_H^{intr}\ge 10^{22}\,cm^{-2}}$.

\section{Unabsorbed type-2 AGN}
\label{unab_agn2}
\begin{figure*}
\begin{center}
\begin{tabular}{ccc}
\includegraphics[angle=-90,width=0.50\textwidth]{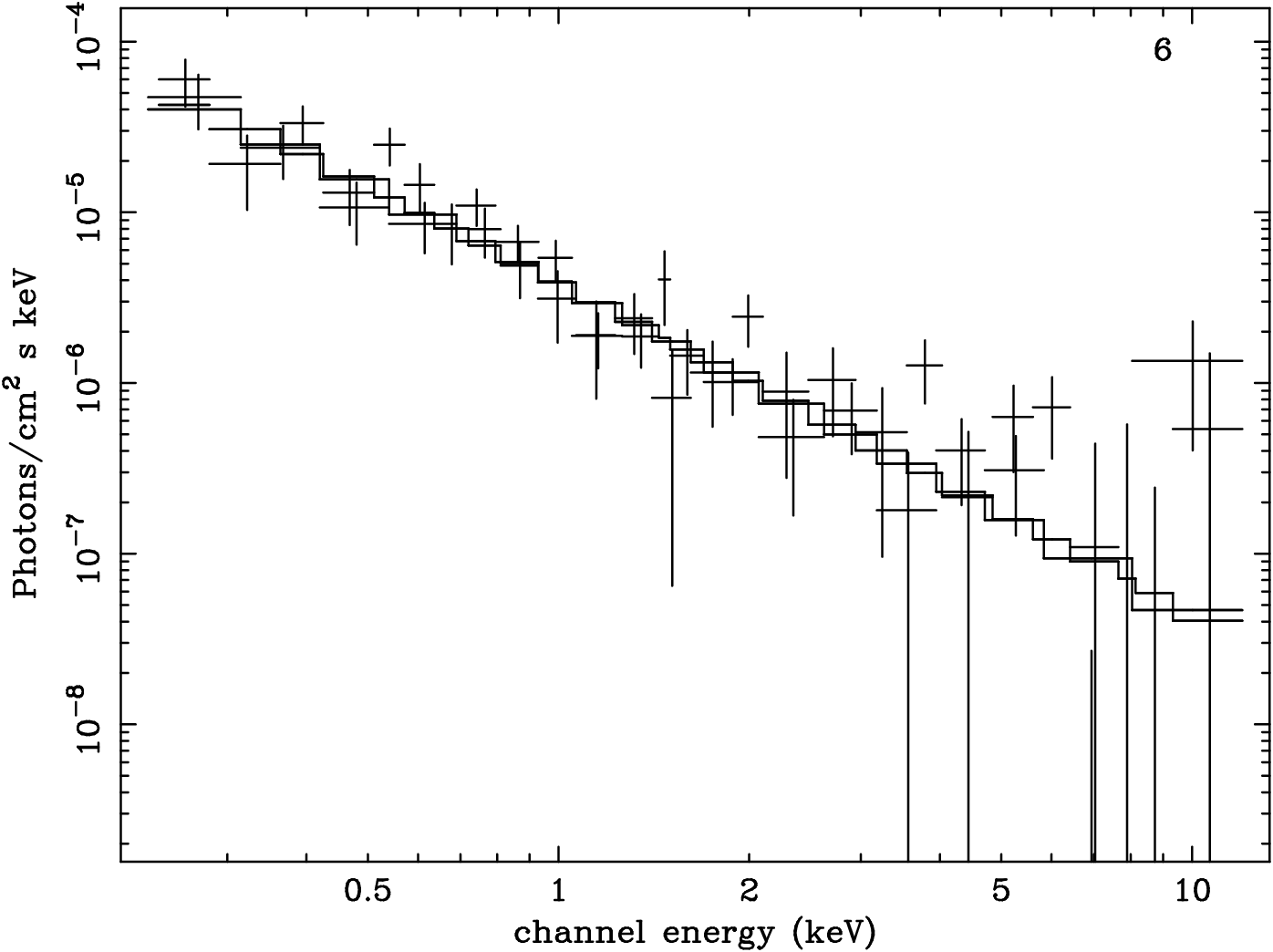} &
\includegraphics[angle=-90,width=0.50\textwidth]{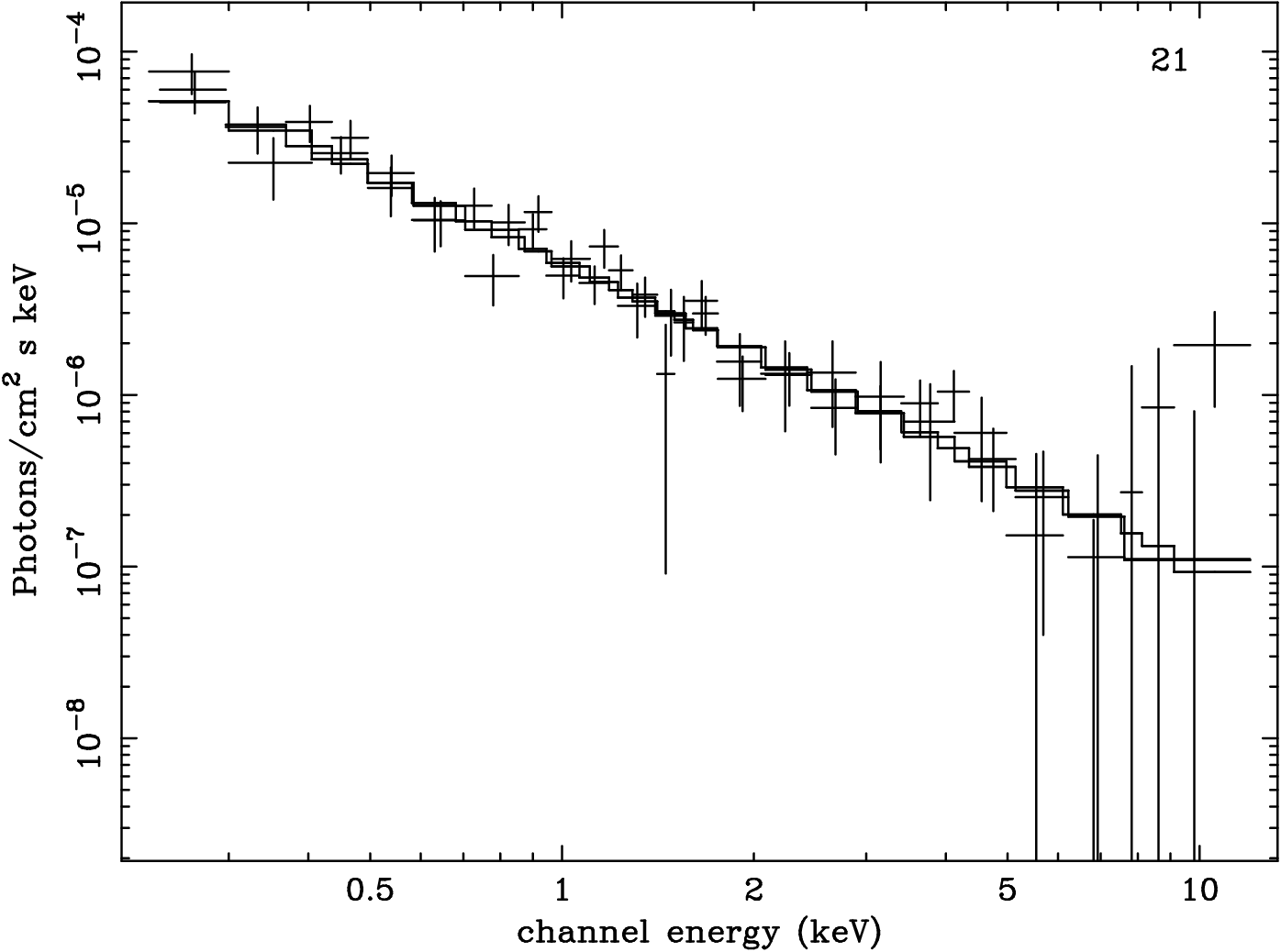}\\
\includegraphics[angle=-90,width=0.50\textwidth]{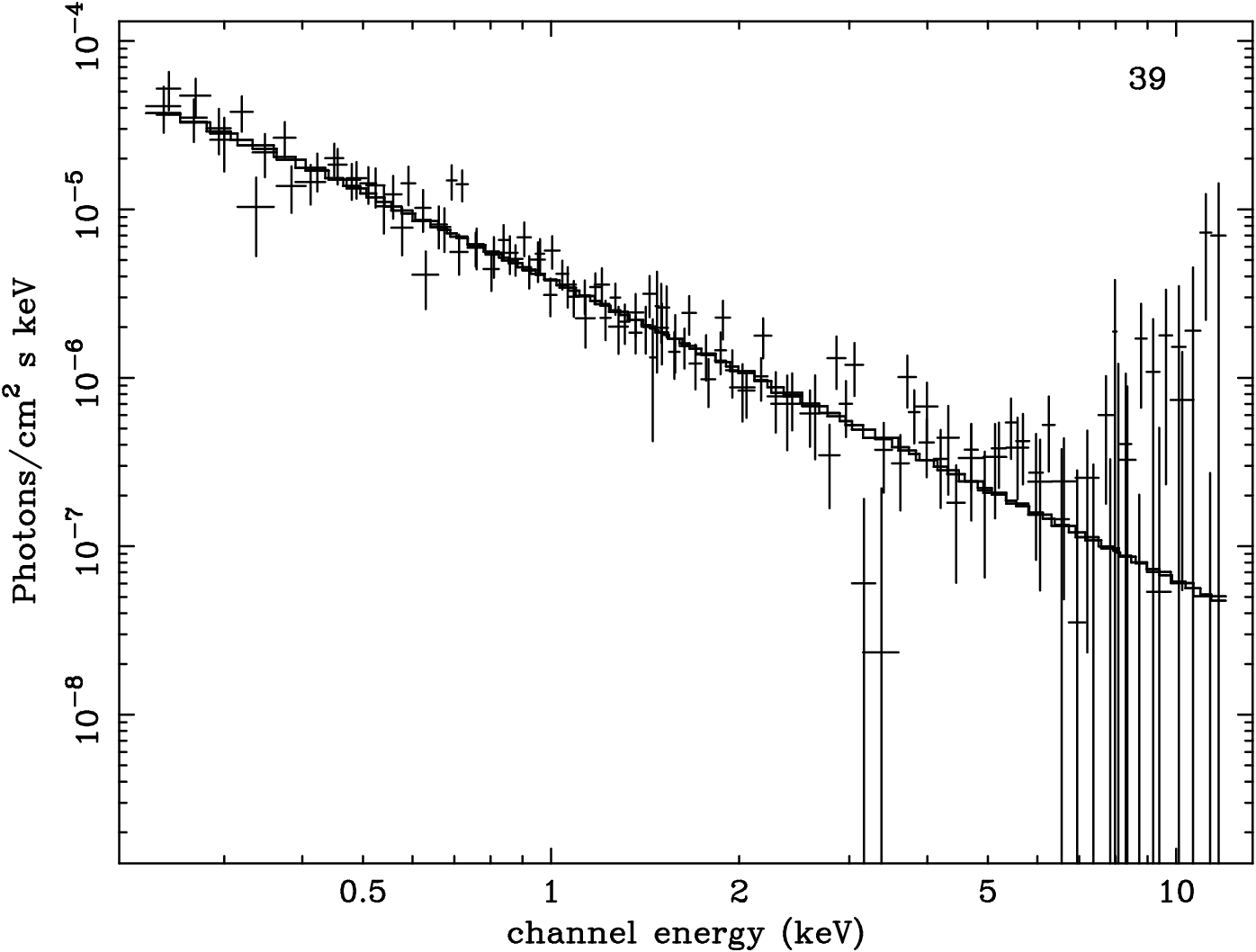} & 
\includegraphics[angle=-90,width=0.50\textwidth]{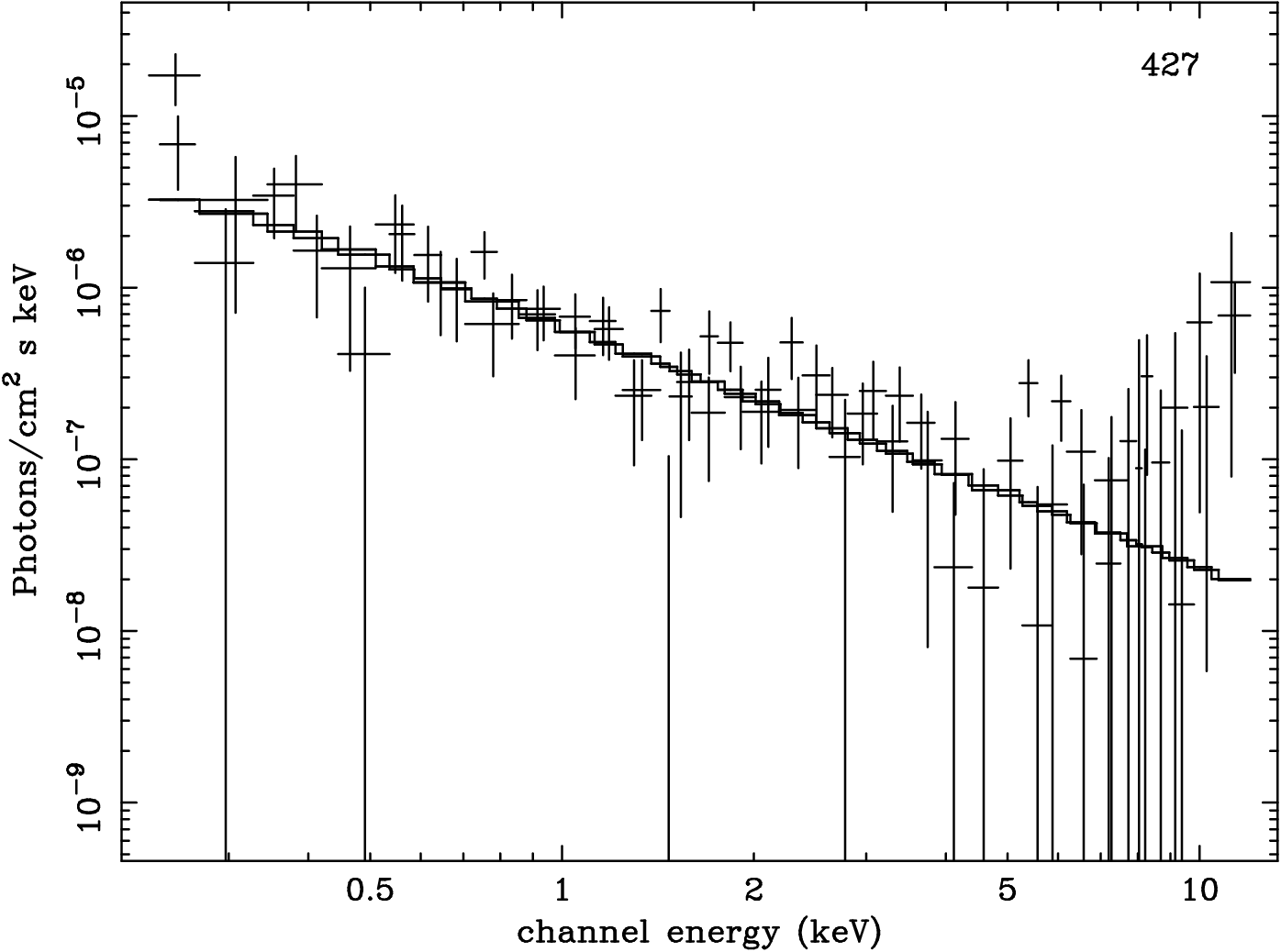}\\
\includegraphics[angle=-90,width=0.50\textwidth]{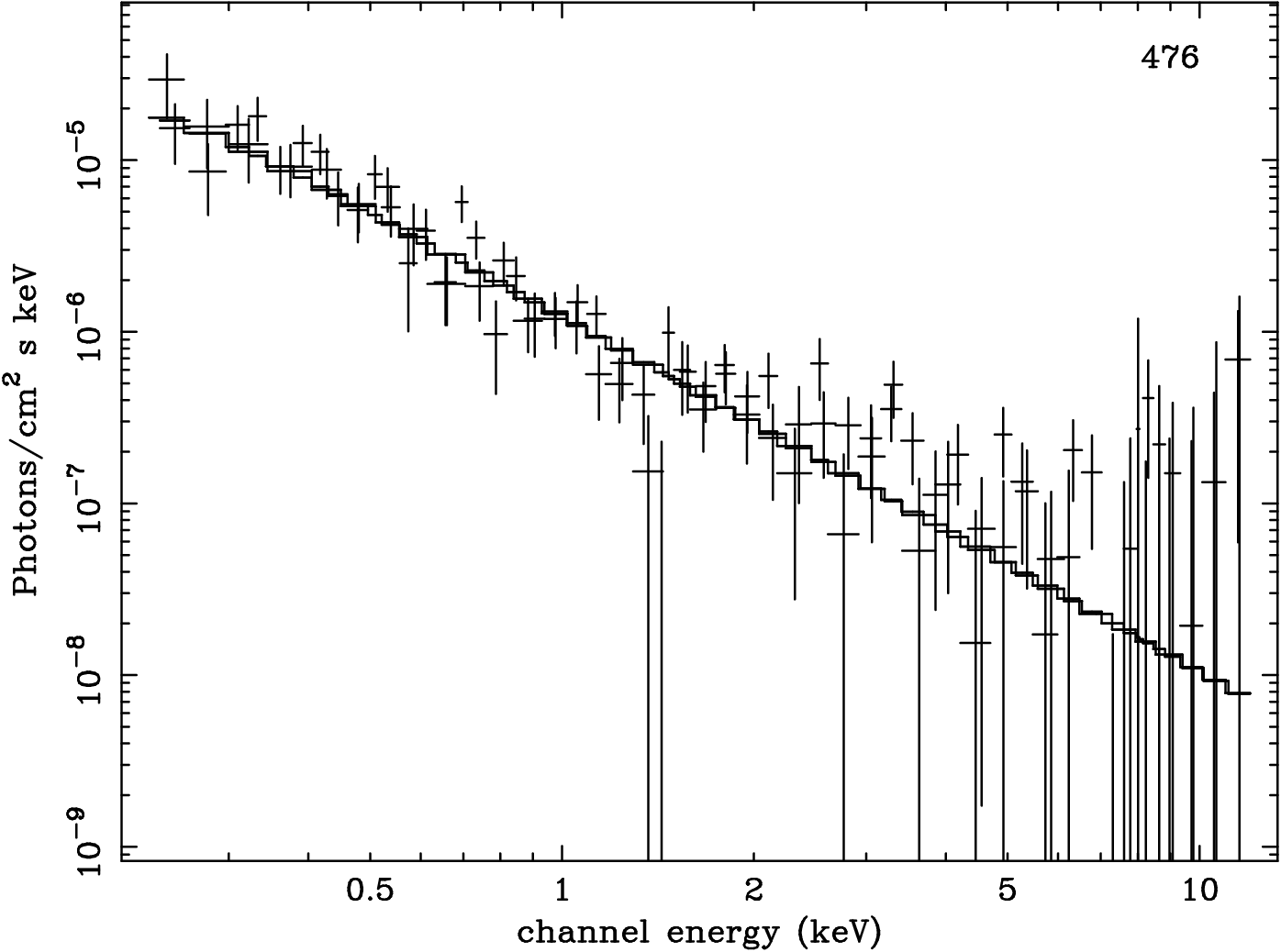} \\ &  
\begin{minipage}{9cm}
\vspace*{-7cm} \caption{Unfolded MOS and pn spectra of the type-2 AGN
without signatures of X-ray absorption. Table~\ref{tabIIa}} lists 
the best fit spectral parameters obtained for these sources.
\label{agnii_unf}
\end{minipage}\\
\end{tabular}
\end{center}
\end{figure*}

We have found 5 objects identified as type-2 AGN but 
with no clear evidence of X-ray absorption in their X-ray spectrum 
($\sim23\%$, see Table.~\ref{tab_absorptions}). 
In Fig.~\ref{agnii_unf}
we show the unfolded spectra of these sources obtained with their best fit 
model (in all the cases a single power law). Several 
authors have found AGN with weak or no broad emission lines in their 
optical spectrum and with unabsorbed X-ray spectra 
(see for example Pappa et al., ~\cite{Pappa2001};
 Panessa et al., ~\cite{Panessa2002}; Barcons et al., ~\cite{Barcons2003}; 
Mateos et al., ~\cite{Mateos2005}; 
Carrera et al., ~\cite{Carrera2004}; Corral et al.~\cite{Corral2005}). 

One possible explanation for these results is 
that the signal to noise of the spectra of these sources is not high enough 
to detect signatures of X-ray absorption (see e.g. Mateos et al., ~\cite{Mateos2005}).
Our current data is however of sufficient quality to detect X-ray absorption 
or X-ray absorption+soft excess to weak levels. 
Moreover, as we see in Fig.~\ref{agnii_unf} the time averaged spectra 
of these sources has enough signal to noise as to detect X-ray absorption with the 
column densities common in type-2 AGN. We have calculated 
the upper limits (at 90\% confidence) to the X-ray absorption in these sources. They are listed in 
Table~\ref{tabIIa}. The values that we obtain are lower than the typical column 
densities found in our absorbed type-2 AGN. 
If these sources are X-ray absorbed, the values for the column density that we have 
found are consistent with arising in absorption from their host galaxy.  
 
Another possibility that might explain the lack of X-ray absorption in these sources is 
X-ray spectral variability. The optical and X-ray observations of these sources have not 
been obtained simultaneously, and the X-ray absorption in these sources might have changed 
with time. Corral et al.~(\cite{Corral2005}) studied the hypothesis of spectral variability
using simultaneous X-ray and optical observations of the Seyfert galaxy Mkn993. 
They found the source to be X-ray unabsorbed but in a type 1.9 optical. Results of a detailed study of X-ray 
flux and spectral variability on scales from months to years of the same sample of sources used for this work, will be 
presented in a forthcoming paper (Mateos et al. 2005b in preparation)
where we show that X-ray variability cannot explain the lack of X-ray 
absorption in our unabsorbed type-2 AGN.
%where we show that the X-ray spectrum of these type-2 X-ray unabsorbed 
%AGN has not varied between different observations. 

We have checked whether these sources can be Compton-thick type-2 AGN. If the torus is 
Compton-thick to optical scattering, even 2-10 keV photons will not be directly seen 
and hence the direct radiation in these sources would be 
completely blocked. In some cases scattered radiation (with no apparent absorption) could be 
the only radiation seen below 10 keV. 
In Compton-thick sources, because the primary radiation is fainter, the equivalent width 
of the K$\alpha$ line increases. Bassani et al.~(\cite{Bassani1999}) 
show a diagram of the EW versus the transmission parameter {\it T}, where {\it T} is 
{$\it S_{\rm X}/S_{\rm [OIII]}$}\footnote { {\it $S_{\rm X}$} is the X-ray 
flux and $S_{\rm [OIII]}$ is the optical flux of the [OIII] $\lambda5007$ emission line. 
[OIII] $\lambda5007$ has been frequently 
used as an isotropic indicator of the intrinsic brightness of the sources.}. This diagram can be used 
to identify Compton-thick sources if the values of EW and {\it T} are known. 
We do not know the value of {$\it S_{\rm [OIII]}$} for our sources, 
however we have measured the EW to check in which part of the diagram our sources fall. 
To measure the EW we added a Gaussian line representing the iron K$\alpha$ emission 
to the spectrum of our sources. We fixed the centroid of the line to 6.4 and the 
width to 0 (the quality of the fits did not improve allowing the line 
parameters to vary). Only in source 21, there might be emission from 
iron K$\alpha$ line (F-test significance $\sim$95\%). In the 
other sources there are no indications of iron K$\alpha$ emission. 
The values of the EW that we have found are listed in column (6) of Table~\ref{tabIIa}. 
In most of the cases we obtained a value of the EW below $\sim$1500, and therefore these sources 
fall outside the region of Compton-thick sources in the Bassani et al.~(\cite{Bassani1999}) 
diagram. However, to confirm our results, 
specially for objects 21 and 476 with EW values above 1000 eV, we need a reliable measurement 
of the [OIII] flux. 

\begin{table}[!ht] \caption[]{X-ray properties of the type-2 AGN for which we 
did not found absorption in their X-ray spectrum.}
  $$ \begin{array}{c c c c c c c}
    \hline \noalign{\smallskip}
    {\rm{ID}} & {\rm redshift} & {\Gamma} & {\rm N_H^{intr}} & {\rm L_{2-10}}& {\rm Equivalent} \\
              &                &          &   (90\%\,{\rm prob.})  & {\rm erg\,s^{-1}}              & {\rm width\,(eV)} \\
    (1) & (2) & (3) & (4) & (5) & (6) \\
    \noalign{\smallskip} \hline \hline
    \noalign{\smallskip}
    {\rm 6}    & 0.528 & 1.94\pm0.20 & \le20.91 & 43.07 & 576 \\
    {\rm 21}   & 0.498 & 1.77\pm0.14 & \le20.62 & 43.28 & 1462 \\
    {\rm 39}   & 0.711 & 1.79\pm0.08 & \le20.66 & 43.44 & 292 \\
    {\rm 427}  & 0.696 & 1.40\pm0.20 & \le21.08 & 42.79 & 249 \\
    {\rm 476}  & 0.607 & 2.07\pm0.17 & \le20.22 & 42.66 & 1393 \\

    \noalign{\smallskip} \hline \end{array} $$
%  \begin{list}{}{}
%\item[$^{\mathrm{a}}$] Strength of the soft excess measured as the ratio of black body and power law luminosities in the 0.5-2 keV band.
%\item[$^{\mathrm{b}}$] Temperature of the black body in eV.
%  {\bf \item[$^{\mathrm{b}}$] 2-10 keV luminosity of the power law component (corrected for soft excess and absorption).
     Columns are as follows: (1) Source X-ray identification number; 
    (2) redshift; (3) $\Gamma$ from best fit model (for all sources the best fit model 
    was a single power law);
    (4) upper limit in intrinsic (rest-frame) X-ray absorption (90\% confidence); 
    (5) logarithm of the 2-10 keV luminosity; 
    (6) rest-frame equivalent width of an emission line centred at 6.4 keV with $\sigma$=0 (the value of the EW 
    was obtained using a Gaussian to fit the emission line)
%\end{list}
\label{tabIIa}
\end{table}

\section{Extragalactic X-ray background}
\label{EX_XRB}
The spectrum of the extragalactic X-ray background (XRB) 
was measured by the HEAO satellite (Marshall et al.~\cite{Marshall1980}) 
from 1-50 keV. At these energies, the XRB spectrum can be reproduced well
by an optically thin plasma of temperature $\sim$40 keV. At low
energies, $\le$15 keV, a good description of the data is obtained 
with a power law of $\Gamma$=1.4. The spectrum of the XRB is 
significantly flatter than the typical spectrum of AGN. A population 
of heavily absorbed AGN, predicted by 
synthesis models of the XRB (e.g. Setti \& Woltjer, ~\cite{Setti1989}; 
Comastri et al., ~\cite{Comastri1995}; Gilli et al., ~\cite{Gilli2001};
Gandhi \& Fabian, ~\cite{Gandhi2003}; 
Ueda et al., ~\cite{Ueda2003}), might account for this discrepancy.

We have carried out a stacking of MOS/pn time averaged spectra of the sources that we 
have analysed. Details on the analysis are given in Appendix~\ref{apendix_A}.
The goal of this analysis was to compare the integrated emission of our sources,  
with an average spectral shape of $\sim$1.92 (see Sec.~\ref{broad band continuum}),  
with the spectrum of the XRB in the 2-7 keV energy band. 

We fitted MOS and pn stacked spectra with {\tt xspec} using a simple power law model. Then we divided the 
measured XRB intensities (in units of ${\rm keV^2\,keV^{-1}\,cm^{-2}\,s^{-1}}$ at 1 keV) by the total 
solid angle covered by {\it XMM-Newton} (0.4 deg$^2$ or 1.2185$\times10^{-4}$ sr) to obtain the 
resolved fraction of XRB in the area surveyed.

\begin{figure}
    \hbox{
    \includegraphics[angle=90,width=0.50\textwidth]{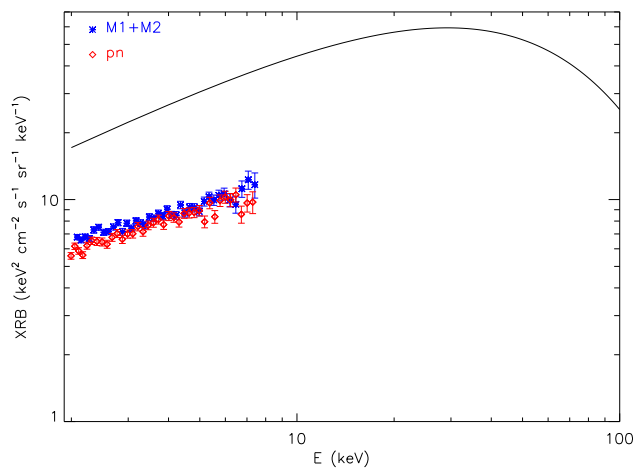}}
    \caption{Spectrum of the extragalactic X-ray background as measured by the HEAO satellite 
     but renormalised to the 2-8 keV intensity observed by De Luca \& Mondeli (\cite{Luca2004}) 
     (solid line). The points show the 2-7 keV stacked spectra of the sources that we have 
     analysed (M1+M2 (diamonds) and pn (stars)).} 
    \label{XRB}
\end{figure}

Fig.~\ref{XRB} shows the total extragalactic XRB spectrum as measured by the HEAO-1 mission (solid 
line) but renormalised to the 2-8 keV intensity observed by De Luca \& Molendi~(\cite{Luca2004}). The points 
show the 2-10 keV XRB spectrum seen by MOS (diamonds) and pn (stars).

We found that the 2-7 keV XRB resolved by our 
sources was best fitted with a power law of $\Gamma=1.59\pm0.03$ and 
N=${\rm 4.78\pm0.15\,keV^2\,keV^{-1}\,cm^{-2}\,s^{-1}\,sr^{-1}}$ (at 1 keV) for MOS data and 
$\Gamma=1.54\pm0.04$ and N=${\rm 4.38\pm0.14\,keV^2\,keV^{-1}\,cm^{-2}\,s^{-1}\,sr^{-1}}$ (at 1 keV) for pn data. 
The integrated contribution of our sources is indeed harder than the spectrum of the 
brightest AGN, but still softer than the XRB at these energies. Since our 
sources have been selected in the 0.2-12 keV band, we probably missed faint absorbed sources. 
Worsley et al.~(\cite{Worsley2004}) did include fainter sources than the present study, and hence they 
reached a significantly higher integrated emission at 1 keV 
(${\rm N=11\pm0.5\,keV^2\,keV^{-1}\,cm^{-2}\,s^{-1}\,sr^{-1}}$).
A further component not detectable by {\it XMM-Newton} arising above 5 keV 
might also be present, as argued by Worsley et al.~(\cite{Worsley2004}).

%\begin{figure}
%    \hbox{
%    \includegraphics[angle=90,width=0.50\textwidth]{hist_hflux_all_lh.ps}}
%    \caption{Histograms of fluxes for the whole sample of sources analysed and
%      for the sources with optical identifications. The fluxes were obtained from the
%      sources best fit model.
%      }
%    \label{hflux_hist}
%\end{figure}
 
%__________________________________________________________________
\section{Discussion and Conclusions}
\label{conclusions}

We have carried out a detailed study of the X-ray spectra of a sample of 123 objects
detected with {\it XMM-Newton} in a deep observation in the {\it Lockman Hole} field.
The EPIC spectra of these sources have all more than 500 background subtracted counts (MOS+pn)
in the energy interval form 0.2-12 keV. Thanks to the good signal to noise of the data
we could study in detail the different spectral components that contribute to the X-ray
emission of AGN in the above energy interval.

\subsection{X-ray continuum shape and intrinsic absorption}
\label{continuum}
 The 0.2-12 keV spectra of many AGN cannot be well reproduced with a
single power law model. We found that this model is the best fit model for
only $\sim$53$\%$ of the sample. The average continuum shape of our 
sources appears to harden at fainter 0.5-2 keV fluxes 
(see e.g. Giacconi et al., \cite{Giacconi2001}; 
Mainieri et al., \cite{Mainieri2002}; Mateos et al., \cite{Mateos2005}), 
but there are some indications that the same effect is
seen for objects detected in the 2-10 keV band (Tozzi et al., \cite{Tozzi2001a};
Streblyanska et al., \cite{Streblyanska2004}). Two hypothesis have been suggested to explain it:
X-ray absorption is more important at fainter fluxes or 
there exists a population of faint sources with intrinsically flatter continuum.

We confirm that this apparent effect is due to absorption. Both because it is not 
seen in the 2-10 keV band, and because it disappears when absorption is taken into 
account. We do not see any evidence for a population of intrinsically harder sources at 
faint fluxes. However, if sources with larger column densities were
found, then we should see the hardening of $\langle \Gamma \rangle$ 
in the 2-10 keV band. Using a single power law model, X-ray absorbed sources
had the faintest 0.5-2 keV fluxes,
but they had the same distribution of 2-10 keV fluxes than unabsorbed sources.
Because of this, the average $\Gamma$ obtained using a single power law model as a function of the 2-10 keV
flux was measured to be harder ($\Gamma \sim 1.7$) than the typical value of $\sim$ 1.9 found in
unabsorbed AGN.

In the hard band we found no
dependence of X-ray absorption or fraction of absorbed sources with flux. 
If observed (not corrected for absorption) 0.5-2 keV fluxes are used, then the column density and
the fraction of absorbed sources increase significantly at fainter fluxes.
However, using unabsorbed 0.5-2 keV fluxes (i.e. fluxes corrected for the effect of absorption)
the correlations are not observed and we obtain the same results than in the 2-10 keV band.
Therefore to study the dependence of $\Gamma$ with the X-ray flux it is important to
first correct the fluxes for the effect of X-ray absorption. 
Another interesting result from our analysis is that allowing the objects 
to be X-ray absorbed we found a value of $\langle \Gamma \rangle \sim 1.9$ at all 0.5-2 keV and 2-10 keV fluxes but
with significant scatter in the points. 

We have carried out a detailed study of the 
X-ray spectra of each individual object. 
We have searched for soft excess emission and for signatures of reflection
 at high energies (Compton reflection). We also studied the presence of
 Fe K$\alpha$ emission. Using for each source the value of $\Gamma$ obtained after 
including all these components to the fitting model we found that the scatter in $\langle \Gamma \rangle$ 
is much smaller, and therefore is mostly due to the presence of other spectral components 
in the X-ray emission.

 However we still expect some intrinsic scatter in $\Gamma$. To calculate the intrinsic
dispersion in the continuum shape of our objects, we assumed that the distribution of
$\Gamma$ could be well represented with a Gaussian. Under this hypothesis, we found our
sources to have an average spectral slope of $\sim$ 1.92 with an intrinsic dispersion 
of $\sim$0.28. 

\subsection{Soft excess emission}
\label{Soft excess} Soft excess was detected in the time averaged spectra of
18 sources. However only in 9 objects (4 type-1 AGN, 4 type-2 AGN and 1 unidentified
source) we could fit the spectral signatures with a black body model (a
Raymond Smith gave an equally good fit). We found the average temperature of
the black body to be $0.09\pm0.01$ keV for type-1 AGN and $0.26\pm0.08$ keV for
type-2 AGN. The average 0.5-2 keV luminosities of the black body were found to be
(in log units) ${\rm 43.42\pm0.43\,erg\,s^{-1}}$ in type-1 AGN and ${\rm
44.11\pm0.44\,erg\,s^{-1}}$ in type-2 AGN. The 0.5-2 keV luminosities of the soft
excess component do not differ significantly between type-1 and type-2 AGN, but our
results seem to indicate that the black body temperatures are slightly higher
in type-2 AGN than in type-1 AGN. This might be due to a higher 
contribution from scattering in type-2 AGN. 
However due to the small number of AGN with soft
excess analysed we can not reach any conclusion. 
The temperatures of the black body are in most cases $\ge$60 eV, 
and therefore the origin of the soft excess component in these sources 
cannot be explained as thermal emission from the accretion disc only. Componization 
of cool disc photons by hot electrons surrounding the accretion disc 
might be an alternative explanation.

In 9 sources (including 1 type-1 AGN, 3 type-2 AGN and 5 sources unidentified) the black body model could not
fit the spectral signatures of the soft excess emission. For these sources we obtained a
good fit with a scattering or partial covering model (two power laws with the same spectral index but different absorptions). 
The average covering fraction of the absorber was found to be $0.82\pm0.06$, which 
means that the scattering fraction in these sources is rather large ($18\pm6\%$).

\subsection{Reprocessed components}
\label{Reprocessed components}
We found one object in our sample with
a flattening in the spectral slope at high energies. The source is still unidentified
so we used a model with two power laws to fit its X-ray emission.

Signatures of an emission line at high energies were found (F-test 
significance $\ge$95\%) in 8 sources
(1 type-1 AGN, 5 type-2 AGN and 2 unidentified sources). Although in 
some sources the profiles of the lines did not appear to be symmetrical, 
the signal to noise 
of the spectra was not high enough to use 
more physical models.
 Therefore
we fitted the lines in all the sources with a Gaussian model. 
For most AGN we found the centroids of the lines to be consistent 
(within the error bars) with being Fe K$\alpha$ at an energy 
of $\sim$6.4 keV (rest frame). Only in one object we found 
the line centroid to be slightly higher than the expected value for 
neutral iron. In this source the line might arise as reflection in an 
ionised accretion disc. 

In two objects we found significant widths for the emission  
lines. In another two sources the centroids of the lines were found at 
energies lower than 6.4 keV. In all these sources the emission lines 
might have relativistic line profiles, and therefore they might 
have been emitted in the inner regions of the black hole accretion disc.

\subsection{X-ray spectra of AGN}
\label{X-ray spectra of AGN}
We found the best fit average spectral slope to be $1.89\pm0.03$ in type-1 AGN and $1.71\pm0.03$ in type-2 AGN.
These values seem to indicate that type-2 AGN have harder spectral slopes than type-1 AGN. However assuming that the values of $\Gamma$ that we have obtained for each individual object follow a 
Gaussian distribution, and allowing the spectral slope to have intrinsic dispersion, 
the significance of type-2 AGN being harder than type-1 AGN is of 1.62$\sigma$.
This small difference in $\langle\Gamma\rangle$ for type-1 
and type-2 AGN might be due to the fact that if the signatures of absorption 
are not very significant, the detected values 
of ${\rm N_H}$ will tend to be lower than the real ones 
and then, the fitted $\Gamma$ will be flatter. 
This effect will be more important for type-2 AGN where we expect more 
sources to be absorbed.

X-ray absorbed objects were found among type-1 AGN ($\sim10\%$) and type-2 AGN ($\sim77\%$). 
We found the fraction of absorbed objects in type-1 and type-2 AGN to be different 
with a significance of $>99.99\%$.
The distribution of absorbing column densities also suggest that type-1 AGN are less absorbed
than type-2 AGN. A comparison with a KS test gave a significance of these distributions being different at 92\% confidence.

We did not see a dependence of the AGN continuum shape with the X-ray luminosity or redshift. We
found the same results for the column density.

\subsection{Unabsorbed type-2 AGN}
\label{concl_unab}
We studied in more detail the X-ray emission of the 5 type-2 AGN with 
unabsorbed MOS and pn X-ray spectra. All these sources have 
spectra with enough signal to noise, hence we should have been able to detect signatures 
of X-ray absorption if they are present. 
We have calculated the upper limit (90\%) in the column density. We found 
that in all cases the values for the column density are significantly 
lower than the typical values found in the absorbed type-2 AGN. If there 
is X-ray absorption in these sources, the low values of the column densities that 
we find could be explained as arising from the host galaxy. We argue that 
spectral variability is unlikely to be at the heart of the apparent X-ray/optical mismatch.   

Finally we do not find compelling evidence that these sources are Compton-thick, although 
the [OIII] flux needs to be measured to reach a firm conclusion.

\onecolumn
\begin{landscape}
\begin{longtable}{lcccccccccccccccccccc}
\caption{Source X-ray properties}\\
%\begin{tabular}{lcccccccccccccccccccc}
\hline
\hline
ID & {\it ROSAT} & RA & Dec & Class & {\it z} & Model & $\Gamma$ & ${\rm log (N_H)}$ & ${\rm {\it S}_{0.5-2}}$ & ${\rm {\it S}_{2-10}}$ & ${\rm log({\it L}_{0.5-2})}$ & ${\rm log({\it L}_{2-10})}$ \\
(1) & (2) & (3) & (4) & (5) & (6) & (7) & (8) & (9) & (10) & (11) & (12) & (13)\\
\hline
\endfirsthead
\hline
\hline
ID & {\it ROSAT} & RA & Dec & Class & {\it z} & Model & $\Gamma$ & ${\rm log (N_H)}$ & ${\rm {\it S}_{0.5-2}}$ & ${\rm {\it S}_{2-10}}$ & ${\rm log({\it L}_{0.5-2})}$ & ${\rm log({\it L}_{2-10})}$ \\
(1) & (2) & (3) & (4) & (5) & (6) & (7) & (8) & (9) & (10) & (11) & (12) & (13)\\
\hline
\endhead
\hline
\endfoot
 607&        -& 10 53 01.86& +57 15 00.69&    --&     -&       SPL&$  1.21_{  0.13}^{  0.13}$&$         _{         }^{         }$& 0.73& 2.81&         &         &\\[0.5ex]
 599&      54A& 10 53 07.46& +57 15 05.84&  AGNI& 2.416&       SPL&$  1.46_{  0.11}^{  0.11}$&$         _{         }^{         }$& 1.02& 2.68&   44.36 &   44.80 &\\[0.5ex]
 400&      13A& 10 52 13.29& +57 32 25.58&  AGNI& 1.873&       SPL&$  1.90_{  0.06}^{  0.06}$&$         _{         }^{         }$& 0.63& 0.84&   44.14 &   44.27 &\\[0.5ex]
  63&        -& 10 52 36.49& +57 16 04.07&    --&     -&       APL&$  2.13_{  0.23}^{  0.32}$&$    20.88_{     0.38}^{     0.24}$& 0.54& 0.66&         &         &\\[0.5ex]
   5&      52A& 10 52 43.30& +57 15 45.95&  AGNI& 2.144&       SPL&$  1.90_{  0.06}^{  0.06}$&$         _{         }^{         }$& 1.60& 2.16&   44.68 &   44.68 &\\[0.5ex]
   6& 504(51D)& 10 51 14.30& +57 16 16.88& AGNII& 0.528&       SPL&$  1.94_{  0.19}^{  0.20}$&$         _{         }^{         }$& 0.87& 1.11&   42.97 &   43.07 &\\[0.5ex]
  16&        -& 10 51 46.64& +57 17 16.02&    --&     -&       APL&$  2.27_{  0.98}^{  2.08}$&$    21.59_{     0.24}^{     0.40}$& 0.22& 0.44&         &         &\\[0.5ex]
  21&      48B& 10 50 45.67& +57 17 32.60& AGNII& 0.498&       SPL&$  1.77_{  0.14}^{  0.14}$&$         _{         }^{         }$& 1.34& 2.18&   43.07 &   43.28 &\\[0.5ex]
  26&        -& 10 52 32.99& +57 17 50.96&    --&     -&       SPL&$  1.09_{  0.20}^{  0.20}$&$         _{         }^{         }$& 0.21& 0.95&         &         &\\[0.5ex]
  31&        -& 10 52 00.34& +57 18 08.24&    --&     -&      CAPL&$  2.34_{  0.63}^{  0.30}$&$    22.29_{     0.16}^{     0.12}$& 0.15& 0.99&         &         &\\[0.5ex]
  41&      46A& 10 51 19.14& +57 18 34.09&  AGNI& 1.640&       APL&$  2.06_{  0.27}^{  0.38}$&$    21.82_{     0.38}^{     0.25}$& 0.53& 0.68&   44.09 &   44.11 &\\[0.5ex]
  39&      45Z& 10 53 19.09& +57 18 53.58& AGNII& 0.711&       SPL&$  1.79_{  0.08}^{  0.08}$&$         _{         }^{         }$& 0.86& 1.37&   43.24 &   43.44 &\\[0.5ex]
  53&      43A& 10 51 04.39& +57 19 23.90&  AGNI& 1.750&       APL&$  1.83_{  0.18}^{  0.22}$&$    21.68_{     0.44}^{     0.24}$& 0.84& 1.43&   44.22 &   44.40 &\\[0.5ex]
  65&        -& 10 52 55.46& +57 19 52.80&    --&     -&       APL&$  1.57_{  0.22}^{  0.27}$&$    21.24_{     0.26}^{     0.15}$& 0.25& 0.86&         &         &\\[0.5ex]
  74&     905A& 10 52 51.13& +57 20 15.70&    --&     -&       SPL&$  1.67_{  0.20}^{  0.21}$&$         _{         }^{         }$& 0.19& 0.35&         &         &\\[0.5ex]
  72&      84Z& 10 52 16.94& +57 20 19.71& AGNII& 2.710*&      APL&$  1.69_{  0.10}^{  0.08}$&$    23.28_{     0.05}^{     0.06}$& 0.85& 4.67&   45.04 &   45.31 &\\[0.5ex]
  85&      38A& 10 53 29.50& +57 21 06.22&  AGNI& 1.145&       SPL&$  2.13_{  0.08}^{  0.09}$&$         _{         }^{         }$& 0.72& 0.69&   43.77 &   43.74 &\\[0.5ex]
  86&        -& 10 53 09.68& +57 20 59.58&  AGNI& 3.420&       SPL&$  1.66_{  0.16}^{  0.17}$&$         _{         }^{         }$& 0.23& 0.45&   44.16 &   44.45 &\\[0.5ex]
  88&      39B& 10 52 09.37& +57 21 05.43&  AGNI& 3.279&       SPL&$  1.73_{  0.12}^{  0.12}$&$         _{         }^{         }$& 0.40& 0.70&   44.41 &   44.65 &\\[0.5ex]
  90&      37A& 10 52 48.09& +57 21 17.43&  AGNI& 0.467&    SPL+SE&$  2.10_{  0.10}^{  0.13}$&$         _{         }^{         }$& 1.56& 1.40&   43.21 &   43.08 &\\[0.5ex]
  96& 814(37G)& 10 52 44.87& +57 21 24.84&  AGNI& 2.832&       SPL&$  2.08_{  0.08}^{  0.09}$&$         _{         }^{         }$& 0.58& 0.60&   44.64 &   44.65 &\\[0.5ex]
 107&        -& 10 52 19.49& +57 22 15.26& AGNII& 0.075&       APL&$  2.02_{  0.28}^{  0.23}$&$    21.34_{     0.21}^{     0.09}$& 0.26& 0.48&   40.78 &   40.83 &\\[0.5ex]
 120&        -& 10 52 25.17& +57 23 07.02&    --&     -&       SPL&$  1.91_{  0.07}^{  0.07}$&$         _{         }^{         }$& 0.62& 0.83&         &         &\\[0.5ex]
 108&        -& 10 50 50.91& +57 22 15.65&    --&     -&      2SPL&$  1.56_{  0.18}^{  0.18}$&$         _{         }^{         }$& 0.78& 8.48&         &         &\\[0.5ex]
 900&        -& 10 54 59.43& +57 22 18.84&    --&     -&       APL&$  1.72_{  0.16}^{  0.15}$&$    21.03_{     0.27}^{     0.14}$& 2.46& 5.83&         &         &\\[0.5ex]
 121&     434B& 10 52 58.08& +57 22 51.95& AGNII& 0.772&       APL&$  1.95_{  0.20}^{  0.25}$&$    21.55_{     0.24}^{     0.16}$& 0.31& 0.52&   43.05 &   43.15 &\\[0.5ex]
 135& 513(34O)& 10 52 54.39& +57 23 43.89&  AGNI& 0.761&       SPL&$  1.78_{  0.04}^{  0.04}$&$         _{         }^{         }$& 1.56& 2.50&   43.56 &   43.77 &\\[0.5ex]
 124&     634A& 10 53 11.72& +57 23 09.07&  AGNI& 1.544&       SPL&$  1.69_{  0.10}^{  0.11}$&$         _{         }^{         }$& 0.35& 0.64&   43.60 &   43.86 &\\[0.5ex]
 125& 607(36Z)& 10 52 19.90& +57 23 07.92&    --&     -&       SPL&$  1.66_{  0.16}^{  0.16}$&$         _{         }^{         }$& 0.21& 0.41&         &         &\\[0.5ex]
 142&        -& 10 52 03.74& +57 23 39.62&    --&     -&       SPL&$  1.81_{  0.21}^{  0.22}$&$         _{         }^{         }$& 0.16& 0.24&         &         &\\[0.5ex]
 133&      35A& 10 50 38.77& +57 23 39.67&  AGNI& 1.439&       SPL&$  2.04_{  0.10}^{  0.10}$&$         _{         }^{         }$& 2.49& 2.75&   44.52 &   44.56 &\\[0.5ex]
 148&      32A& 10 52 39.66& +57 24 32.83&  AGNI& 1.113&    SPL+SE&$  2.12_{  0.03}^{  0.03}$&$         _{         }^{         }$& 7.72& 7.45&   44.93 &   44.74 &\\[0.5ex]
 166&        -& 10 52 31.98& +57 24 30.82&    --&     -&       APL&$  1.44_{  0.24}^{  0.19}$&$    22.19_{     0.16}^{     0.14}$& 0.09& 1.48&         &         &\\[0.5ex]
 156&        -& 10 51 54.59& +57 24 09.28& AGNII**& 2.365&       APL&$  2.40_{  0.45}^{  0.57}$&$    22.65_{     0.25}^{     0.20}$& 0.22& 0.27&   44.49 &   44.29 &\\[0.5ex]
 163&      33A& 10 51 59.88& +57 24 26.31&  AGNI& 0.974&       APL&$  1.52_{  0.22}^{  0.32}$&$    21.57_{     0.49}^{     0.25}$& 0.22& 0.63&   42.97 &   43.35 &\\[0.5ex]
 168&      31A& 10 53 31.72& +57 24 56.19&  AGNI& 1.956&       SPL&$  1.93_{  0.03}^{  0.03}$&$         _{         }^{         }$& 3.30& 4.22&   44.92 &   45.03 &\\[0.5ex]
 172&        -& 10 53 15.71& +57 24 50.84& AGNII&  1.17&       APL&$  2.26_{  0.40}^{  0.63}$&$    23.22_{     0.18}^{     0.17}$& 0.08& 0.83&   44.11 &   44.00 &\\[0.5ex]
 183&      82A& 10 53 12.27& +57 25 08.28&  AGNI&  0.96&       SPL&$  1.88_{  0.07}^{  0.07}$&$         _{         }^{         }$& 0.57& 0.78&   43.39 &   43.53 &\\[0.5ex]
 176&      30A& 10 52 57.25& +57 25 08.77&  AGNI& 1.527&       SPL&$  2.21_{  0.05}^{  0.05}$&$         _{         }^{         }$& 1.08& 0.92&   44.29 &   44.22 &\\[0.5ex]
 179&        -& 10 52 31.64& +57 25 03.93&    --&     -&      CAPL&$  2.07_{  0.52}^{  0.26}$&$    22.42_{     0.22}^{     0.14}$& 0.06& 0.56&         &         &\\[0.5ex]
 174&        -& 10 51 20.63& +57 24 58.24&    --&     -&       SPL&$  1.37_{  0.14}^{  0.14}$&$         _{         }^{         }$& 0.30& 0.90&         &         &\\[0.5ex]
 186&        -& 10 51 49.93& +57 25 25.13& AGNII& 0.676&       APL&$  1.70_{  0.12}^{  0.11}$&$    22.21_{     0.07}^{     0.08}$& 0.54& 2.44&   43.38 &   43.64 &\\[0.5ex]
 171&      28B& 10 54 21.22& +57 25 45.40& AGNII& 0.205&       APL&$  1.73_{  0.03}^{  0.02}$&$    21.33_{     0.02}^{     0.02}$&33.50&84.89&   43.75 &   44.00 &\\[0.5ex]
 200&        -& 10 53 46.81& +57 26 07.77&    --&     -&       APL&$  2.37_{  0.26}^{  0.36}$&$    20.93_{     0.32}^{     0.24}$& 0.32& 0.28&         &         &\\[0.5ex]
 187&        -& 10 50 47.96& +57 25 22.71&    --&     -&       SPL&$  1.98_{  0.08}^{  0.08}$&$         _{         }^{         }$& 1.15& 1.38&         &         &\\[0.5ex]
 191&      29A& 10 53 35.03& +57 25 44.13&  AGNI& 0.784&       SPL&$  1.82_{  0.02}^{  0.02}$&$         _{         }^{         }$& 4.20& 6.35&   44.04 &   44.22 &\\[0.5ex]
 199&        -& 10 52 25.28& +57 25 51.27&    --&     -&       APL&$  1.80_{  0.21}^{  0.28}$&$    21.99_{     0.14}^{     0.14}$& 0.16& 1.14&         &         &\\[0.5ex]
 217&        -& 10 51 11.60& +57 26 36.67&    --&     -&       APL&$  1.87_{  0.12}^{  0.11}$&$    20.80_{     0.31}^{     0.17}$& 0.65& 1.10&         &         &\\[0.5ex]
 214&        -& 10 53 15.09& +57 26 30.65&    --&     -&       APL&$  1.88_{  0.19}^{  0.18}$&$    21.41_{     0.13}^{     0.09}$& 0.24& 0.63&         &         &\\[0.5ex]
 222&        -& 10 53 51.67& +57 27 03.64& AGNII& 0.917&       APL&$  1.88_{  0.10}^{  0.09}$&$    21.45_{     0.12}^{     0.09}$& 1.04& 1.76&   43.69 &   43.84 &\\[0.5ex]
2020&      27A& 10 53 50.19& +57 27 11.61&  AGNI& 1.720&       SPL&$  2.23_{  0.07}^{  0.07}$&$         _{         }^{         }$& 0.97& 0.80&   44.39 &   44.30 &\\[0.5ex]
 226&        -& 10 51 20.49& +57 27 03.47&    --&     -&       SPL&$  1.88_{  0.10}^{  0.10}$&$         _{         }^{         }$& 0.50& 0.70&         &         &\\[0.5ex]
 243&        -& 10 51 28.14& +57 27 41.55&    --&     -&      CAPL&$  1.82_{  0.17}^{  0.19}$&$    22.02_{     0.08}^{     0.08}$& 0.51& 3.49&         &         &\\[0.5ex]
 254&     486A& 10 52 43.37& +57 28 01.49& AGNII& 1.210&       APL&$  1.76_{  0.13}^{  0.13}$&$    22.18_{     0.10}^{     0.08}$& 0.32& 0.91&   43.59 &   43.81 &\\[0.5ex]
 261&      80A& 10 51 44.63& +57 28 08.89&  AGNI& 3.409&       SPL&$  1.63_{  0.08}^{  0.08}$&$         _{         }^{         }$& 0.45& 0.92&   44.44 &   44.74 &\\[0.5ex]
 259&        -& 10 53 05.60& +57 28 12.50& AGNII& 0.792&    APL+SE&$  1.22_{  0.18}^{  0.14}$&$    23.24_{     0.13}^{     0.12}$& 0.07& 3.39&   43.32 &   43.90 &\\[0.5ex]
 270&     120A& 10 53 09.28& +57 28 22.65&  AGNI& 1.568&    SPL+SE&$  2.17_{  0.09}^{  0.06}$&$         _{         }^{         }$& 0.80& 0.72&   44.32 &   44.12 &\\[0.5ex]
 267&     428E& 10 53 24.54& +57 28 20.65&  AGNI& 1.518&       APL&$  1.74_{  0.10}^{  0.10}$&$    21.27_{     0.40}^{     0.12}$& 0.53& 0.97&   43.81 &   44.05 &\\[0.5ex]
 287&     821A& 10 53 22.04& +57 28 52.76&  AGNI& 2.300&       SPL&$  1.67_{  0.13}^{  0.14}$&$         _{         }^{         }$& 0.20& 0.39&   43.74 &   44.02 &\\[0.5ex]
 268&        -& 10 53 48.09& +57 28 17.75&    --&     -&       APL&$  1.83_{  0.20}^{  0.19}$&$    21.81_{     0.11}^{     0.13}$& 0.24& 1.21&         &         &\\[0.5ex]
 277&      25A& 10 53 44.85& +57 28 42.24&  AGNI& 1.816&       SPL&$  1.74_{  0.04}^{  0.04}$&$         _{         }^{         }$& 1.18& 2.03&   44.31 &   44.55 &\\[0.5ex]
 272&      26A& 10 50 19.40& +57 28 13.99& AGNII& 0.616&       APL&$  2.39_{  0.17}^{  0.17}$&$    21.58_{     0.13}^{     0.10}$& 1.39& 1.38&   43.62 &   43.43 &\\[0.5ex]
 290&     901A& 10 52 52.74& +57 29 00.81& AGNII& 0.204&    APL+SE&$  1.86_{  0.22}^{  0.39}$&$    23.60_{     0.14}^{     0.07}$& 0.14& 1.79&   42.69 &   42.83 &\\[0.5ex]
 369&        -& 10 51 06.50& +57 15 31.92&    --&     -&       SPL&$  1.96_{  0.11}^{  0.11}$&$         _{         }^{         }$& 1.99& 2.46&         &         &\\[0.5ex]
 300&     426A& 10 53 03.64& +57 29 25.56&  AGNI& 0.788&      CAPL&$  1.98_{  0.15}^{  0.18}$&$    22.09_{     0.22}^{     0.16}$& 0.50& 0.87&   43.32 &   43.40 &\\[0.5ex]
 306&        -& 10 52 06.84& +57 29 25.43& AGNII& 0.708&       APL&$  1.38_{  0.20}^{  0.24}$&$    22.20_{     0.17}^{     0.17}$& 0.16& 1.03&   42.76 &   43.23 &\\[0.5ex]
 321&      23A& 10 52 24.74& +57 30 11.40&  AGNI& 1.009&       SPL&$  1.70_{  0.08}^{  0.08}$&$         _{         }^{         }$& 0.39& 0.71&   43.22 &   43.48 &\\[0.5ex]
 326&     117Q& 10 53 48.80& +57 30 36.09& AGNII&  0.78&       APL&$  1.53_{  0.05}^{  0.06}$&$    21.96_{     0.06}^{     0.06}$& 1.38& 5.49&   43.71 &   44.08 &\\[0.5ex]
 350&        -& 10 52 41.65& +57 30 39.97&    --&     -&       SPL&$  1.28_{  0.14}^{  0.14}$&$         _{         }^{         }$& 0.14& 0.50&         &         &\\[0.5ex]
 332&      77A& 10 52 59.16& +57 30 31.81&  AGNI& 1.676&       SPL&$  1.62_{  0.04}^{  0.04}$&$         _{         }^{         }$& 1.08& 2.22&   44.14 &   44.45 &\\[0.5ex]
 411&      53A& 10 52 06.02& +57 15 26.41& AGNII& 0.245&      CAPL&$  1.63_{  0.19}^{  0.22}$&$    22.18_{     0.16}^{     0.13}$& 0.85& 4.84&   42.63 &   42.94 &\\[0.5ex]
2024&        -& 10 54 10.68& +57 30 56.73&    --&     -&       SPL&$  2.02_{  0.07}^{  0.07}$&$         _{         }^{         }$& 0.82& 0.93&         &         &\\[0.5ex]
 343&        -& 10 50 41.22& +57 30 23.31&    --&     -&       SPL&$  1.54_{  0.18}^{  0.19}$&$         _{         }^{         }$& 0.44& 1.01&         &         &\\[0.5ex]
 342&      16A& 10 53 39.62& +57 31 04.89&  AGNI& 0.586&    SPL+SE&$  2.26_{  0.05}^{  0.04}$&$         _{         }^{         }$& 4.84& 3.81&   43.92 &   43.78 &\\[0.5ex]
 351&        -& 10 51 46.39& +57 30 38.14&    --&     -&       SPL&$  1.87_{  0.14}^{  0.15}$&$         _{         }^{         }$& 0.21& 0.30&         &         &\\[0.5ex]
 353&      19B& 10 51 37.27& +57 30 44.43&  AGNI& 0.894&       SPL&$  1.72_{  0.10}^{  0.10}$&$         _{         }^{         }$& 0.40& 0.71&   43.12 &   43.37 &\\[0.5ex]
 354&      75A& 10 51 25.25& +57 30 52.33&  AGNI& 3.409&       SPL&$  1.75_{  0.07}^{  0.08}$&$         _{         }^{         }$& 0.71& 1.20&   44.71 &   44.94 &\\[0.5ex]
 358&      17A& 10 51 03.86& +57 30 56.65&  AGNI& 2.742&       SPL&$  1.78_{  0.07}^{  0.08}$&$         _{         }^{         }$& 1.13& 1.83&   44.72 &   44.93 &\\[0.5ex]
 355&        -& 10 52 37.33& +57 31 06.67& AGNII& 0.708&       APL&$  1.60_{  0.12}^{  0.12}$&$    21.17_{     0.28}^{     0.17}$& 0.36& 0.86&   42.87 &   43.19 &\\[0.5ex]
 385&      14Z& 10 52 42.37& +57 32 00.64& AGNII& 1.380&       APL&$  1.83_{  0.08}^{  0.08}$&$    22.33_{     0.04}^{     0.04}$& 0.76& 2.10&   44.16 &   44.33 &\\[0.5ex]
 364&      18Z& 10 52 28.36& +57 31 06.57&  AGNI& 0.931&       SPL&$  1.79_{  0.13}^{  0.13}$&$         _{         }^{         }$& 0.23& 0.37&   42.94 &   43.14 &\\[0.5ex]
 901&        -& 10 50 05.55& +57 31 09.01&    --&     -&       SPL&$  1.64_{  0.12}^{  0.14}$&$         _{         }^{         }$& 0.79& 1.58&         &         &\\[0.5ex]
 902&      73C& 10 50 09.12& +57 31 46.29&  AGNI& 1.561&       SPL&$  1.50_{  0.14}^{  0.14}$&$         _{         }^{         }$& 0.78& 1.94&   43.88 &   44.28 &\\[0.5ex]
 377&        -& 10 52 52.11& +57 31 38.02&    --&     -&       APL&$  1.81_{  0.21}^{  0.21}$&$    21.87_{     0.12}^{     0.11}$& 0.17& 0.92&         &         &\\[0.5ex]
 384&        -& 10 53 21.63& +57 31 49.44&    --&     -&       APL&$  2.22_{  0.26}^{  0.19}$&$    21.33_{     0.17}^{     0.11}$& 0.26& 0.40&         &         &\\[0.5ex]
 387&      15A& 10 52 59.78& +57 31 56.69&  AGNI& 1.447&       SPL&$  2.27_{  0.11}^{  0.11}$&$         _{         }^{         }$& 0.37& 0.28&   43.78 &   43.67 &\\[0.5ex]
 394&        -& 10 52 51.40& +57 32 02.03& AGNII& 0.664&       APL&$  1.85_{  0.22}^{  0.22}$&$    21.24_{     0.45}^{     0.24}$& 0.17& 0.29&   42.55 &   42.71 &\\[0.5ex]
 406&     828A& 10 53 57.16& +57 32 44.00&  AGNI& 1.282&       SPL&$  2.12_{  0.10}^{  0.10}$&$         _{         }^{         }$& 0.62& 0.60&   43.82 &   43.80 &\\[0.5ex]
 419&        -& 10 54 00.46& +57 33 22.19&    --&     -&       APL&$  1.81_{  0.17}^{  0.23}$&$    21.62_{     0.11}^{     0.12}$& 0.41& 1.54&         &         &\\[0.5ex]
 407&      12A& 10 51 48.69& +57 32 50.07& AGNII& 0.990&      CAPL&$  1.25_{  0.10}^{  0.10}$&$    22.69_{     0.11}^{     0.12}$& 0.57& 4.89&   43.64 &   44.20 &\\[0.5ex]
 424&        -& 10 52 37.93& +57 33 22.65& AGNII& 0.707&    APL+SE&$  1.70_{  0.18}^{  0.32}$&$    22.95_{     0.12}^{     0.12}$& 0.22& 2.36&   43.50 &   43.74 &\\[0.5ex]
 427&        -& 10 52 27.88& +57 33 30.65& AGNII& 0.696&       SPL&$  1.40_{  0.20}^{  0.20}$&$         _{         }^{         }$& 0.14& 0.39&   42.33 &   42.79 &\\[0.5ex]
 430&      11A& 10 51 08.19& +57 33 47.06&  AGNI& 1.540&       APL&$  1.42_{  0.08}^{  0.08}$&$    21.58_{     0.19}^{     0.13}$& 1.24& 3.83&   44.09 &   44.53 &\\[0.5ex]
 458&        -& 10 51 06.22& +57 34 36.67&    --&     -&      CAPL&$  1.34_{  0.16}^{  0.39}$&$    22.31_{     0.20}^{     0.16}$& 0.28& 4.43&         &         &\\[0.5ex]
 442&     805A& 10 53 47.28& +57 33 50.41&  AGNI& 2.586&       SPL&$  1.76_{  0.14}^{  0.15}$&$         _{         }^{         }$& 0.33& 0.55&   44.12 &   44.34 &\\[0.5ex]
 443&        -& 10 52 36.89& +57 33 59.80& AGNII& 1.877&       APL&$  2.05_{  0.34}^{  0.42}$&$    23.22_{     0.18}^{     0.16}$& 0.17& 0.97&   44.42 &   44.46 &\\[0.5ex]
 474&        -& 10 51 28.13& +57 35 04.20&    --&    --&       SPL&$  1.98_{  0.11}^{  0.11}$&$         _{         }^{         }$& 0.47& 0.56&         &         &\\[0.5ex]
 451&        -& 10 52 07.87& +57 34 17.48&    --&     -&       APL&$  1.94_{  0.21}^{  0.18}$&$    21.31_{     0.15}^{     0.11}$& 0.29& 0.62&         &         &\\[0.5ex]
 450&     477A& 10 53 05.98& +57 34 26.70&  AGNI& 2.949&       SPL&$  2.03_{  0.10}^{  0.10}$&$         _{         }^{         }$& 0.44& 0.49&   44.53 &   44.57 &\\[0.5ex]
 453&     804A& 10 53 12.24& +57 34 27.39&  AGNI& 1.213&       SPL&$  2.05_{  0.14}^{  0.14}$&$         _{         }^{         }$& 0.31& 0.34&   43.44 &   43.47 &\\[0.5ex]
 456&       9A& 10 51 54.30& +57 34 38.66&  AGNI& 0.877&       SPL&$  2.01_{  0.05}^{  0.05}$&$         _{         }^{         }$& 1.25& 1.44&   43.68 &   43.74 &\\[0.5ex]
 491&        -& 10 51 41.91& +57 35 56.00&    --&     -&       APL&$  2.18_{  0.25}^{  0.23}$&$    21.24_{     0.25}^{     0.15}$& 0.33& 0.48&         &         &\\[0.5ex]
 469&        -& 10 54 07.21& +57 35 24.89&    --&     -&       SPL&$  2.14_{  0.04}^{  0.04}$&$         _{         }^{         }$& 3.86& 3.62&         &         &\\[0.5ex]
 475&       6A& 10 53 16.51& +57 35 52.23&  AGNI& 1.204&       SPL&$  1.78_{  0.02}^{  0.02}$&$         _{         }^{         }$&10.21&16.61&   44.85 &   45.06 &\\[0.5ex]
 476&     827A& 10 53 03.43& +57 35 30.80& AGNII& 0.607&       SPL&$  2.07_{  0.17}^{  0.17}$&$         _{         }^{         }$& 0.27& 0.29&   42.64 &   42.66 &\\[0.5ex]
 505&     104A& 10 52 41.54& +57 36 52.85& AGNII& 0.137&      CAPL&$  1.72_{  0.08}^{  0.09}$&$    21.69_{     0.08}^{     0.11}$& 1.36& 3.96&   42.04 &   42.29 &\\[0.5ex]
 504&        -& 10 54 26.22& +57 36 49.05&    --&     -&    APL+SE&$  1.98_{  0.10}^{  0.19}$&$    22.66_{     0.06}^{     0.06}$& 1.01&42.52&         &         &\\[0.5ex]
 511&        -& 10 53 38.50& +57 36 55.47& AGNII& 0.704&    APL+SE&$  2.62_{  0.29}^{  0.73}$&$    22.81_{     0.14}^{     0.15}$& 0.19& 0.96&   43.97 &   43.58 &\\[0.5ex]
 518&        -& 10 53 36.33& +57 37 32.14&    --&    --&       APL&$  2.38_{  0.19}^{  0.52}$&$    21.09_{     0.20}^{     0.14}$& 0.48& 0.47&         &         &\\[0.5ex]
 523&        -& 10 51 29.98& +57 37 40.71&    --&     -&       SPL&$  1.94_{  0.09}^{  0.09}$&$         _{         }^{         }$& 0.92& 1.17&         &         &\\[0.5ex]
 529&        -& 10 51 37.30& +57 37 59.11&    --&     -&      CAPL&$  1.98_{  0.21}^{  0.20}$&$    21.93_{     0.20}^{     0.23}$& 1.51& 2.88&         &         &\\[0.5ex]
 532&     801A& 10 52 45.36& +57 37 48.69&  AGNI& 1.677&       SPL&$  2.32_{  0.20}^{  0.21}$&$         _{         }^{         }$& 0.22& 0.16&   43.74 &   43.60 &\\[0.5ex]
 527&       5A& 10 53 02.34& +57 37 58.62&  AGNI& 1.881&       SPL&$  1.91_{  0.05}^{  0.06}$&$         _{         }^{         }$& 1.13& 1.50&   44.40 &   44.53 &\\[0.5ex]
 537&        -& 10 50 50.04& +57 38 21.79&    --&     -&       SPL&$  2.40_{  0.08}^{  0.08}$&$         _{         }^{         }$& 3.04& 1.95&         &         &\\[0.5ex]
 548&     832A& 10 52 07.53& +57 38 41.40&  AGNI**& 2.730&       SPL&$  1.79_{  0.14}^{  0.15}$&$         _{         }^{         }$& 0.30& 0.48&   44.15 &   44.13 &\\[0.5ex]
 557&        -& 10 52 07.75& +57 39 07.49&    --&     -&       APL&$  2.01_{  0.17}^{  0.36}$&$    20.94_{     0.45}^{     0.22}$& 0.33& 0.49&         &         &\\[0.5ex]
 553&       2A& 10 52 30.06& +57 39 16.81&  AGNI& 1.437&       APL&$  2.08_{  0.04}^{  0.04}$&$    21.20_{     0.14}^{     0.12}$& 2.87& 3.14&   44.62 &   44.64 &\\[0.5ex]
 555&        -& 10 51 52.07& +57 39 09.41&    --&     -&       SPL&$  1.84_{  0.12}^{  0.12}$&$         _{         }^{         }$& 0.49& 0.72&         &         &\\[0.5ex]
 594&        -& 10 52 48.40& +57 41 29.14&    --&     -&       SPL&$  1.59_{  0.24}^{  0.24}$&$         _{         }^{         }$& 0.38& 0.82&         &         &\\[0.5ex]
2045&        -& 10 52 04.47& +57 41 15.65&    --&     -&       SPL&$  1.89_{  0.09}^{  0.09}$&$         _{         }^{         }$& 0.94& 1.28&         &         &\\[0.5ex]
 584&        -& 10 52 06.28& +57 41 25.53&    --&     -&       SPL&$  2.17_{  0.08}^{  0.07}$&$         _{         }^{         }$& 1.61& 1.45&         &         &\\[0.5ex]
 591&        -& 10 52 23.17& +57 41 24.62&    --&    --&       SPL&$  2.08_{  0.14}^{  0.14}$&$         _{         }^{         }$& 0.59& 0.61&         &         &\\[0.5ex]
 601&        -& 10 51 15.91& +57 42 08.59&    --&     -&       SPL&$  1.84_{  0.19}^{  0.20}$&$         _{         }^{         }$& 1.30& 1.92&         &         &\\[0.5ex]
\hline
\hline
\label{ress_fits_all}
\end{longtable}
Columns are as follows: (1) XMM-Newton identification number; (2) {\it ROSAT}
identification number; (3) Right ascension (J2000); (4) Declination (J2000); 
(5) Optical class from optical spectroscopy; (6) Source redshift;
(7) Best fit model of the X-ray spectrum of each object: SPL: single power law, APL:
absorbed power law, SE: Soft-excess emission, CAPL: absorption with partial covering absorber, 2SPL:
two power laws;
(8) Slope of the broad band continuum emission; (9) Logarithm of the
hydrogen column density of the X-ray absorber (observed ${\rm N_H}$ 
if the source is still unidentified or rest-frame absorption if the source is 
identified; 
(10) and (11) 0.5-2 and 2-10 keV flux in units of ${\rm 10^{-14}\,erg\,cm^{-2}\,s^{-1}}$ obtained from the best fit model; 
(12) and (13) logarithm of the 0.5-2 and 2-10 keV luminosities (corrected for 
absorption) for the identified sources 
      obtained from the best fit model. 
\\
\\
The source coordinates RA and DEC, are not the X-ray positions from the 
{\it Lockman Hole} catalogue (H. Brunner et al. 2005, in preparation), 
but the centres of the regions used to extract the spectra of the objects.\\
$^{*}$ Source photometric redshift.
\\
$^{**}$ BAL QSO
\end{landscape}
\twocolumn

\begin{acknowledgements}
SM acknowledges support from a Universidad de Cantabria fellowship.
XB, FJC and MTC acknowledge financial support from the Spanish Ministerio
de Educaci\'on y Ciencia, under project ESP2003-00812.
Some of the data presented here were obtained at the W. M. Keck Observatory, 
which is operated as a scientific partnership among the California Institute 
of Technology, the University of California, and the National Aeronautics and 
Space Administration. The Observatory was made possible by the generous financial 
support of the W. M. Keck Foundation. We also thank the anonymous referee for his/her suggestions that helped to improve the manuscript considerably.
\end{acknowledgements}

\appendix
\section{Stacking of spectra of {\it Lockman Hole} sources}
\label{apendix_A}
To calculate the contribution to the Cosmic X-ray background from our sources, MOS and 
pn spectra have been stacked.

We have kept MOS and pn 
data separately because of the different instrumental responses and to provide support to the results. 
In addition, as it was explained in Sec.~\ref{X-ray spectral analysis}, MOS and pn time averaged spectra 
were not necessary built using the same set of observations, hence the normalisations of the XRB spectrum
obtained with MOS and pn might be slightly different. 

At the time of this study there were calibration uncertainties 
between MOS and pn data at energies below $\sim$1 keV (pn gives higher fluxes below 0.7 keV by 10-15\% with 
respect to MOS, see Kirsh et al.~\cite{Kirsh2004}). We also found our data to be rather uncertain 
at energies $\ge$ 7keV. Because of the rapid decrease in the effective area of 
the X-ray detectors at energies $\ge$ 5keV the signal to noise of the data becomes very low. 
In addition we know that the particle background is very important at high energies.
Therefore we restricted our analysis to the 2-7 keV energy interval where we know that 
our results will not be affected by calibration problems or inaccurate background subtraction.

MOS and pn stacked spectra were obtained with the following procedure: 
suppose we want to add two spectra having exposure times {\tt t1} and {\tt t2}, backscales 
{\tt b1} and {\tt b2} and response matrices {\tt rsp1} and {\tt rsp2}, and that  
the corresponding background extraction regions have exposure 
times {\tt tb1} and {\tt tb2} and backscales {\tt bb1} and {\tt bb2}:

\begin{enumerate}
\item Add spectra: The total backscale will be {\tt b1+b2}\footnote{Note 
that we want to extract the XRB from the regions where we have the sources, 
    and therefore, we are accumulating counts in an increasing solid angle.} and the total 
exposure time {\tt (t1$\times$b1+t2$\times$b2)/(b1+b2)}.
\item Add background files: The total backscale will be {\tt bb1+bb2} and the total 
exposure time {\tt (tb1$\times$bb1+tb2$\times$bb2)/(bb1+bb2)}.
\item Add response matrices. The combined response matrix will be 
  {\tt (t1$\times$b1)$\times$rsp1+(t2$\times$b2)$\times$rsp2/(t1$\times$b1+t2$\times$b2)}.
\end{enumerate}
This procedure can be trivially extended to stack any number of sources.

\end{document}